\documentclass[twocolumn]{aastex631}
\usepackage{color}
\usepackage{amsmath}

\newcommand{\Ha}{\makebox{H$\alpha$}}
\newcommand{\Hb}{\makebox{H$\beta$}}

\newcommand{\HeI}{He\,{\sc i}}

\newcommand{\halpha}{H{$\alpha$}}

\def \OIII {[O\,{\sc iii}]}
\newcommand{\OIIIa}{[O{\sevenrm\,III}]\,$\lambda$4959}
\newcommand{\OIIIb}{[O{\sevenrm\,III}]\,$\lambda$5007}
\newcommand{\OIIIab}{[O{\sevenrm\,III}]\,$\lambda\lambda$4959,5007}

\newcommand{\NIIa}{[N{\sevenrm\,II}]\,$\lambda$6548}
\newcommand{\NIIb}{[N{\sevenrm\,II}]\,$\lambda$6584}

   \font\sevenrm=cmr7 scaled 1000
\newcommand{\sersic}{S\'{e}rsic}

\newcommand{\SIII}{[S{\sevenrm III}]}

\usepackage{soul}

\begin{document}

\title{NEXUS: A Spectroscopic Census of Broad-line AGNs and Little Red Dots at $3\lesssim z\lesssim 6$}

\author[0000-0001-5105-2837]{Ming-Yang Zhuang}
\email{mingyang@illinois.edu}
\affiliation{Department of Astronomy, University of Illinois Urbana-Champaign, Urbana, IL 61801, USA}

\author[0000-0002-1605-915X]{Junyao Li}
\affiliation{Department of Astronomy, University of Illinois Urbana-Champaign, Urbana, IL 61801, USA}

\author[0000-0003-1659-7035]{Yue Shen}
\email{shenyue@illinois.edu}
\affiliation{Department of Astronomy, University of Illinois Urbana-Champaign, Urbana, IL 61801, USA}
\affiliation{National Center for Supercomputing Applications, University of Illinois Urbana-Champaign, Urbana, IL 61801, USA}

\author[0000-0001-6052-4234]{Xiaojing Lin}
\affiliation{Department of Astronomy, Tsinghua University, Beijing 100084, China}
\affiliation{Steward Observatory, University of Arizona, 933 N Cherry Ave, Tucson, AZ 85721, USA}

\author[0000-0003-3509-4855]{Alice E.~Shapley}
\affiliation{Department of Physics \& Astronomy, University of California, Los Angeles, 430 Portola Plaza, Los Angeles, CA 90095, USA}

\author[0000-0002-7633-431X]{Feige Wang}
\affiliation{Department of Astronomy, University of Michigan, 1085 S. University Ave., Ann Arbor, MI 48109, USA}

\author[0000-0003-4202-1232]{Qiaoya Wu}
\affiliation{Department of Astronomy, University of Illinois Urbana-Champaign, Urbana, IL 61801, USA}

\author[0000-0002-6893-3742]{Qian Yang}
\affiliation{Center for Astrophysics $\vert$ Harvard \& Smithsonian, 60 Garden Street, Cambridge, MA 02138, USA}

\begin{abstract}

We present a spectroscopic sample of 23 broad-line AGNs (BLAGNs) at $3\lesssim z\lesssim 6$ selected using F322W2+F444W NIRCam/WFSS grism spectroscopy of the central 100\,${\rm arcmin^2}$ area of the NEXUS survey. Among these BLAGNs, 15 are classified as Little Red Dots (LRDs) based on their rest-frame UV-optical spectral slopes and compact morphology. The number density of LRDs is $\sim 10^{-5}\,{\rm cMpc^{-3}}$, with a hint of declining towards the lower end of the probed redshift range. These BLAGNs and LRDs span broad \halpha\ luminosities of $\sim 10^{42.2}-10^{43.7}\,{\rm erg\,s^{-1}}$, black hole masses of $\sim 10^{6.3}-10^{8.4}\,M_\odot$, and Eddington ratios of $\sim 0.1-1$ (median value 0.4), though the black hole mass and Eddington ratio estimates carry large systematic uncertainties. Half of the LRDs show strong Balmer absorption, suggesting high-density gas surrounding the line-emitting region. We detect extended (hundreds of parsec) rest-frame UV-optical emission from the host galaxy in the majority of these LRDs, which contributes significantly or even dominantly to their total UV emission. This host emission largely accounts for the peculiar UV upturn of the LRD spectral energy distribution. We also measure the small-scale ($\lesssim 1\,{\rm cMpc}$) clustering of these BLAGNs and LRDs by cross-correlating with a photometric galaxy sample. Extrapolating the power-law two-point correlation function model to large linear scales, we infer a linear bias of {$3.30_{-2.04}^{+2.88}$} and typical halo masses of a few $\times 10^{11}\,h^{-1}M_\odot$ for BLAGNs at the sample median redshift of $z\sim 4.5$. However, the inferred linear bias and halo masses of LRDs, while formally consistent with those for BLAGNs at $\sim 1.5\sigma$, appear too large to be compatible with their space density, suggesting LRDs may have strong excess clustering on small scales. 
\end{abstract}


\section{Introduction} \label{sec:intro}

Recent JWST observations of the high-redshift (e.g., $z>3$) Universe have revealed a remarkable population of red compact galaxies with apparently broad Balmer emission lines with FWHM exceeding $\sim 1000\,{\rm km\,s^{-1}}$ \citep[e.g.,][]{Labbe+2025ApJ, Matthee+24,Greene+2024, Kocevski_LRD_selection, Hainline+2025_LRD_selection, Leung2024, WangBJ2025lrd, Labbe2024, Kokorev2024z4LRD, Taylor+2024, ZhangJY2025, DEugenio2025, Taylor2025, Killi2024}, dubbed ``little red dots'' (LRDs). The nature of LRDs is not fully understood, but a popular view is that most of them are accreting supermassive black holes (SMBHs), given the broad line widths. Albeit with significant uncertainties of their bolometric luminosity estimates, LRDs seem to dominate the population of broad-line AGNs (BLAGNs) with bolometric luminosities of $\sim10^{44} \sim 10^{46}\ \rm erg\ s^{-1}$ \citep{Greene+2024, ZhangJY2025}, and are much more abundant than luminous quasars in the same redshift range.  

In addition to their compact morphology, LRDs are characterized by a peculiar spectral energy distribution (SED) shape, with bending rest-frame UV and optical slopes ($\beta_{\rm UV}$ and $\beta_{\rm opt}$). This SED shape is likely due to a combination of the following factors: intrinsic properties of the accreting SMBH, strong dust reddening or gas-enshrouded black hole accretion in the nucleus region, star formation in the host galaxy, and scattered blue AGN continuum \citep{Casey2024, Greene+2024, Perez-Gonzalez+2024, Rinaldi+2024, Setton2024, Chen+2025a, Inayoshi&Maiolino2025, LiZR2025, Naidu_BHstar_2025, Volonteri2025, WangBJ2025lrd, Naidu_BHstar_2025}. Their compact sizes generally imply small host galaxies. The black hole mass estimates of LRDs currently carry large systematic uncertainties, as they are estimated using broad \halpha\ single-epoch virial black hole mass estimators based on more massive, and lower-redshift AGNs \citep{Greene_Ho_2005,Shen_2013, Bertemes2025}. Nevertheless, these single-epoch mass estimates suggest LRDs contain SMBHs with masses of $\sim 10^5-10^{8}\,M_\odot$ and Eddington ratios of $\sim 0.1-1$, which place LRDs significantly above the local BH mass--stellar mass ($M_{\rm BH}-M_{*}$) relation \citep{Harikane+2023, Pacucci2023, Maiolino+2024_JADES_BLAGN, Chen+2025a, Li2025a, Li_Ruancun+2025, SunY2025}. If these BH masses are reliable, it would imply LRDs are growing SMBHs that are overmassive relative to their host galaxy stellar mass, making them good candidates to support the heavy seed scenario of SMBH formation at high redshift \citep{Bogdan2024, Natarajan2024}. However, it is unclear if LRDs are a representative population of growing SMBHs at high redshift, as flux-limited pencil-beam JWST surveys would preferentially miss lower-luminosity, and hence undermassive black holes \citep[e.g.,][]{Lauer_2007,Li2025a,Li2025b}. 

Two recent studies (\citealt{Rusakov2025, Naidu_BHstar_2025}) further argued, based on the profile of the broad Balmer lines, that the broadening of the Balmer lines in LRDs could be significantly complicated by scattering (either electron scattering \citep{Laor_2006} or Balmer scattering). This would imply the virial BH masses in LRDs are severely overestimated by orders of magnitude. These arguments are countered in \citet{JADES_BLAGN} based on the observed line profiles in both Balmer and Paschen lines in an example LRD. Regardless, the Balmer line profile in LRDs is often \citep[e.g., $>30\%$;][]{Lin+2024} observed to be asymmetric and show absorption, which would normally imply high gas density in order to produce Balmer absorption \citep{Hall_2007,Inayoshi&Maiolino2025, deGraaff2025}. A coherent interpretation is yet to emerge to explain the peculiar line profiles of LRDs, compared with normal broad-line AGNs. 

There are also studies that question the AGN nature of LRDs \citep{Baggen2024, Kokubo2024, Bellovary2025}. One argument is the lack of variability of LRDs using multi-epoch observations \citep{Kokubo2024, Tee2025}. However, there are also reported cases of LRDs either varying in Balmer line equivalent width \citep{Furtak2025} or in continuum \citep{ZhangZJ2024}. Currently time-domain observations of LRDs are still limited in baseline and sample statistics, preventing a definitive test for the general LRD population. This variability argument needs to be revisited with more time-domain data on LRDs \citep{nexus}. Another argument is based on the lack of X-ray \citep{Yue2024xray, Ananna2024} and radio detection \citep{Perger2025}. However, given the unconstrained accretion flow properties and nuclear obscuration of LRDs, it is difficult to use the X-ray argument to rule out accreting SMBHs for LRDs \citep{Inayoshi2024xray, Madau2024, Lambrides2024, Maiolino2025}.   

If LRDs are accreting SMBHs in nature, they clearly represent a population of significant interest to many key questions regarding the seeding scenarios of SMBHs, the co-evolution of SMBHs and galaxies, and the physics of accretion in the high-redshift Universe \citep{Bhowmick2024, Jeon2025, Inayoshi2025firstbh}. In this work, we contribute to this topic using photometric and spectroscopic data from the NEXUS program \citep{nexus}. As a multi-cycle Treasury JWST program, NEXUS is obtaining cadenced JWST NIRCam imaging and WFSS spectroscopy over a contiguous $\sim 0.1\,{\rm deg^2}$ field around the North Ecliptic Pole (NEXUS-Wide) for three yearly epochs during 2024-2028. A central $\sim 50\,{\rm arcmin^2}$ area (NEXUS-Deep) will be further revisited by NIRCam imaging and NIRSpec MSA spectroscopy for 18 epochs with a cadence of $2$~months. The technical details of NEXUS are described in \citet{nexus}, and the first observations in NEXUS-Wide are described in \citet{nexus-edr}. Here we use the imaging and NIRCam/WFSS spectroscopy from the Early Data Release (EDR) of NEXUS to study the LRDs therein. The final NEXUS-Wide observations will produce a sample size $\sim 4$ times larger than the present sample. 

This paper is organized as follows. Section~\ref{sec2} describes the selection criteria of the BLAGN and LRD samples. Section~\ref{sec3} presents the properties of the BLAGNs, including the spectral properties of the LRDs, physical properties of the accreting SMBH, properties of their potential underlying host galaxy and close environment, abundance of BLAGNs, and their clustering properties. Section~\ref{sec4} discusses the nature of LRDs, including the origin of their rest-UV emission, and the implications of their measured abundance and clustering. We adopt a flat $\Lambda$CDM cosmology with $H_0=70\,{\rm km\,s^{-1}Mpc^{-1}}$, $\Omega_{\rm m}= 0.3$, and $\Omega_{\rm \Lambda}= 0.7$, and a \citet{Chabrier2003} initial mass function (IMF) for stellar population analysis. Comoving distances are in units of ${\rm cMpc}$ or $h^{-1} {\rm cMpc}$.

\section{Data and Samples}\label{sec2}

\subsection{NEXUS EDR Data}

NEXUS has two area/depth tiers (Wide and Deep), where NIRCam/WFSS spectroscopy is performed for the $\sim 0.1\,{\rm deg^2}$ Wide area and NIRSpec/MSA spectroscopy (with PRISM) is performed for the central $\sim 50\,{\rm arcmin^2}$ area \citep{nexus}. In addition to spectroscopy, NEXUS also obtains NIRCam and MIRI imaging in different filters depending on the tier. However, MIRI imaging only covers a fraction of the primary NEXUS footprint since it is conducted as coordinated parallel \citep{nexus}. While formally there are three annual epochs for the entire Wide area, the first Wide epoch is split in two observations \citep{nexus-edr}, with the first observation conducted in September 2024 that covers the central $\sim 100\,{\rm arcmin^2}$ of the Wide area.  

In this work we use data from the first partial epoch of the NEXUS-Wide observations \citep[NEXUS EDR,][]{nexus-edr}. These data include six-band NIRCam imaging (F090W, F115W, F150W, F200W, F356W, F444W), and F322W2+F444W WFSS spectroscopy for the central NEXUS-Wide area. WFSS is generally sensitive to line emitters but not the continuum given the faint continuum magnitude of high-redshift line emitters. Therefore we only focus on spectroscopic line emitters in this work. Nevertheless, we construct a photometric source catalog in NEXUS using all available photometry.

\subsection{Redshift Determination}
NEXUS-Wide NIRCam/WFSS observations utilize a combination of F322W2 and F444W filters in the long wavelength channel to cover a broad wavelength range from $\sim$2.4 to 5 micron with grism spectroscopy. In this paper, we make use of the coadded 1D spectra of each source produced by combining 1D spectra optimally extracted from individual 2D spectra presented in NEXUS early data release \citep{nexus-edr}. We generate initial line-only spectra by removing background and continuum using median filters \citep[see][for details]{ASPIRE}. These initial line-only spectra are fed to the package \texttt{unfold\_jwst} (Wang et~al., in preparation) for emission line identification and redshift determination by cross-correlating detected emission lines with a theoretical emission line template \citep{Lin+2024}. Robust spectroscopic redshifts ($z_{\rm spec}$) are obtained if (1) multiple emission lines ($\geq2$) are detected (5$\sigma$) and matched with the template; or (2) only one emission line is detected but consistent with the photometric redshift ($z_{\rm phot}$). 

The photometric redshift $z_{\rm phot}$ for NEXUS sources is derived using \texttt{EAZY} \citep{EAZY} with the \texttt{agn\_blue\_sfhz\_13} template set. In addition to the six-band NIRCam photometry from NEXUS, we also include Subaru Hyper Suprime-Cam (HSC) photometry from the HEROES survey \citep{HEROES}, which includes five broad filters $grizy$ and two narrow filters NB816 and NB921, in the photo-z estimation. Finally, we obtain 363 spectroscopically confirmed objects at $z>2.5$. The details of the full photometric and spectroscopic redshift samples will be presented in a forthcoming paper.

\begin{figure*}[t]
\centering
  \includegraphics[width=\textwidth]{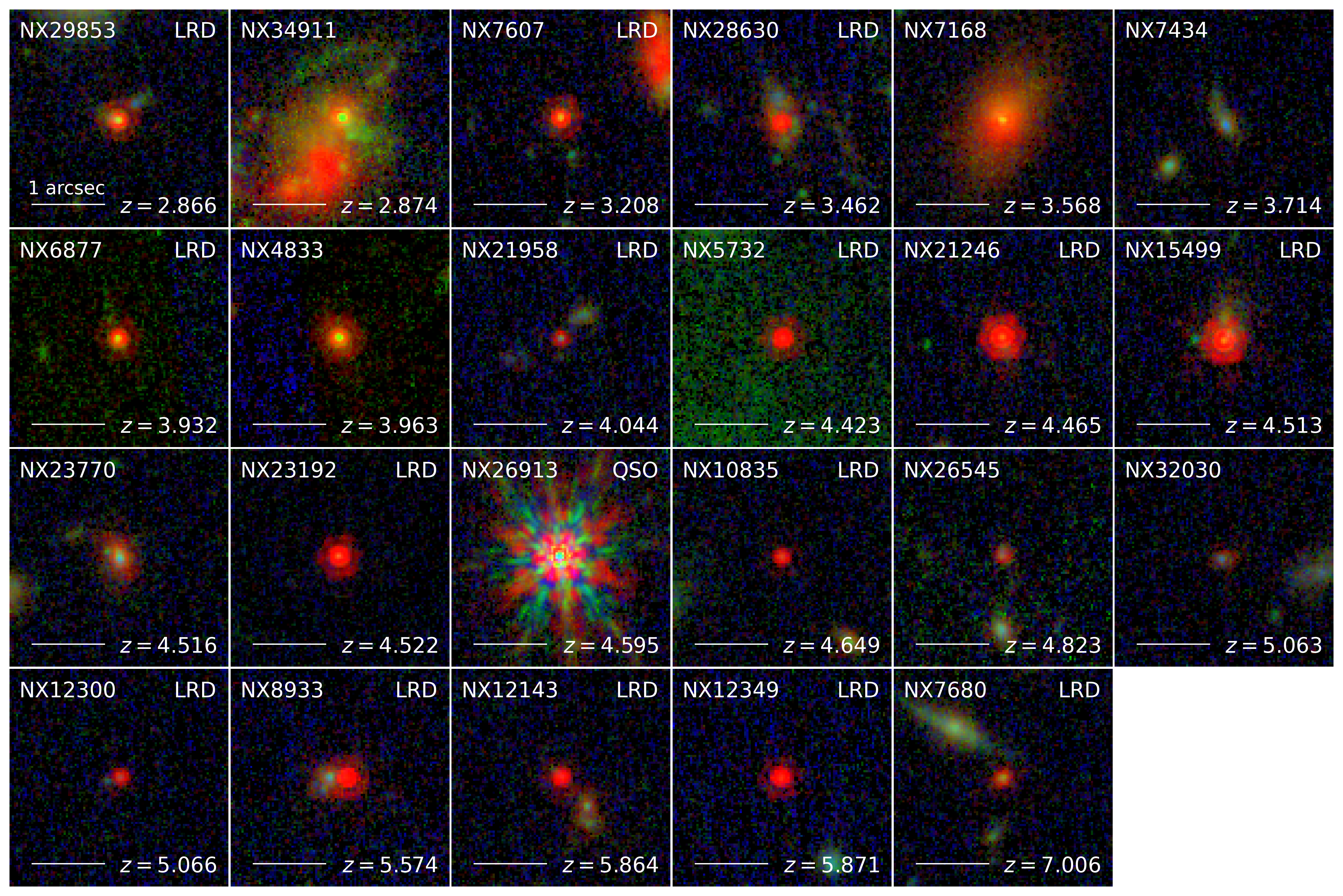}
  \caption{Composite color (F090W+F200W+F444W) stamps of broad-line emitters identified in NEXUS field using NIRCam/WFSS spectra in the central 100 arcmin$^2$ area. Each stamp has a size of 101 by 101 pixels (3\farcs03 by 3\farcs03), with source ID and redshift labeled at the top-left and bottom-right corners, respectively. A 1-arcsec scalebar is shown at the lower-left corner. Classified LRDs are labeled in the top-right corner. 
  \label{fig:BLE_stamp}}
\end{figure*}

\begin{figure*}[t]
\centering
\includegraphics[width=\textwidth]{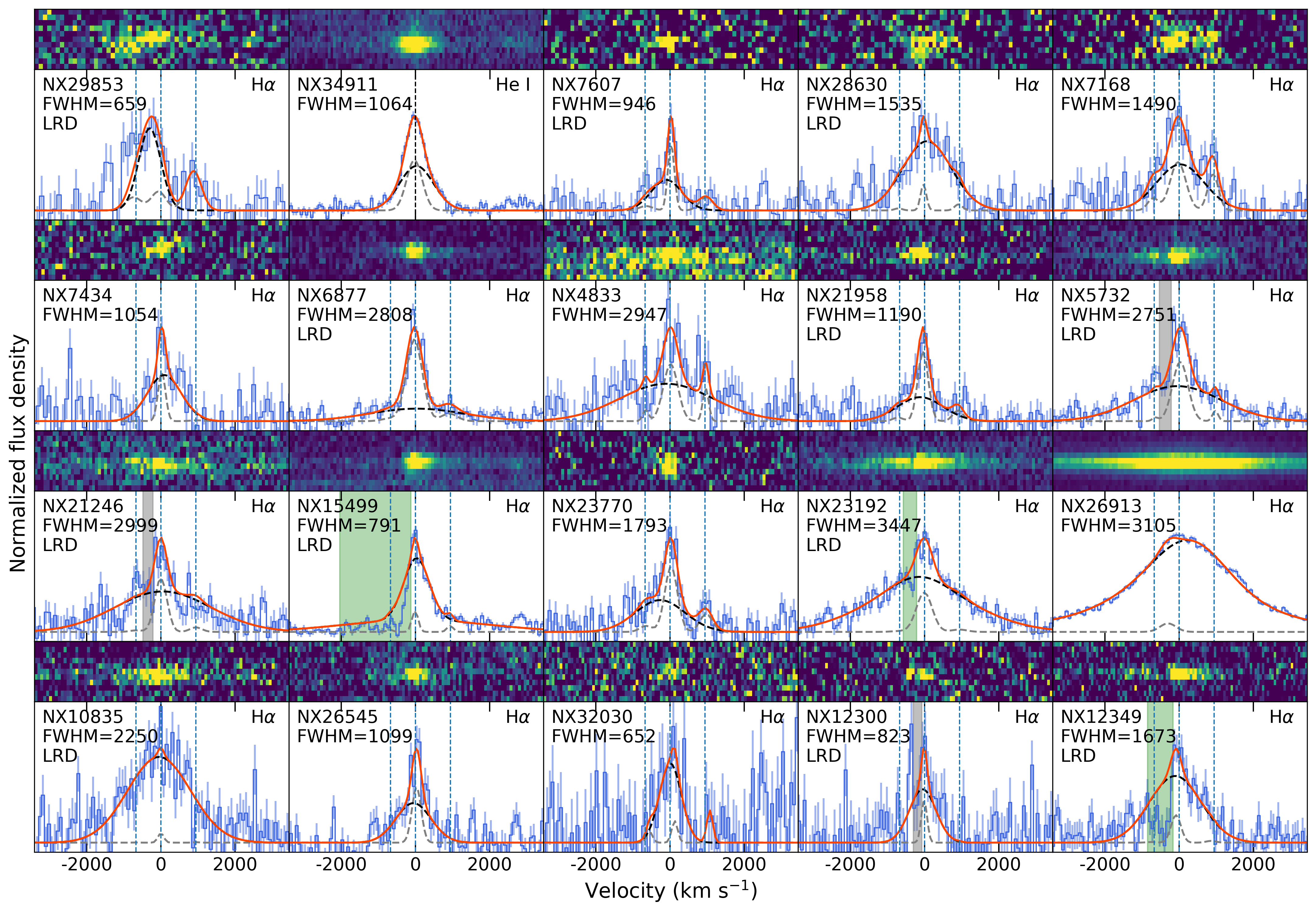}
\caption{2D and 1D NIRCam/WFSS grism spectra of broad-line emitters with background and continuum removed. Spectra are zoomed in around \Ha, except for NX34911, which is centered on \HeI. Errorbars indicate the 1$\sigma$ uncertainty of the data. Gray and black dashed curves represent best-fit narrow and broad components, respectively, while red solid curve represents the total emission lines (narrow + broad). Source ID and the FWHM of the broad component in unit of km s$^{-1}$ are shown in the upper-left corner. {We visually identify \Ha\ absorption features (shaded vertical bands) in six objects: NX5732, NX12300, NX12349, NX15499, NX21246, and NX23192, but only correct their impact on the fit in three objects (highlighted in green).} The extended blueward emission in NX29853 is masked during the fit. The systemic redshift of NX26913 is based on the peak of the overall \Ha\ profile.
\label{fig:BLE_spec}}
\end{figure*}

\begin{figure}[t]
\centering
  \includegraphics[width=0.45\textwidth]{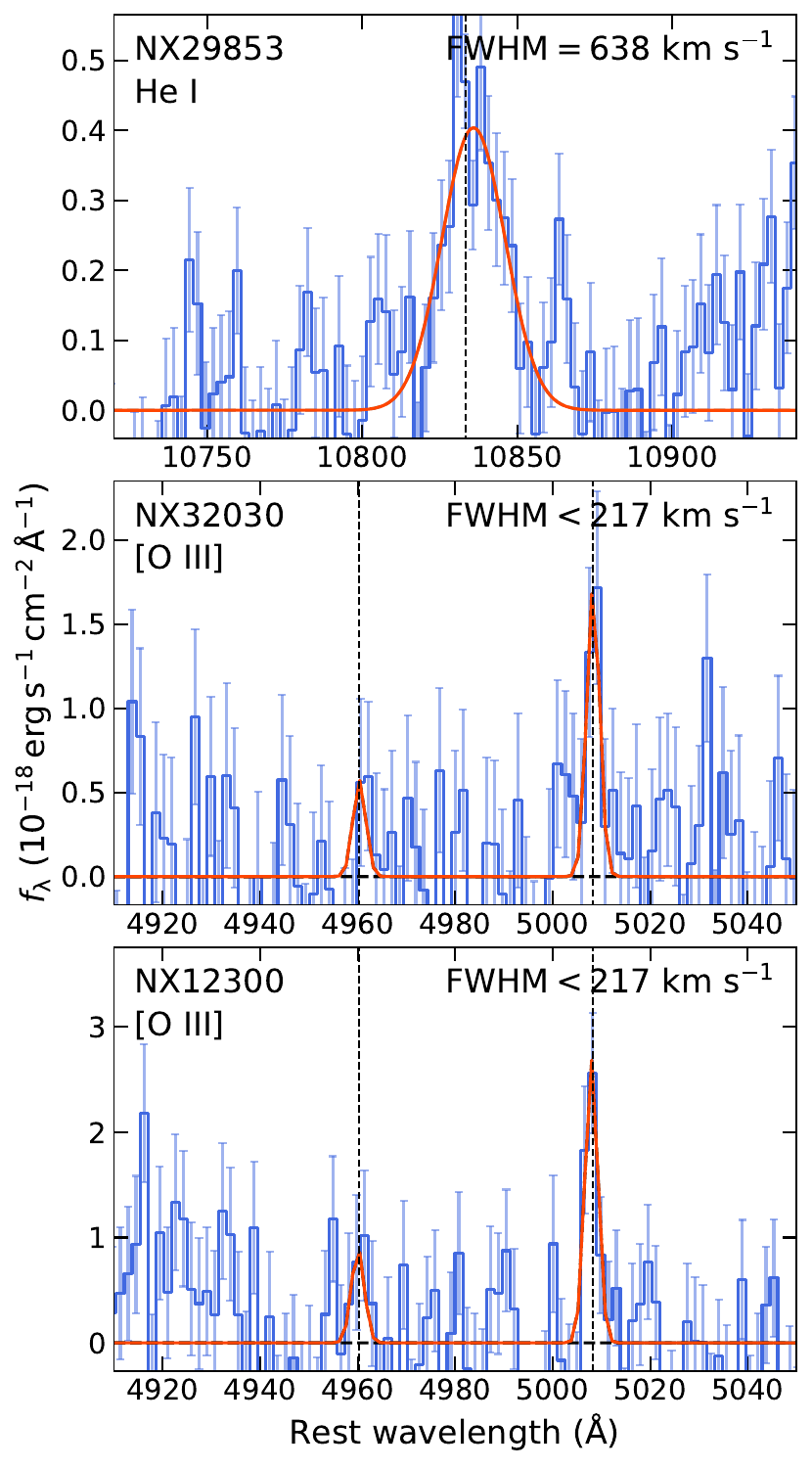}
  \caption{Additional emission lines detected in NIRCam/WFSS grism spectra of the three BLAGNs with FWHM$<900$ km s$^{-1}$. Symbols are the same as Figure~\ref{fig:BLE_spec}. The intrinsic line width of \OIII\ in NX32030 and NX12300 is unknown since it is smaller than instrumental broadening. 
  \label{fig:BLE_spec2}}
\end{figure}

\begin{deluxetable*}{lcccCCCCcClc}\label{tab1}
\tabletypesize{\scriptsize}
\caption{Broad-line Emitter Sample}
\centering
\tablehead{
\colhead{ID} & \colhead{R.A.} & \colhead{Decl.} & \colhead{$z_{\rm spec}$} & \colhead{$m_{\rm F444W}$} & \colhead{$\log\, L_{\rm broad}$} & \colhead{$f_{\rm broad}$} & \colhead{FWHM$_{\rm broad}$} & \colhead{Line} & \colhead{$\log\,M_{\rm BH}$} & \colhead{Type} & \colhead{Selection}\\
\nocolhead{} & \colhead{deg} & \colhead{deg} & \nocolhead{} & \colhead{AB mag} & \colhead{${\rm erg\,s^{-1}}$} & \colhead{$10^{-18}\,{\rm erg\,s^{-1}\,cm^{-2}}$} & \colhead{km\,s$^{-1}$} & \nocolhead{} & \colhead{$M_{\odot}$} & \nocolhead{} & \nocolhead{}\\
\colhead{(1)} & \colhead{(2)} & \colhead{(3)} & \colhead{(4)} & \colhead{(5)} & \colhead{(6)} & \colhead{(7)} & \colhead{(8)} & \colhead{(9)} & \colhead{(10)} & \colhead{(11)} & \colhead{(12)}
}
\startdata
NX29853 & 268.525957 & 65.227706 & 2.866 & 23.60 \pm 0.01 & 42.61 \pm 0.07 &   59.47 \pm 14.09 &  659 \pm 332 & H$\alpha$ & 6.49 & BLAGN (LRD) & 12\\
NX34911 & 268.428710 & 65.203777 & 2.874 & 22.23 \pm 0.01 & 42.33 \pm 0.03 &   30.65 \pm  2.36 & 1064 \pm  67 & \HeI & \nodata & BLAGN & 12 \\
NX7607  & 268.432893 & 65.147652 & 3.208 & 23.79 \pm 0.01 & 42.45 \pm 0.12 &   30.75 \pm  8.01 &  946 \pm 219 & H$\alpha$ & 6.73 & BLAGN (LRD) & 2 \\
NX28630 & 268.505621 & 65.233145 & 3.462 & 23.40 \pm 0.01 & 42.90 \pm 0.05 &   73.56 \pm  8.43 & 1535 \pm 307 & H$\alpha$ & 7.38 & BLAGN (LRD) & 12\\
NX7168  & 268.483161 & 65.147721 & 3.568 & 21.90 \pm 0.00 & 42.61 \pm 0.08 &   34.60 \pm  6.82 & 1490 \pm 406 & H$\alpha$ & 7.21 & BLAGN & 12 \\
NX7434  & 268.380805 & 65.146778 & 3.714 & 25.00 \pm 0.03 & 42.48 \pm 0.07 &   23.51 \pm  3.64 & 1054 \pm 274 & H$\alpha$ & 6.84 & BLAGN & 12 \\
NX6877  & 268.442554 & 65.144320 & 3.932 & 24.04 \pm 0.01 & 42.83 \pm 0.08 &   45.61 \pm  7.78 & 2808 \pm 539 & H$\alpha$ & 7.88 & BLAGN (LRD*) & 2 \\
NX4833  & 268.522059 & 65.135414 & 3.963 & 23.53 \pm 0.01 & 42.81 \pm 0.07 &   43.04 \pm  6.44 & 2947 \pm 282 & H$\alpha$ & 7.92 & BLAGN & 12 \\
NX21958 & 268.426066 & 65.267830 & 4.044 & 25.47 \pm 0.03 & 42.45 \pm 0.10 &   17.95 \pm  4.17 & 1190 \pm 285 & H$\alpha$ & 6.94 & BLAGN (LRD) & 1 \\
NX5732  & 268.541898 & 65.138808 & 4.423 & 23.67 \pm 0.01 & 43.17 \pm 0.02 &   75.90 \pm  4.33 & 2751 \pm 256 & H$\alpha$ & 8.03 & BLAGN (LRD) & 12 \\
NX21246 & 268.511349 & 65.270607 & 4.465 & 22.49 \pm 0.01 & 43.09 \pm 0.05 &  61.71 \pm  7.60 & 2999 \pm 354 & H$\alpha$ & 8.07 & BLAGN (LRD) & 12 \\
NX15499 & 268.486881 & 65.187833 & 4.513 & 22.21 \pm 0.01 & 43.64 \pm 0.02 &  214.31 \pm  7.45 &  791 \pm  54 & H$\alpha$ & 7.13 & BLAGN (LRD) & 2\\
NX23770 & 268.542071 & 65.260276 & 4.516 & 23.83 \pm 0.01 & 42.74 \pm 0.05 &   26.53 \pm  3.15 & 1793 \pm 296 & H$\alpha$ & 7.44 & BLAGN & 12 \\
NX23192 & 268.522548 & 65.262645 & 4.522 & 23.43 \pm 0.01 & 43.48 \pm 0.02 &  147.26 \pm  5.47 & 3447 \pm 158 & H$\alpha$ & 8.37 & BLAGN (LRD) & 12 \\
NX26913 & 268.666877 & 65.243642 & 4.595 & 19.67 \pm 0.01 & 44.58 \pm 0.01 & 1776.81 \pm 11.86 & 3105 \pm  35 & H$\alpha$ & 8.80 & BLAGN (QSO) & 12 \\
NX10835 & 268.437675 & 65.167485 & 4.649 & 25.08 \pm 0.02 & 43.03 \pm 0.04 &   48.87 \pm  4.93 & 2250 \pm 472 & H$\alpha$ & 7.78 & BLAGN (LRD) & 1 \\
NX26545 & 268.520329 & 65.246470 & 4.823 & 25.33 \pm 0.03 & 42.61 \pm 0.08 &   17.10 \pm  3.17 & 1099 \pm 190 & H$\alpha$ & 6.94 & BLAGN & 12 \\
NX32030 & 268.639098 & 65.216381 & 5.063 & 25.76 \pm 0.05 & 42.19 \pm 0.16 &    5.78 \pm  2.07 &  652 \pm 122 & H$\alpha$ & 6.28 & BLAGN & 1 \\
NX12300 & 268.472995 & 65.174101 & 5.066 & 25.26 \pm 0.02 & 42.42 \pm 0.13 &    9.81 \pm  2.55 &  823 \pm 314 & H$\alpha$ & 6.59 & BLAGN (LRD) & 12 \\
NX8933  & 268.615474 & 65.155855 & 5.574 & 23.21 \pm 0.01 & \nodata & \nodata & \nodata & \nodata & \nodata & BLAGN* (LRD*) & 3 \\
NX12143 & 268.428338 & 65.173501 & 5.864 & 24.44 \pm 0.01 &  \nodata &  \nodata &  \nodata & \nodata & \nodata & BLAGN* (LRD*) & 3 \\
NX12349 & 268.568556 & 65.174341 & 5.871 & 23.83 \pm 0.01 & 43.13 \pm 0.03 &   35.36 \pm  2.80 & 1673 \pm 173 & H$\alpha$ & 7.56 & BLAGN (LRD) & 12\\
NX7680 & 268.393574 & 65.147946 & 7.006 & 24.93 \pm 0.02 &  \nodata &  \nodata &  \nodata & \nodata & \nodata & BLAGN* (LRD*) & 3\\
\enddata
\tablecomments{Column (1): source ID. Columns (2) and (3): J2000 right ascension and declination. Column (4): spectroscopic redshift. Column (5): F444W magnitude. Column (6-8): luminosity, flux, and FWHM of the broad component(s). FWHM has been corrected for instrumental broadening using the estimate in \citet{2017JATIS...3c5001G}. Column (9): name of the broad line. Column (10): BH mass derived from Equation~\ref{eq:MBH}. Column (11): source classification with subclass indicated in the parentheses; tentative classifications are marked with an *. Column (12): BLAGN selection flag among the three criteria described in Sec~\ref{sec:blagn_sample}. A BLAGN can be selected by more than one criteria. }
%}
\end{deluxetable*}

\subsection{Broad-line Emitter Selection}\label{sec:blagn_sample}

In this work we focus on the population of broad-line emitters since our primary interest is on accreting SMBHs. We further
assume that these broad-line emitters are of AGN nature (i.e., BLAGNs), though this is still being debated, in particular for the LRD population (Section~\ref{sec:intro}). Following most of the recent works on LRDs, we define LRDs as a subset of BLAGNs, with detailed definitions described in Section~\ref{sec:lrd_sample}. Below we first describe our procedure of measuring line widths and initial selection of the BLAGN parent sample. 

For each identified line emitter, we fit the emission lines with a series of Gaussian models. For the spectral fitting, we first re-subtract the local continuum in small spectral windows around \Hb+\OIII, H$\alpha$, \SIII, \HeI, and Pa$\beta$ in the optimally-extracted coadded 1D spectrum. In general, we fit a linear function to a $\pm$4000-6000 ${\rm km s^{-1}}$ spectral region with respect to the line center to avoid over-subtraction of the broad wing. For objects contaminated by a strong non-linear continuum from nearby bright sources ($\lesssim$21.5 mag), we use a smaller spectral region for continuum fitting. We then use the continuum-free spectra to fit the emission lines in several line windows (H$\beta$, H$\alpha$, \SIII\ and Pa$\beta$) with Gaussian profiles in logarithmic wavelength space. 

For emission line fitting, we adopt both a one-Gaussian model and a two-Gaussian model to fit broad H$\beta$, \OIIIab, H$\alpha$, \HeI, and Pa$\beta$ with potential broad components. We restrict the Full-width-at-half-maximum (FWHM) in the one-Gaussian model to be $>187.5\, {\rm km\, s^{-1}}$ (corresponds to $R=1600$) and $<2000\, {\rm km\, s^{-1}}$. In the two-Gaussian model, the FWHM of the narrow component is constrained to be smaller than $500\, {\rm km\, s^{-1}}$ and also smaller than the FWHM of the broad component, while the upper limit of the broad component FWHM is set to be $3000\, {\rm km\, s^{-1}}$. We constrain the narrow components within each line region to have the same velocity offset and line width. The flux ratios between \NIIa\ and \NIIb, \OIIIa\ and \OIIIb\ are fixed to be 1:3. The uncertainty of our spectral fits is estimated through a Monte Carlo approach, where 100 mock spectra are generated using the spectral errors. The semi-amplitude of the 16th and 84th percentiles from fits to the mock spectra are taken as the measured $1\sigma$ uncertainty.

Our initial BLAGN candidates are then selected if they satisfy at least one of the following three criteria:
\begin{enumerate}
    \item[(1)] The FWHM in the single-Gaussian fit is greater than 600 km s$^{-1}$; 
    
    \item[(2)] The Bayesian information criterion (BIC $=\chi^2+k\, \ln (n)$, where $\chi^2=\rm \Sigma\frac{(data-model)^2}{error^2}$, $k$ is the number of free parameters, and $n$ is the number of data points) parameter \citep{BIC} favors the two-Gaussian model over the one-Gaussian model with $\Delta{\rm BIC}>10$, and the broad-component FWHM is at least 100 km s$^{-1}$ broader than that of the narrow component and other narrow lines including \OIII$\lambda5007$ and \SIII$\lambda9532$ if present; 
    
    \item[(3)] The object exhibits photometric spectral shapes and morphology typical of LRDs but broad \halpha\ falls out of the WFSS bandpass (only three objects; described below). This last criterion implicitly assumes that LRDs are a subset of BLAGNs, and by virtue objects selected by criterion (3) alone are still candidates. 

\end{enumerate}

An initial sample of 90 candidates satisfying any of these three criteria are selected. We then perform visual inspection of their NIRCam images and 2D grism spectra to exclude spurious broad-line emitters. The broad lines in spurious objects are mainly due to kinematic-morphological broadening along the dispersion direction, including off-nucleus star-forming clumps and close companions. 

{The resulting BLAGN sample consists of 23 objects at $2.9\lesssim z \lesssim 7$ (all but one are within $2.9\lesssim z \lesssim 5.9$). Figure~\ref{fig:BLE_stamp} shows their $\sim3\times3\arcsec$ composite color stamps (F090W+F200W+F444W), and Table~\ref{tab1} summarizes their properties and selection criteria. Many objects are selected by more than one criteria. Among this sample, 17 objects satisfy criterion (1); 17 objects satisfy criterion (2) and in most cases the broad-line FWHM is much larger than the narrow-line FWHM: NX29853 has the smallest FWHM difference between the broad and narrow components of $\sim200$ km s$^{-1}$ -- this object is further discussed below as it also has one of the lowest broad-line FWHMs. Three objects satisfy criterion (3).}

Among the 23 BLAGNs, 19 show broad \Ha, and the two lowest-redshift objects (NX29853 and NX34911) show broad \HeI. The remaining three objects {(NX8933, NX12143, and NX7680)} are selected by criterion (3) only, i.e., they do not have \Ha\ coverage in WFSS (their $z_{\rm spec}$ is based on \OIII) but show characteristic LRD features with bending rest-UV-optical SED shapes and a compact F444W size (see Section~\ref{sec:lrd_sample} for details). Therefore, they are included in the final sample but are marked as BLAGN/LRD candidates. The \Ha\ or \HeI\ line profiles of these BLAGNs (with coverage) are shown in Fig.~\ref{fig:BLE_spec}. 

{Our selection criterion (1) of single-Gaussian-fit FWHM $>600$ km s$^{-1}$ is more lenient than the often adopted criterion of $\gtrsim 1000\, {\rm km\,^{-1}}$ \citep[e.g.,][]{Harikane+2023, Greene+2024, Matthee+24, Lin+2025_C3D_BLAGN} for BLAGNs. Several works also adopted a smaller FWHM threshold for selecting BLAGNs. For instance, \citet{Taylor+2024} set a lower limit of 700 km s$^{-1}$; \citet{JADES_BLAGN} set a lower limit of 800 km s$^{-1}$; \citet{Maiolino+2024_JADES_BLAGN} require the FWHM of the broad component to be at least two times that of the narrow component. As a result, these works reported several BLAGNs with broad component FWHM as low as 700 km s$^{-1}$.}

{Since lower-mass BHs tend to exhibit narrower broad lines, the moderately high spectral resolution of NIRCam/WFSS ($R\sim1600$; corresponding to 187 km s$^{-1}$) makes it possible to identify these lower-mass BHs. However, confirming their BLAGN nature requires further scrutiny. Figure~\ref{fig:BLE_spec2} shows other emission lines detected in the WFSS spectra of the three BLAGN candidates with broad-component FWHM $<$ 900 km s$^{-1}$. NX29853 has a broad \HeI\ emission line with a FWHM similar to that of the \Ha, supporting its BLAGN classification. \OIII\ in NX32030 and NX12300 is not resolved by WFSS and is significantly narrower than the broad component of \Ha\ after accounting for measurement uncertainties. A fourth object, NX15499, also has a broad FWHM $<$ 900 km s$^{-1}$ but lacks coverage of other emission lines in NIRCam/WFSS. Nonetheless, the prominent absorption in the blue wing of the \Ha\ line strongly suggests that NX15499 is a genuine broad-line emitter. Based on these additional arguments, we classify these objects as BLAGNs.} 

Among the 23 BLAGNs (including 3 candidates), 15 are further classified as LRDs as detailed in Section~\ref{sec:lrd_sample}. One particular BLAGN (NX26913) is a quasar with a much higher luminosity than the rest of the sample. This quasar is excluded from most of our statistical analyses below. For the sake of discussion, we explicitly use the term ``LRD'' to refer to this subset of BLAGNs in our spectroscopic sample.

Figure~\ref{fig:BLE_spec} shows the 2D grism spectra, 1D spectra, and spectral decomposition of our sample. Basic information and properties of the broad component of the emission lines are also compiled in Table~\ref{tab1}. For the final emission line measurements, we use up to three Gaussians to fit the broad component. Notably, we visually identify \Ha\ absorption in NX5732, NX12300, NX12349, NX15499, NX21246, and NX23192, which is commonly found in LRDs \citep[e.g.,][]{Matthee+24, Maiolino+2024_JADES_BLAGN, Kocevski_LRD_selection, Lin+2024}. This amounts to half (6/12) of the LRD sample with \Ha\ coverage. Due to the limited signal-to-noise ratio of our grism spectra and complications in the remaining objects, we only attempt to correct the effect of absorption in NX15499, NX12349, and NX23192, in which we mask the spectral region affected by absorption and use the remaining part to recover the line profile. FWHM of the broad component is calculated by summing up all the associated Gaussian functions. Note that the line luminosity and FWHM of NX15499 may be underestimated since the blue side of the line peak is significantly affected by absorption and our fitting may be biased without independent redshift measurement from other lines (e.g., \OIII).

\begin{figure*}
  \includegraphics[width=0.55\textwidth]{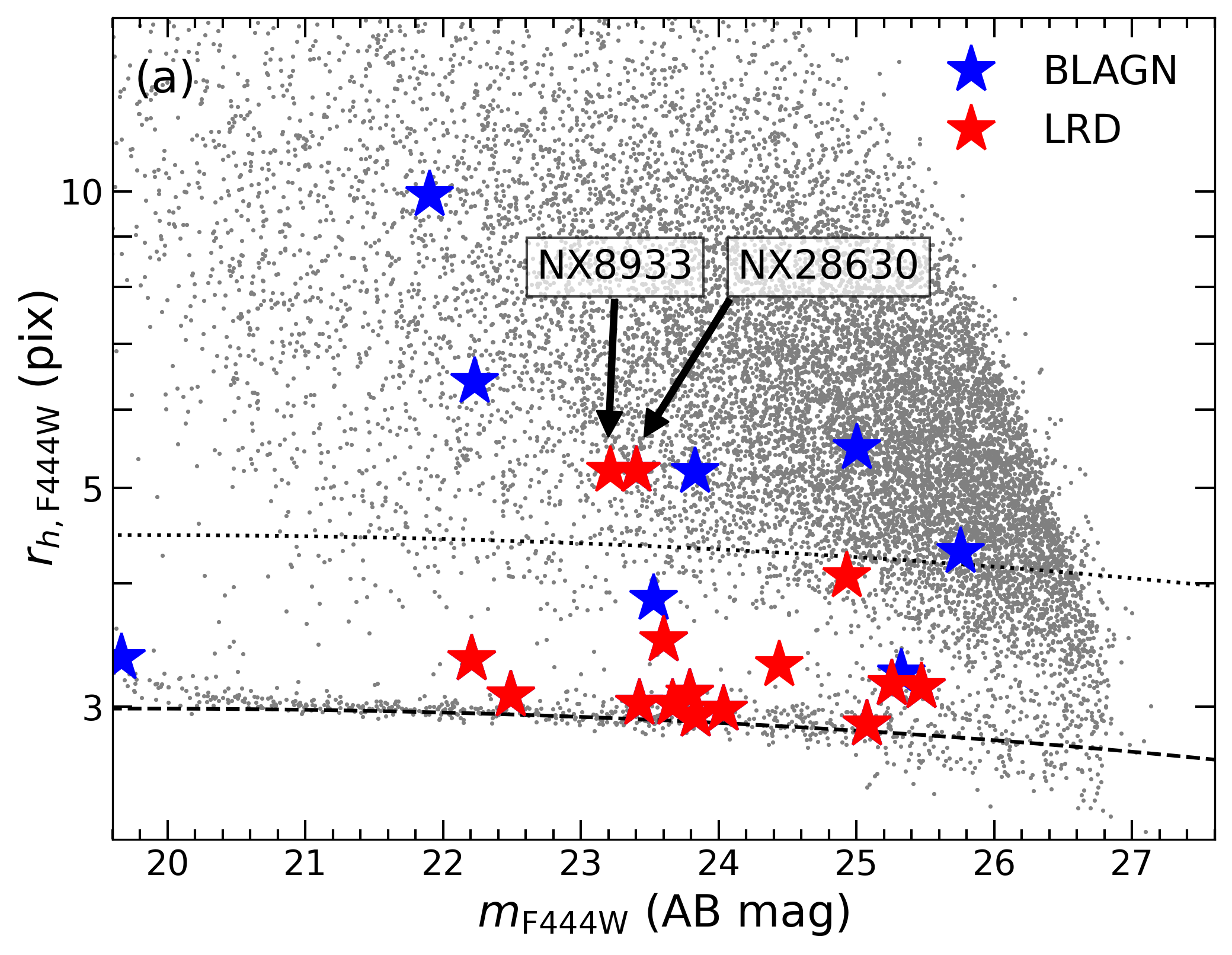}
  \includegraphics[width=0.43\textwidth]{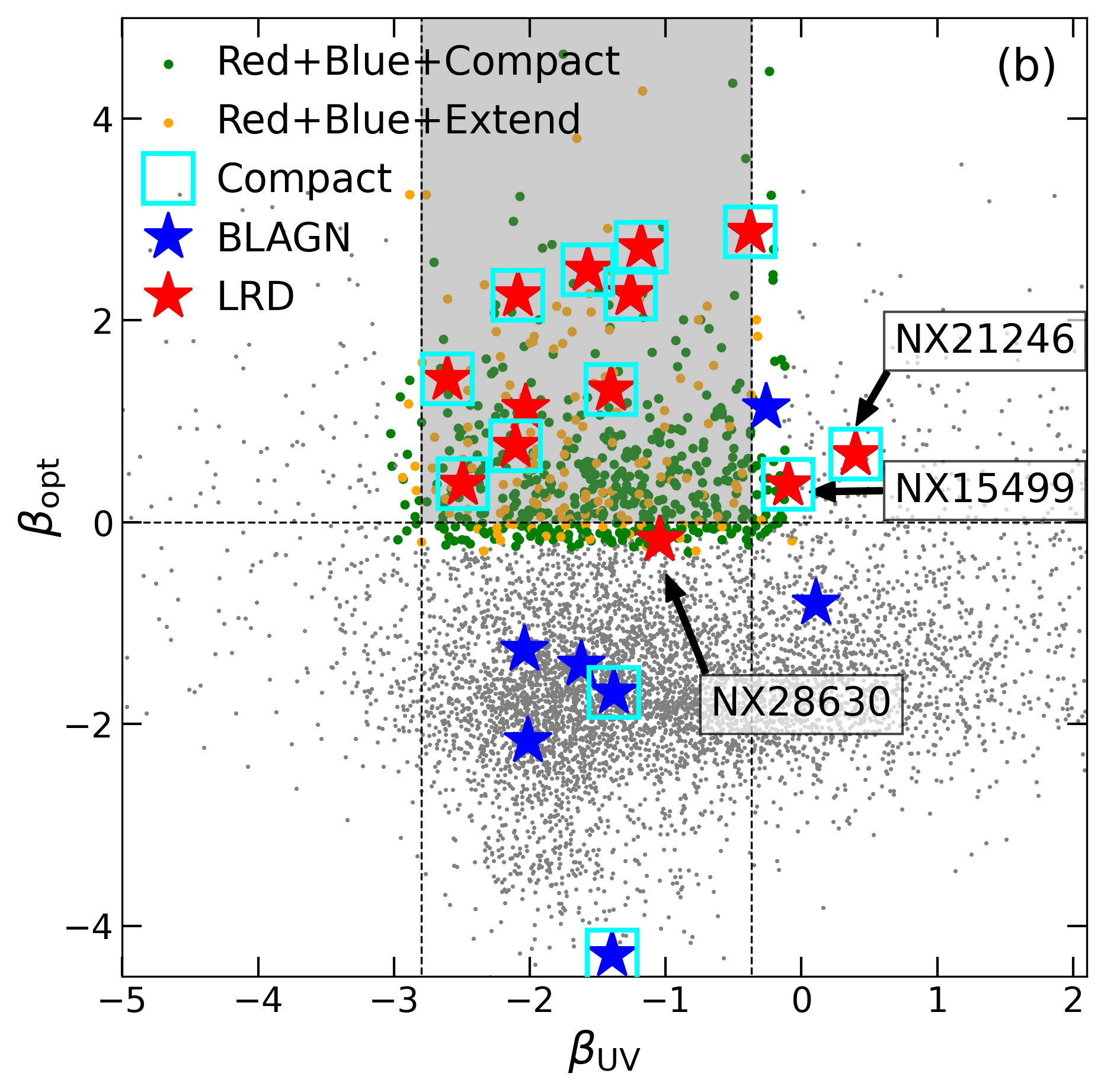}
  \caption{(a) Half-light radius in the F444W filter ($r_{h, \rm F444W}$) versus F444W magnitude and (b) rest-UV slope ($\beta{\rm UV}$) versus rest-optical slope ($\beta{\rm opt}$) at $z\geq2$ for bright objects (F444W$_{\rm SNR}>$12) in NEXUS EDR. Each pixel has a scale of 0\farcs03. Blue and red stars and gray dots represent BLAGNs, LRDs, and all the objects, respectively. Dashed and dotted curves in the panel (a) indicates the locus of stars and our cut to select compact objects, respectively. Gray shaded region in panel (b) indicates our selection criteria of sources with blue and red continuum slopes. Green and orange dots represent objects with red and blue continuum that are compact and extended, respectively. Compact BLAGNs are highlighted with cyan squares. The labeled LRDs (NX8933, NX28630, NX21246, NX15499) are further discussed in Section~\ref{sec:lrd_sample}.
  \label{fig:LRD_phot}}
\end{figure*}

\begin{figure*}
\centering
  \includegraphics[width=\textwidth]{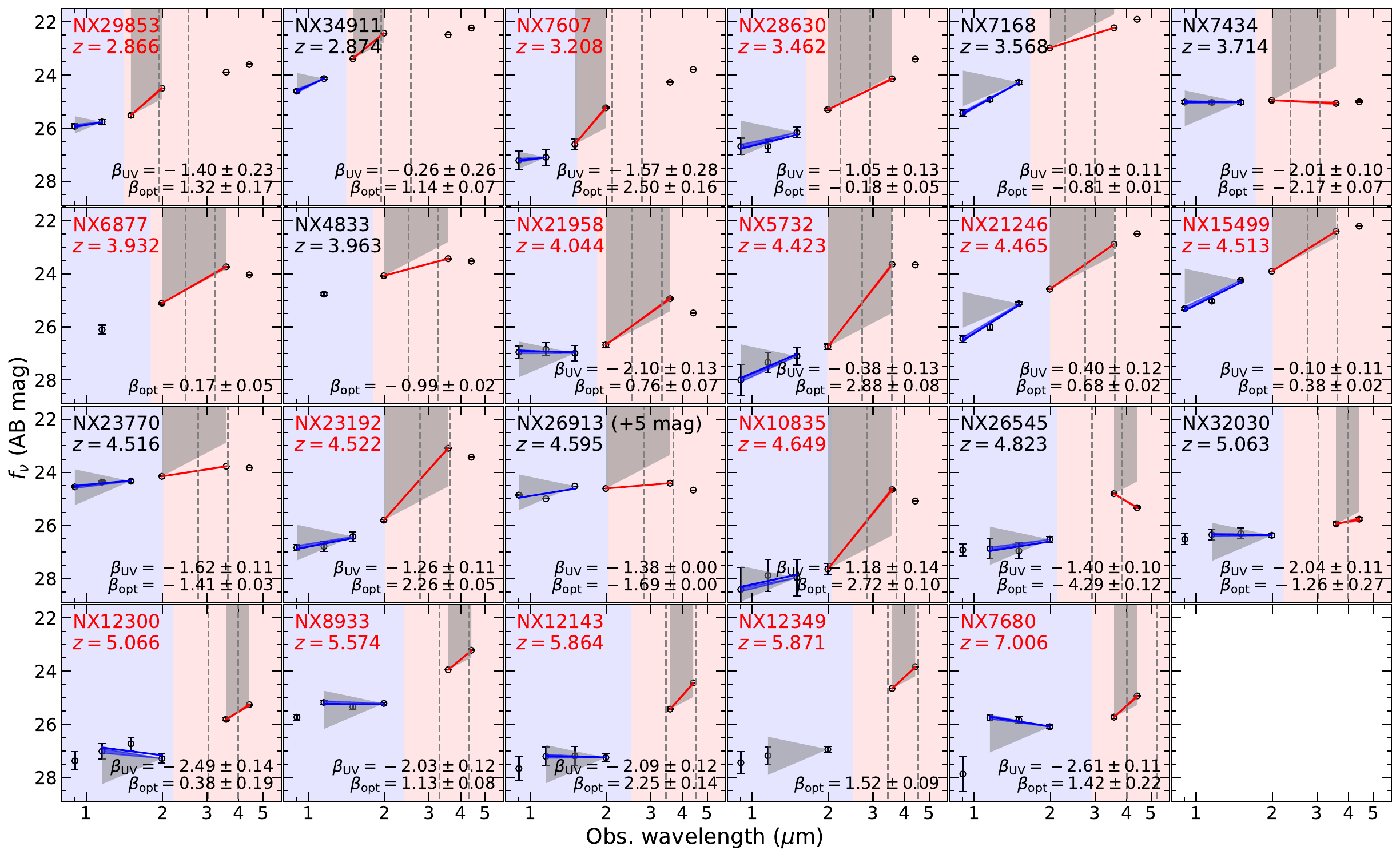}
  \caption{Spectral energy distributions (SEDs) of NEXUS BLAGN sample. Object name and redshift are labeled at the top-left corner, with LRDs marked in red fonts. Best-fit spectral slopes and their 1$\sigma$ uncertainties are shown at the lower-right corner and illustrated as blue (red) solid lines for $\beta_{\rm UV}$ ($\beta_{\rm opt}$). Shaded gray areas represent the LRD selection criteria based on spectral slopes from \citet{Kocevski_LRD_selection}. Vertical dashed lines represent the locations of \OIIIb\ and \Ha. We add 5 mag to the quasar NX26913 for illustration purposes.
  \label{fig:BLAGN_SED}}
\end{figure*}

\subsection{Classification of LRDs}\label{sec:lrd_sample}

Qualitatively LRDs refer to broad-line AGNs with compact morphology (typically measured in F444W), blue/flat rest-UV spectral slopes, and red rest-optical spectral slopes. The composite color images in Figure~\ref{fig:BLE_stamp} show the characteristic appearances of LRDs. In the literature, more quantitative criteria have been used to select LRD candidates from photometry, and confirmed subsequently with spectroscopy for redshift and broad emission lines. 

Early LRD samples were mainly selected using NIRCam colors and size in the F444W filter \citep[color-based selection; e.g.,][]{Greene+2024, Labbe+2025ApJ}. The widely-adopted \citet{Greene+2024} selection criteria are: 
\[
\rm F115W - F150W < 0.8\ \&
\]
\[
\rm F200W - F277W > 0.7\ \&
\]
\[
\rm F200W - F356W > 1.0 
\]
for $4<z<6$, and 
\[
\rm F150W - F200W < 0.8\ \&
\]
\[
\rm F277W - F356W > 0.7\ \&
\]
\[
\rm F356W - F444W > 1.0 
\]
for $z>6$, with a size cut of 
\[
f_{\rm F444W(0\farcs4)}/f_{\rm F444W(0\farcs2)} < 1.7,
\]
where $f_{\rm F444W(0\farcs4)}$ and $f_{\rm F444W(0\farcs2)}$ represent the flux density within 0\farcs4- and 0\farcs2-diameter aperture. 

To account for rest-wavelength coverage shifts of NIRCam filters over a wide redshift range, \citet{Kocevski_LRD_selection} modified the above color-based selection method by utilizing rest-frame optical and UV spectral slopes $\beta_{\rm opt}$ and $\beta_{\rm UV}$ (spectral slope-based selection). These spectral slopes are derived by fitting a linear function
\[
m_i=-2.5\,(\beta+2)\,\log(\lambda_i)+ {\rm const.} 
\]
to a set of photometric bands blueward and redward of the Balmer break at 3645\AA, where $m_i$ is the AB magnitude measured in the $i$th filter with an effective wavelength of $\lambda_i$. They translate the color cuts to constraints in the spectral slopes: $-2.8<\beta_{\rm UV}<-0.37$ and $\beta_{\rm opt}>0$. {The lower limit on $\beta_{\rm UV}$ reduces the contamination from brown dwarfs and the upper limit on $\beta_{\rm UV}$ reduces the contamination from dusty star-forming galaxies in photometric LRD selection.} They also update the compact morphology cut by selecting objects with half-light radius ($r_h$) measured before deconvolution from \texttt{SExtractor} \citep{1996A&AS..117..393B} in the F444W filter within 1.5 times that of the stellar locus.

LRDs selected from the above two methods are not identical. By compiling a large sample of photometrically-selected and spectroscopically-confirmed LRDs from the literature, \citet{Hainline+2025_LRD_selection} find that a significant fraction of spectral slope-selected LRDs fall outside the color-based selection box. While the spectral slope method captures most color-selected LRDs and the majority of spectroscopically-confirmed sources, one spectroscopically-confirmed LRD from the UNCOVER survey \citep{Greene+2024} lies outside the spectral slope selection box, exhibiting a relatively red UV spectral slope ($\beta_{\rm UV}\approx0$). Their results suggest that although the spectral slope-based selection is effective at reducing contamination in photometric LRD samples, applying strict cuts on spectral slopes will not yield a complete sample and may exclude a subsample of LRDs, particularly those with relatively red UV slopes.

In this work we adopt the spectral slope-based selection criteria in \citet{Kocevski_LRD_selection} to classify photometric LRDs. We apply the strict spectral slope cuts to the full NEXUS sample to select photometric LRD candidates based on $z_{\rm phot}$, spectral slopes, and F444W size. These photometric LRDs are for comparison and plotting purposes (e.g., Fig.~\ref{fig:LRD_phot}) only, and are not used in our statistical analyses. We defer a more detailed investigation on the efficiency and completeness of photometric LRD selection to a future paper with more NEXUS data. 

For LRDs identification in our spectroscopic BLAGN sample, we adopt soft spectral slope cuts, allowing objects slightly beyond the cuts to be classified as LRDs, to account for the incompleteness from strict spectral slope cuts. In particular given our spectroscopic confirmation, contamination from stars and dusty star-forming galaxies that have $\beta_{\rm UV}$ outside the canonical \citet{Kocevski_LRD_selection} box is not a concern. In practice, however, we do find that most of our spectroscopic LRDs fall within the $\beta_{\rm UV}$ boundaries in the spectral slope-based selection (Figure~\ref{fig:LRD_phot} right). 

For size measurements, we fit a second-order polynomial function to the locus of bright stars ($20<m_{\rm F444W}<27$ mag) to determine the magnitude-dependent size cut (Figure~\ref{fig:LRD_phot}a). We select compact objects that are less than 1.5 times larger than the size of stars (Figure~\ref{fig:LRD_phot}a). 

To measure $\beta_{\rm UV}$ and $\beta_{\rm opt}$, we use different sets of filters for different redshift ranges to account for bandpass shifting. This enables more reliable identification of LRDs over a wider redshift range than the simple color-color cut method. Table~\ref{tab2} shows the filter sets adopted in this work in four redshift bins. We select sources that are detected with a signal-to-noise ratio (SNR) greater than 12 in the F444W filter and adopt an error floor of 0.05 mag for continuum slope measurements. To account for measurement uncertainties, we allow the spectral slopes to deviate from the defined boundaries by $<1\sigma$ to be still classified as an LRD. We note that $\sim 25$\% of all photometric objects and three spectroscopic BLAGNs in NEXUS do not have $\beta_{\rm UV}$ or $\beta_{\rm opt}$ measurements owing to the non-uniform sky coverage of different filters in the NEXUS-Wide EDR data. 

{The formal photometric selection criteria are then:}
\begin{enumerate}
\item F444W$_{\rm SNR}>$12; 
\item $-2.8<\beta_{\rm UV}<-0.37$ (blue); 
\item $\beta_{\rm opt}>0$ (red); 
\item $r_{h}<1.5\times r_{h,\, \rm star}$ in the F444W filter (compact); 
\item $z_{\rm phot}$$\geq2$.
\end{enumerate}

Figure~\ref{fig:LRD_phot} shows the distribution of our spectroscopically-confirmed BLAGNs in the $r_h$ versus $m_{\rm F444W}$ plane and $\beta_{\rm opt}$ versus $\beta_{\rm UV}$ plane. The SEDs of the BLAGN sample are shown in Figure~\ref{fig:BLAGN_SED}. Directly applying the above selection criteria to the spectroscopic BLAGN sample presented in Section~\ref{sec:blagn_sample} results in nine LRDs. We further classify the following six BLAGNs as LRDs even though they do not formally satisfy the full set of photometric criteria: 
\begin{itemize}

\item NX8933 and NX28630: {These two objects do not formally satisfy the compactness criteria (Figure~\ref{fig:LRD_phot}a) due to the presence of blended close companions in the F444W filter (Figure~\ref{fig:BLE_stamp}). The companions of NX28630 show clumpy morphology in the F200W filter but exhibit smooth extended emission in the F444W filter. The compactness criterion is satisfied after decomposing the LRD (nucleus+host) and its close neighbors. In addition, NX28630 has a $\beta_{\rm opt}=-0.18\pm0.05$, slightly below the cut of $\beta_{\rm opt}>0$. However, its nucleus, modeled by a point spread function (PSF) model, has a $\beta_{\rm opt}=1.86\pm0.10$ (see Section~\ref{sec3.2}).}

\item NX12349: This object does not have a formal $\beta_{\rm UV}$ measurement due to the lack of F150W coverage. However, its overall UV slope, as inferred from the F090W, F115W, and F200W filters, satisfy the $\beta_{\rm UV}$ selection criteria (Figure~\ref{fig:BLAGN_SED}).

\item NX6877: This object does not have a formal $\beta_{\rm UV}$ measurement due to the lack of F090W and F150W coverage. We consider this object as a candidate LRD given its red optical spectral slope and compactness. 

\item NX21246 and NX15499: These objects have slightly red UV slope with $\beta_{\rm UV}\approx0$. Such objects, while relatively rare compared to the overall LRD population, are reported as LRDs from UNCOVER \citep{Greene+2024} with the presence of broad \Ha. We therefore classify these two objects as LRDs as well.
\end{itemize} 

Our final selection results in 15 LRDs within our spectroscopic BLAGN sample. These LRDs are red in their rest-optical SEDs, compact (after removing close companions in a few cases), and mostly have blue spectral slopes in rest-UV. {To improve sample statistics, we include the few tentative classifications in our statistical analyses below. However, excluding these tentative classifications does not qualitatively change our conclusions. }

\begin{deluxetable}{Ccc}\label{tab2}
\caption{Filters for continuum slope estimate versus redshift}
\tablehead{
\colhead{Redshift} & \colhead{$\beta_{\rm UV}$} & \colhead{$\beta_{\rm opt}$}
}
\startdata
2\leq z<3.25 & F090W,F115W & F150W,F200W \\
3.25\leq z<4.75 & F090W,F115W,F150W & F200W,F356W \\
4.75\leq z<8 & F115W,F150W,F200W & F356W,F444W \\
z\geq 8 & F150W,F200W & F356W,F444W \\
\enddata
\end{deluxetable}

\section{Properties of BLAGNs}\label{sec3}

\begin{figure}
  \includegraphics[width=0.5\textwidth]{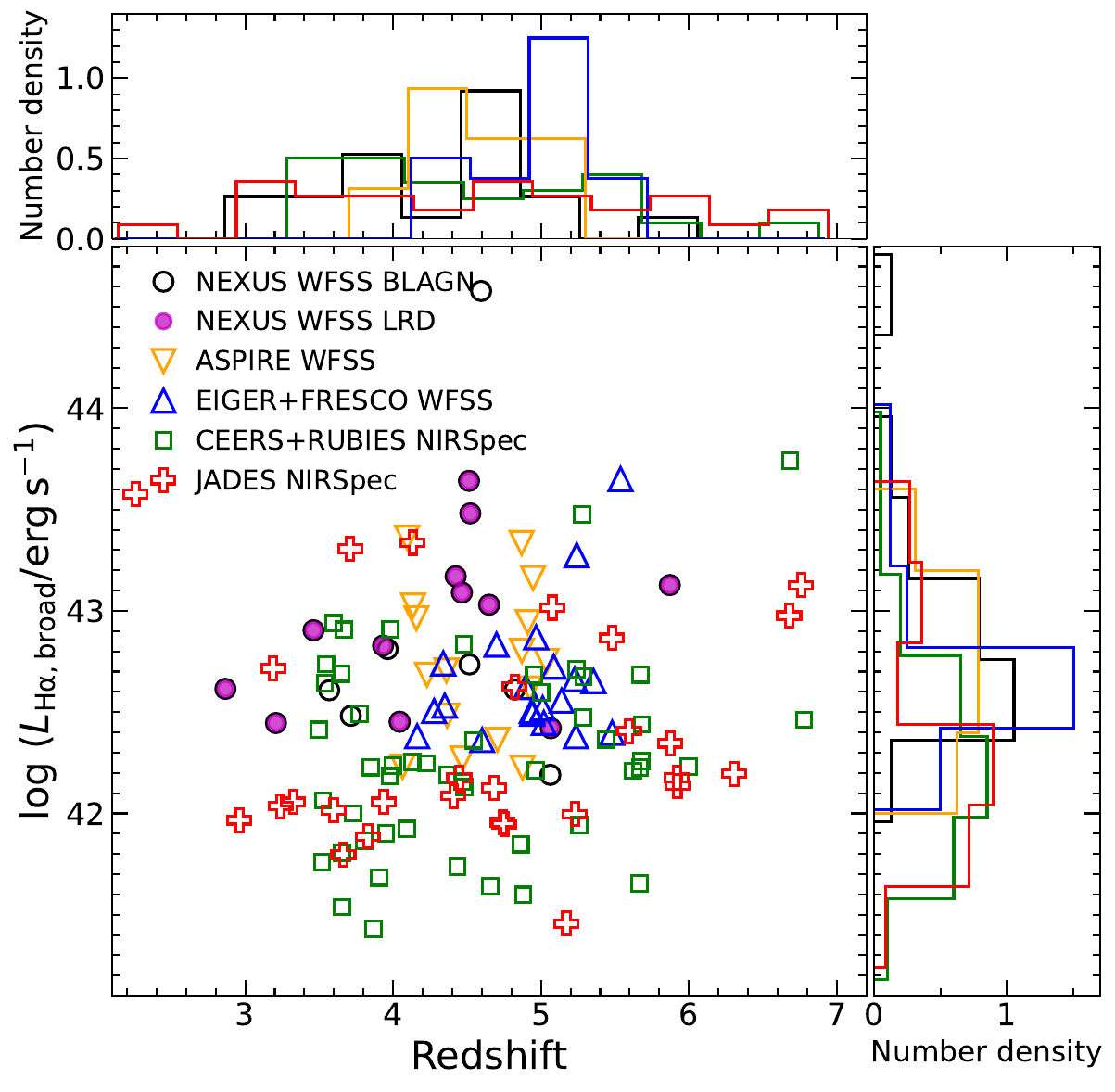}
  \caption{Luminosity of the broad component of \Ha\ ($L_{\rm H\alpha,\, broad}$) versus redshift. Open black circles, orange down triangles, blue up triangles, green squares, and red pluses represent BLAGNs from NEXUS survey, ASPIRE survey \citep{Lin+2024}, EIGER and FRESCO surveys \citep{Matthee+24}, CEERS and RUBIES surveys \citep{Taylor+2024}, and JADES survey \citep{JADES_BLAGN}, respectively. Filled magenta circles highlight the LRDs in NEXUS survey. Upper and right panels show the histograms of redshift and $L_{\rm H\alpha,\, broad}$, respectively.
  \label{fig:LHa_z}}
\end{figure}

\begin{figure}
  \includegraphics[width=0.5\textwidth]{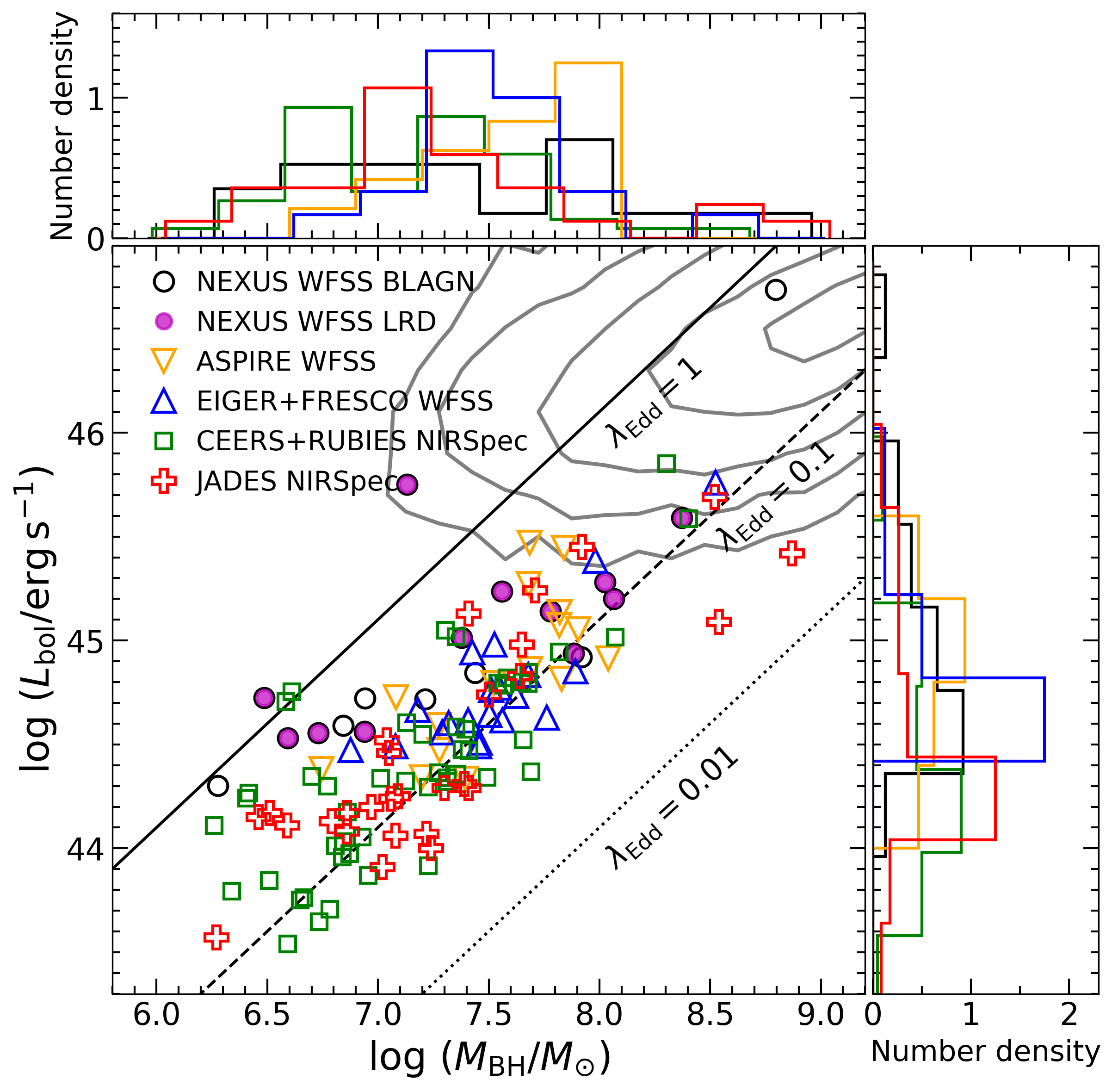}
  \caption{Bolometric luminosity ($L_{\rm bol}$) versus black hole mass ($M_{\rm BH}$). Symbols are the same as Figure~\ref{fig:LHa_z}. The gray contours represent the distribution of $3<z<6$ quasars from SDSS DR16Q catalog \citep{Wu_DR16Q}, with levels enclosing 16\%, 50\%, 84\%, 95\%, and 99\% of the population from the center outward. Solid, dashed, and dotted lines indicate Eddington ratio ($\lambda_{\rm Edd}$) of 1, 0.1, and 0.01, respectively. Typical uncertainty of 0.5 dex for $M_{\rm BH}$ is not shown here for clarity. 
  \label{fig:Lbol_MBH}}
\end{figure}

\begin{figure*}[t]
\centering
  \includegraphics[width=\textwidth]{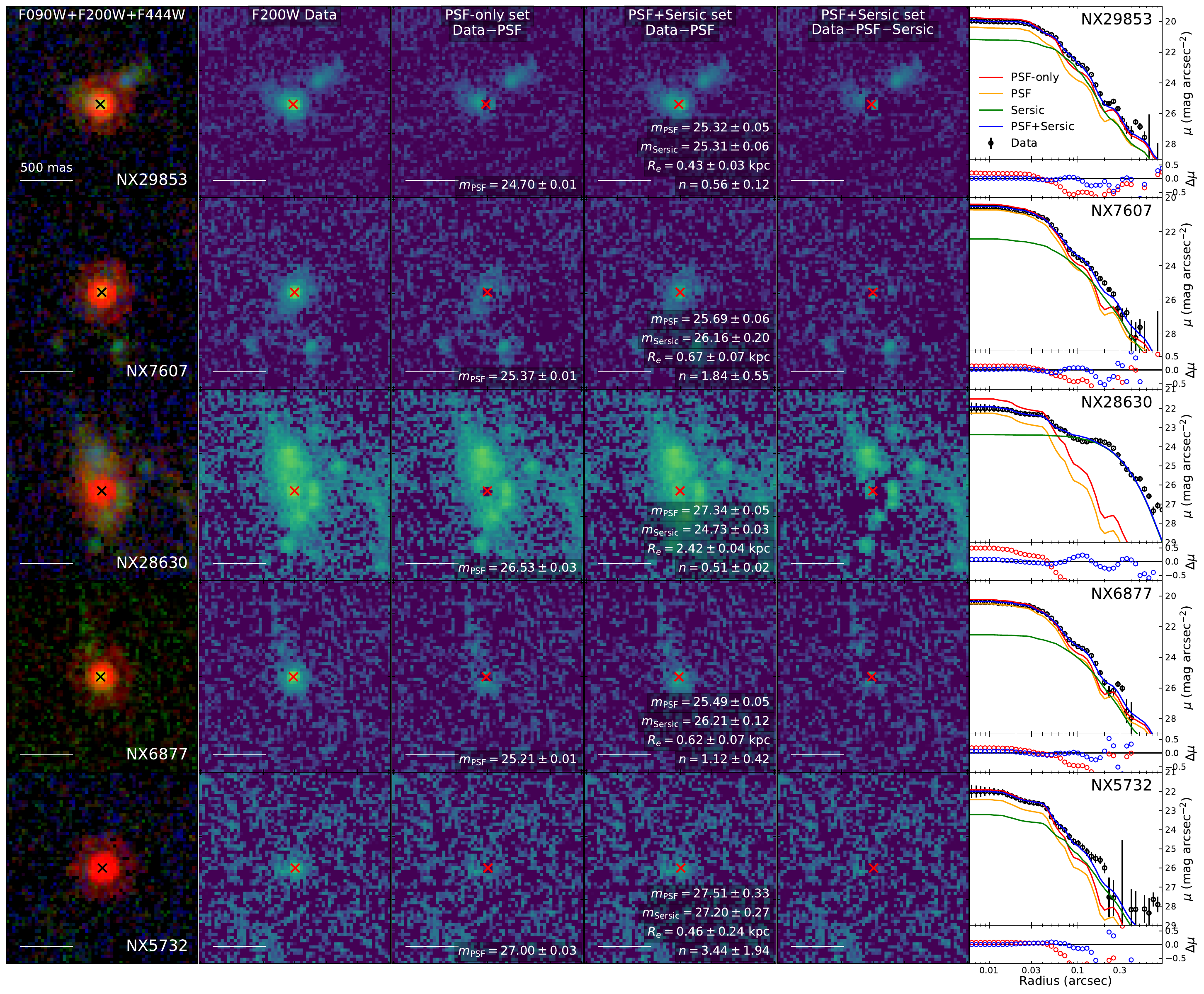}
  \caption{AGN-host image decomposition results of LRDs with significant host emission detections. Columns from left to right show composite color (F090W+F200W+F444W) stamp, F200W data, F200W data minus the best-fit PSF model from the PSF-only model set, F200W data $-$ best-fit PSF model from the PSF+\sersic\ model set, F200W data minus the best-fit PSF model and \sersic\ model from the PSF+\sersic\ model set, and the radial surface brightness profiles, respectively. Radial surface brightness profiles of the data (open black circle), PSF model from the PSF-only model set (PSF-only; red curve), PSF model from the PSF+\sersic\ model set (PSF; orange curve), \sersic\ model from the PSF+\sersic\ model set (\sersic; green curve), and PSF+\sersic\ model from the PSF+\sersic\ model set (PSF+\sersic; blue curve) are shown in the upper panel of the last column, while residuals of data$-$PSF-only (blue circle) and data$-$[PSF+\sersic] (red circle) shown in the lower panel. Best-fit parameters and their 1$\sigma$ uncertainty from the PSF-only model set and PSF+\sersic\ model set are shown in the third and fourth columns, respectively. 500 milli-arcsec (mas) scale bars are shown at the lower-left corner of each image. Note that uncertainties shown here only include nominal fitting errors.
  \label{fig:img_decomp1}}
\end{figure*}

\begin{figure*}[t]
\figurenum{8}
\centering
  \includegraphics[width=\textwidth]{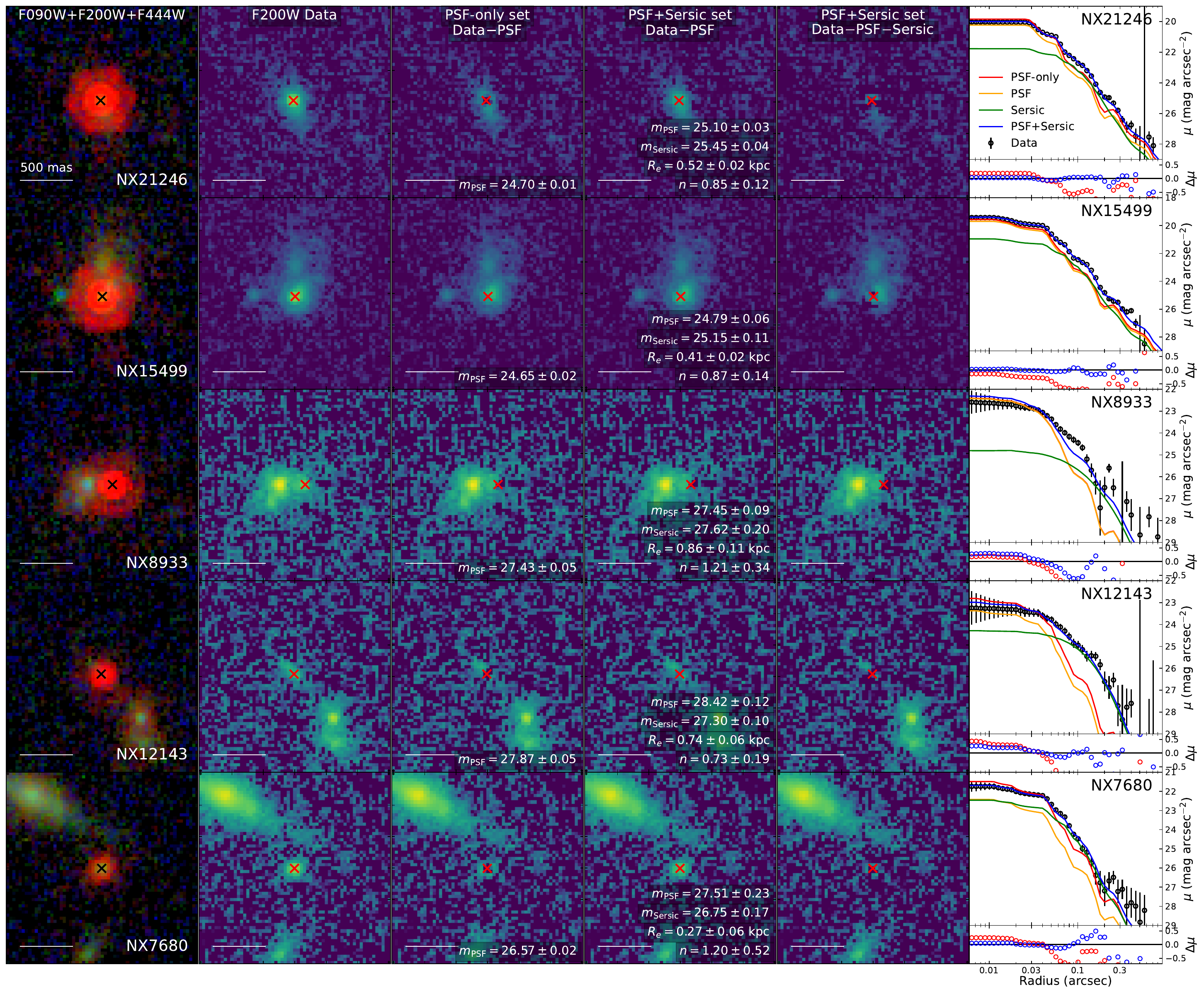}
  \caption{Continued.}
\end{figure*}

\subsection{Physical properties of the accreting SMBH}\label{sec3.1}

Figure~\ref{fig:LHa_z} compares the redshift and luminosity of the broad component of \Ha\ ($L_{\rm H\alpha,\, broad}$) of NEXUS BLAGNs with those from other surveys, including BLAGNs selected using NIRCam/WFSS spectroscopy (WFSS BLAGNs) from EIGER and FRESCO surveys \citep{Matthee+24} and from ASPIRE survey \citep{Lin+2024}, BLAGNs selected using NIRSpec spectroscopy (NIRSpec BLAGNs) from the CEERS and RUBIES surveys \citep{Taylor+2024} and the JADES survey \citep{JADES_BLAGN}. Thanks to the broader wavelength coverage from the combination of F322W2 and F444W grisms, the NEXUS BLAGN sample spans a wider redshift range ($3\lesssim z \lesssim 6$) compared to other WFSS BLAGN samples ($4\lesssim z \lesssim 5.6$). These WFSS BLAGNs have $L_{\rm H\alpha,\, broad}$ within $42.2 \lesssim \log \, (L_{\rm H\alpha,\, broad}/{\rm erg\, s^{-1}}) \lesssim 43.7$ (excluding the quasar NX26913), while about half of the NIRSpec BLAGNs are much fainter with $41.4 \lesssim \log \, (L_{\rm H\alpha,\, broad}/{\rm erg\, s^{-1}}) \lesssim 42.2$.

We adopt the \Ha-based virial black hole mass estimator calibrated in \citet{Reines2015} to estimate black hole mass ($M_{\rm BH}$):
\begin{equation}\label{eq:MBH}
\begin{split}
\log\left( \frac{M_{\rm BH}}{M_{\odot}} \right) 
= 6.57 & + 0.47\, \log\left( \frac{L_{\rm H\alpha,\,broad}}{10^{42}\, \rm erg\, s^{-1}} \right) \\
& + 2.06\, \log\left( \frac{\rm FWHM}{1000\, \rm km\, s^{-1}} \right),
\end{split}
\end{equation}
where $L_{\rm H\alpha,\,broad}$ and FWHM represent the luminosity and instrumental broadening-corrected FWHM of the broad component of \Ha, respectively. We correct instrumental broadening to the line profile using the wavelength-dependent estimate in \citet{2017JATIS...3c5001G}. Note that we do not correct $L_{\rm H\alpha,\,broad}$ for extinction. We estimate the bolometric luminosity ($L_{\rm bol}$) using the calibration, $L_{\rm bol}=130\times L_{\rm H\alpha,\, broad}$, from \citet{Stern&Laor}. The Eddington ratio ($\lambda_{\rm Edd}$) is derived from the ratio between $L_{\rm bol}$ and Eddington luminosity $L_{\rm Edd}=1.26\times10^{38}\, (M_{\rm BH}/M_{\odot})$ erg s$^{-1}$.

As shown in Figure~\ref{fig:Lbol_MBH}, NEXUS BLAGNs follow similar distributions as those from other JWST surveys in the $L_{\rm bol}$ versus $M_{\rm BH}$ plane. Compared to $3<z<6$ quasars from the SDSS DR16Q catalog \citep{Wu_DR16Q}, BLAGNs discovered by JWST have systematically lower $M_{\rm BH}$, as expected. The vast majority of the BLAGNs are located between $\lambda_{\rm Edd}=1$ and $0.01$. Interestingly, one LRD, NX15499, seems to accrete at super-Eddington rate ($\lambda_{\rm Edd}\approx3$), albeit with great uncertainties in the $M_{\rm BH}$ estimate. Its strong absorption features bluewards of \Ha\ may originate from high density gas in the broad-line region and are consistent with strong outflows seen in highly accreting BHs \citep{2019ApJ...880...67J, Matthee+24, Inayoshi&Maiolino2025}. However, other LRDs with absorption features tend to have moderate Eddington ratio $\lambda_{\rm Edd}\approx0.1-0.7$ with a median of 0.14.

\subsection{Host galaxy and close environment}\label{sec3.2}

We study the potential underlying host galaxy of the BLAGNs with a focus on LRDs by performing AGN-host image decomposition. We perform simultaneous multi-band AGN-host image decomposition using \texttt{GALFITM} \citep{Haussler+2013MNRAS} in all six NIRCam filters.  \texttt{GALFITM} is a multi-wavelength version of \texttt{GALFIT} \citep{Peng+2002AJ, Peng+2010AJ} and takes into account the wavelength-dependent galaxy structure. We largely follow the procedures described in previous works utilizing \texttt{GALFITM} in AGN-host decomposition \citep[e.g.,][]{Zhuang&Ho2022, Zhuang&Ho2023, Zhuang+2024}. We adopt two sets of models, one with only one PSF model (PSF-only) to fit mainly the nucleus and the other one with one PSF model to fit the nucleus and one \sersic\ model to fit the (potential) host galaxy (PSF+\sersic). In the latter model set, we assume that the PSF and \sersic\ models have the same centroid, both effective radius ($R_e$) and \sersic\ index ($n$) follow a linear wavelength-dependent relation, ellipticity and position angle are constant across wavelength, and magnitudes of the \sersic\ component and the PSF component are free to vary. Blended companions are modeled using a \sersic\ model and fitted at the same time with BLAGNs. Note that the amplitude of the PSF model may be overestimated in the PSF+\sersic\ model set due to the degenerate between the nucleus and the compact unresolved substructure of the host.

Although the F444W filter is dominated by unresolved emission in most of the objects (see Appendix), we detect significant extended emission in ten LRDs and seven non-LRD BLAGNs across bluer filters. {Here extended emission refers to those centered at the location of the nucleus, rather than from deblended close companions.} Figure~\ref{fig:img_decomp1} shows our image decomposition results in the F200W for LRDs with detected extended emission. We can clearly see the residual extended emission after subtracting the PSF-only model in the Data$-$PSF map. Their radial surface brightness profiles also show clear deviations from that of the PSF, suggesting contribution from extended emission. The additional \sersic\ component from PSF+\sersic\ model significantly improves the fits. We find that the underlying resolved emission in these ten LRDs is extended ($0.6\pm0.2$ kpc) and disk-like ($n\approx1$) and contributes approximately half of the total emission in the F200W filter. 

Interestingly, most (8/10) of the LRDs with detected underlying extended emission have companions or show asymmetric morphology, suggesting they may be undergoing mergers and interactions. Six LRDs (NX29853, NX7607, NX28630, NX15499, NX8933, NX12143) have close companions, while NX6877 and NX21246 show asymmetric morphology. In particular, the companions in NX8933 and NX12143 have the \OIII\ doublet detected in the NIRCam/WFSS spectroscopy, placing them at the same redshift of the LRDs. 

\begin{figure}
\centering
  \includegraphics[width=0.5\textwidth]{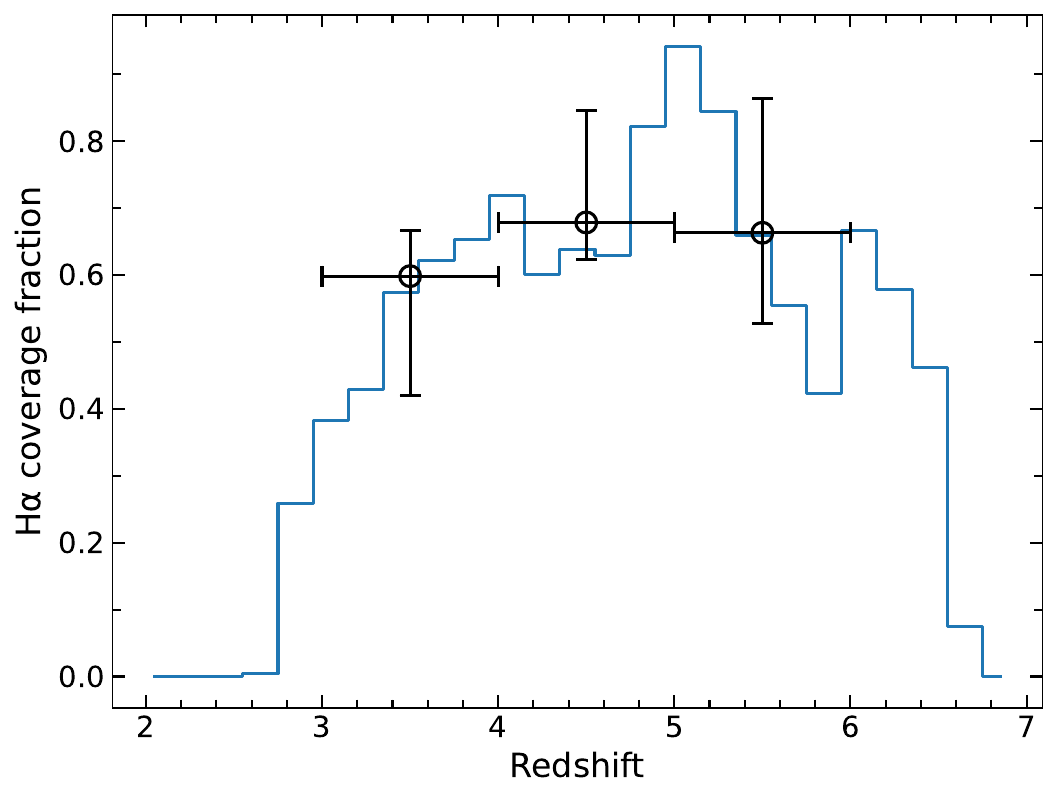}
  \caption{The fraction of sources with \Ha\ covered by NIRCam/WFSS F322W2 or F444W grism as a function of redshift in the NEXUS field. We only consider objects with F444W$_{\rm SNR}>12$. Open black circles represent the median values in the $3\leq z<4$, $4\leq z<5$, and $5\leq z\leq 6$ bins, respectively. Horizontal errorbars indicate the range of the redshift bins, while vertical errorbars indicate 16th and 84th percentiles of the histogram values within this redshift bin. 
  \label{fig:completeness}}
\end{figure}

\begin{figure}
\centering
  \includegraphics[width=0.5\textwidth]{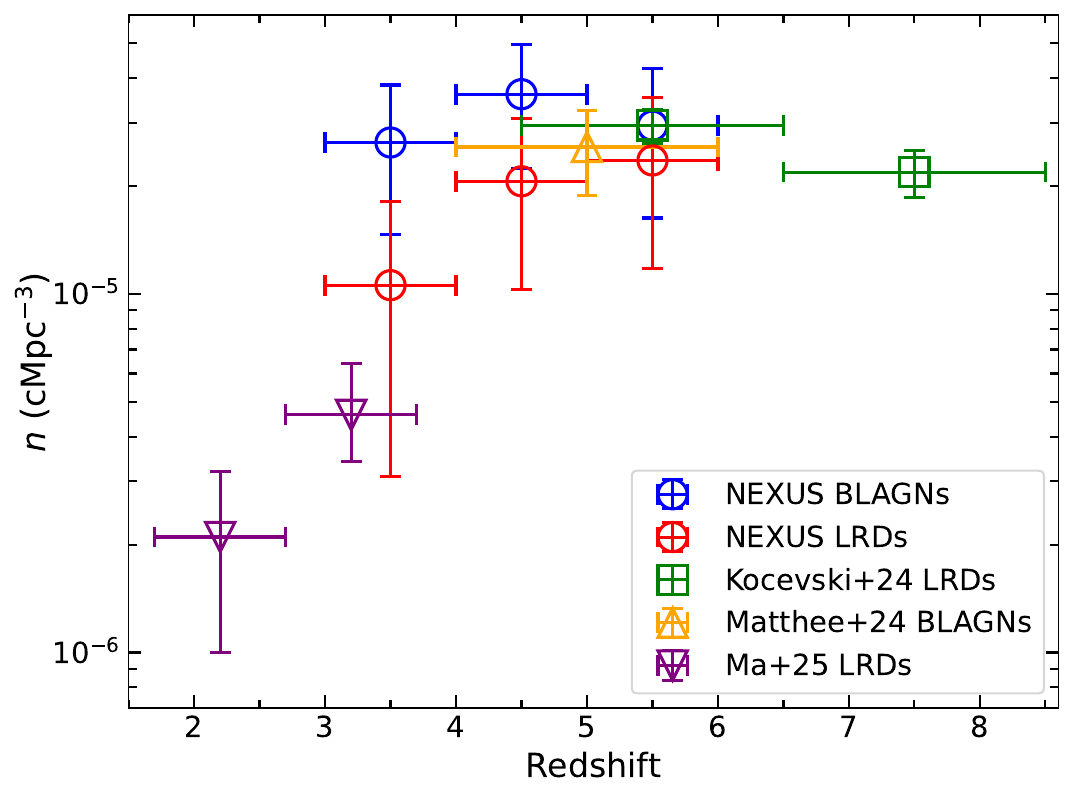}
  \caption{The number density as a function of redshift for BLAGNs and LRDs. The blue and red circles represent the values of BLAGNs and LRDs in this paper. Green squares, orange triangles, and purple triangles represent the measurements from JWST LRDs in \citet{Kocevski_LRD_selection}, JWST BLAGNs in \citet{Matthee+24}, and ground-based LRDs in \citet{Ma+2025}, respectively. The JWST BLAGNs/LRDs samples (including the NEXUS samples) are restricted to sources with $M_{\rm UV}<-18.5$, roughly matching the luminosity threshold of ground-based LRDs with absolute magnitude at rest-frame 5500\AA, $M_{5500}<-20.5$. Vertical errorbars indicate 1$\sigma$ uncertainties in number densities and horizontal errorbars indicate the redshift range of each bin. 
  \label{fig:Num_density}}
\end{figure}

\begin{deluxetable*}{cCCCCCC}\label{tab:N density}
\caption{Number densities of BLAGNs and LRDs}
\tablehead{
\colhead{Redshift} & \colhead{$M_{\rm UV}$ [BLAGN]} & \colhead{$N$ [BLAGN]} & \colhead{$n$ [BLAGN]} & \colhead{$M_{\rm UV}$ [LRD]} & \colhead{$N$ [LRD]} & \colhead{$n$ [LRD]}\\
\nocolhead{} & \colhead{$\rm (AB\, mag)$} & \nocolhead{} & \colhead{$\rm(cMpc^{-3})$} & \colhead{$\rm (AB\, mag)$} & \nocolhead{} & \colhead{$\rm(cMpc^{-3})$}\\
\colhead{(1)} & \colhead{(2)} & \colhead{(3)} & \colhead{(4)} & \colhead{(5)} & \colhead{(6)} & \colhead{(7)}
}
\startdata
$3\leq z <4$ & [$-18.0$, $-19.8$, $-20.7$] & 6\ (5) &  $2.65\pm1.18\times10^{-5}$ & [$-18.0$, $-19.5$, $-19.5$] & 3\ (2) & $1.06\pm0.75\times10^{-5}$\\
$4\leq z <5$  & [$-17.6$, $-19.4$, $-21.2$] & 8\ (7) & $3.61\pm1.36\times10^{-5}$ & [$-17.6$, $-19.4$, $-20.4$] & 6\ (4) & $2.06\pm1.03\times10^{-5}$\\
$5\leq z \leq6$ & [$-18.6$, $-19.3$, $-19.5$] & 5\ (5) & $2.95\pm1.32\times10^{-5}$ & [$-18.6$, $-19.2$, $-19.4$] & 4\ (4) & $2.36\pm1.18\times10^{-5}$\\
$3\leq z \leq6$ & [$-17.6$, $-19.5$, $-21.2$] & 19\ (17) & $3.15\pm0.80\times10^{-5}$ & [$-17.6$, $-19.3$, $-20.4$] & 13\ (10) & $1.85\pm0.61\times10^{-5}$
\enddata
\tablecomments{Column (1): Redshift range. Column (2): The minimum, median, and maximum values of $M_{\rm UV}$ of BLAGNs. Column (3): The number of BLAGNs, with the number of sources with $M_{\rm UV}<-18.5$ mag shown in parentheses. Column (4): Comoving number density of sources with $M_{\rm UV}<-18.5$ mag. The number densities have been corrected for the incompleteness of \Ha\ coverage in NIRCam/WFSS spectroscopy. The uncertainties of $n$ are Poisson counting errors. Columns (5--7) are the same as Columns (2--4) but for LRDs.}
\end{deluxetable*}

\subsection{Abundance}\label{sec3.3}
Given the relatively small number of objects, we will present a crude estimate of number density of BLAGNs in this paper, and defer a more comprehensive luminosity function measurement to a future paper with all BLAGNs in the complete NEXUS-Wide field (400 arcmin$^2$). We quantify the ``incompleteness'' of our NIRCam/WFSS by calculating the fraction of objects with \Ha\ falling outside the wavelength coverage of NIRCam/WFSS spectra. Since the wavelength coverage of NIRCam/WFSS depends on the location on the detector, we obtain a rough estimate by averaging over all sources as a function of redshift, assuming they are randomly distributed across the NEXUS field-of-view. Note that we only consider whether or not \Ha\ is covered by NEXUS NIRCam/WFSS and do not take sensitivity and broad-line detectability into account \citep[see e.g.,][]{Matthee+24, Lin+2025_C3D_BLAGN}. Moreover, contamination from bright objects in WFSS along or across the dispersion direction may further reduce the number of identified spectroscopic BLAGNs. Therefore, our estimate here should be considered a lower limit of the BLAGN number density.

As shown in Figure~\ref{fig:completeness}, the \Ha\ coverage fraction peaks at $z\approx5$ and declines towards both lower and higher redshifts. At $3\leq z \leq 6$, the average \Ha\ coverage fraction is $\gtrsim 50\%$. The comoving volume over the 100 arcmin$^2$ NEXUS-Wide EDR and in three redshift bins, $3\leq z<4$, $4\leq z<5$, and $5\leq z\leq 6$, are 3.16$\times10^5$ cMpc$^2$, 2.86$\times10^5$ cMpc$^2$, and 2.56$\times10^5$ cMpc$^2$, respectively. We estimate the absolute UV magnitude at rest-frame 1500\AA\ ($M_{\rm UV}$) using the total flux from the PSF+\sersic\ model in the F090W filter\footnote{We use F115W instead for NX4833 and NX6877 as they are not covered by F090W in early NEXUS observations.} after removing the contribution from close (blended) companions. NEXUS BLAGNs span a range of $-21\lesssim M_{\rm UV} \lesssim -18$ mag, except for the quasar NX26913 with $M_{\rm UV}=-26.2$ mag. NX26913 is excluded from the following number density estimate. 

Table~\ref{tab:N density} presents the number densities of NEXUS BLAGNs and LRDs separately, in four redshift bins ($3\leq z<4$, $4\leq z<5$, $5\leq z\leq 6$, and $3\leq z\leq 6$). To facilitate a fair comparison across redshift bins and with literature measurements, we limit the measurement of number densities to relatively bright sources with $M_{\rm UV}<-18.5$ mag. Figure~\ref{fig:Num_density} compares the number densities of NEXUS BLAGNs and LRDs with literature measurements, including BLAGNs \citep{Matthee+24} and LRDs \citep{Kocevski_LRD_selection} selected from JWST data, and $z<3.7$ LRDs selected from ground-based data \citep{Ma+2025}. The ground-based LRD sample is restricted to objects with absolute magnitude at rest-frame 5500~\AA, $M_{5500}<-20.5$ mag. For reference, our LRDs with the $M_{\rm UV}<-18.5$ mag cut have $M_{5500}\lesssim-20.4$ mag, roughly matching the luminosity range of the low-$z$ LRD sample in \citet{Ma+2025}. At $4<z<6$, our number densities are consistent with those of BLAGNs and LRDs discovered by JWST in other fields. We find a tentative declining LRD number density towards lower redshift within our sample, consistent with the global trend by combining with the samples in \citet{Kocevski_LRD_selection} and \citet{Ma+2025}, albeit with large uncertainties. These results remain qualitatively the same if we exclude the few candidates from the sample (see Section~\ref{sec:blagn_sample} and Table~\ref{tab1}).

\subsection{Clustering}\label{sec3.4}

\begin{figure}
  \includegraphics[width=0.48\textwidth]{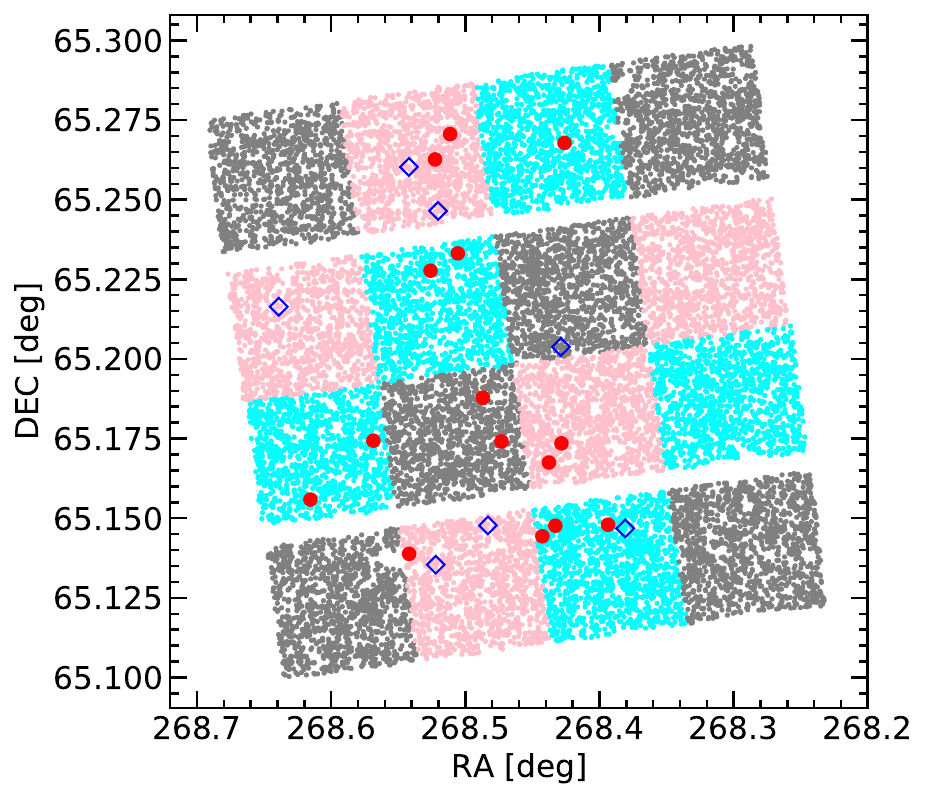}
  \caption{The NEXUS EDR area divided into 16 jackknife regions for clustering analysis. The underlying dots are the full NEXUS photometric galaxy sample, and the red and blue points are the spectroscopic LRDs and remaining BLAGNs. }
  \label{fig:clustering_area}
\end{figure}

Figure~\ref{fig:clustering_area} shows the spatial distributions of our spectroscopic non-LRD BLAGNs (blue open diamonds) and LRDs (red filled circles), where the footprint of the survey is divided into regions with approximately equal area. As discussed earlier, the selections of these BLAGNs and LRDs are by no means complete. However, the sample completeness does not affect the clustering measurements when cross-correlating with a uniform photometric galaxy sample, and the measured signal corresponds to the average signal over the redshift distribution of the target BLAGN/LRD samples. 

To measure the clustering of our spectroscopically confirmed BLAGN/LRD samples, we rely on the NEXUS photometric galaxy sample to cross-correlate with the sparse BLAGN/LRD samples. We restrict the photometric galaxy sample to $22<m_{\rm F444W}<26$ and $3<z_{\rm phot}<6$, resulting in a total of 864 galaxies. The galaxy magnitude cut significantly reduces catastrophic redshift errors due to the non-detection of faint optical magnitudes from HSC, though some low-$z$ interlopers still scatter into the high-$z$ tail of the distribution. The redshift range of the galaxy sample well overlaps with those of the BLAGN and LRD samples for meaningful cross-correlation measurements. In the clustering measurements we exclude the quasar (NX26913) since its luminosity is much higher than the rest of the BLAGN sample. We further exclude NX7680 since it lies significantly beyond the redshift range of the photometric galaxy sample. All other BLAGNs/LRDs are included in the clustering analysis even if their spectroscopic redshifts fall slightly below the photo-z range of the cross-correlation galaxy sample. The BLAGN and LRD clustering samples include 21 and 14 objects, respectively, with a median redshift of $\left<z\right>\approx 4.5$ for both samples.  

\begin{figure}
  \includegraphics[width=0.48\textwidth]{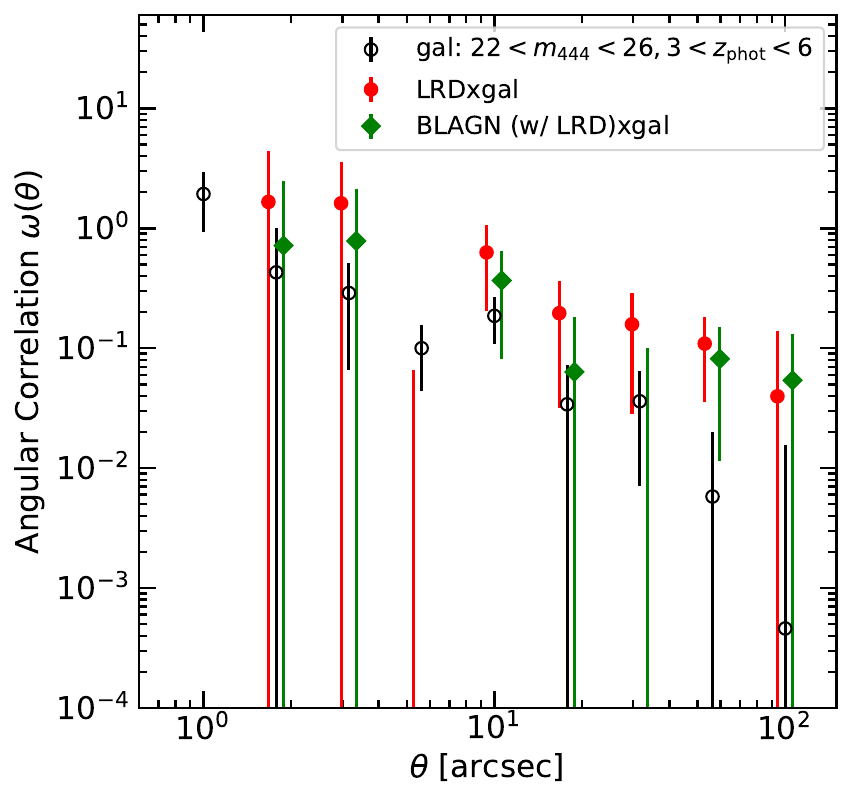}
  \caption{Angular correlation functions for different samples, measured over $1\lesssim \theta\lesssim 100$\arcsec. Albeit with large error bars, the LRD sample appears to have stronger clustering than the underlying photometric galaxy sample. Best-fit power-law models are summarized in Table~\ref{tab:clustering}.}
  \label{fig:acf_ccf}
\end{figure}

We measure the angular cross-correlation function (CCF) between the BLAGN/LRD sample and the photometric galaxy sample, and present the results in Figure~\ref{fig:acf_ccf}. For comparison, the auto-correlation function of the galaxy sample is shown in black open circles in Figure~\ref{fig:acf_ccf}. For the galaxy ACF, we use the Landy-Szalay estimator \citep{Landy_Szalay}. For the cross-correlation, we use the simple estimator $QG/QR-1$, where $QG$ and $QR$ are the normalized numbers of AGN-galaxy and AGN-random pairs in each angular separation bin. We use the standard jackknife resampling method to estimate the uncertainties of the ACF and CCF measurements, using the 16 jackknife samples illustrated in Figure~\ref{fig:clustering_area}. We found that the jackknife errors are substantially larger than Poisson pair counting errors, and provide more conservative uncertainty estimation for the clustering measurements. Because the covariance matrix from the jackknife samples is very noisy given our sample size, we only use the diagonal elements in the covariance matrix to constrain our model fits.  
We detect significant clustering signals for the BLAGN and LRD samples over angular scales of $1\arcsec-100\arcsec$ that are well probed by our survey area. Interestingly, the LRD sample appears to have stronger clustering than the BLAGN sample (see Table~\ref{tab:clustering} below), albeit with large uncertainties. 

Given the limited survey area from the first partial epoch of NEXUS, we are unable to measure the large-scale (i.e., $\gtrsim$ few cMpc) two-point correlation function. We therefore assume that the small-scale clustering we measure is a simple power-law extrapolation from the large-scale two-point correlation function, $\omega(\theta)=(\theta/\theta_0)^{-\beta}=A_0\theta^{-\beta}$, where the angular separation $\theta$ is in units of radian. We only fit to angular bins with $\theta<120\arcsec$ to mitigate edge effects from survey boundaries. Because the correlation function measurements are noisy given the small sample size, we fix the best-fit power-law slope to $\beta=0.8$ for all samples. Table~\ref{tab:clustering} summarizes the clustering measurements for different samples. The small integral constraint (IC) of the correlation function is neglected in subsequent analyses, since the scale of interest here is well within the survey area and the IC term is negligible.

We convert the angular clustering results to the real-space correlation function using Limber's equation \citep{Limber}. The derived correlation length occurs at scales beyond these probed by our clustering sample, thus extrapolation depends on the choice of the power-law slope. {Our choice of a fixed slope $\beta=0.8$ roughly matches the measured large-scale clustering of galaxies at similar redshifts \citep[e.g.,][]{He_etal_2018}. However, we caution that the derived correlation length $r_0$ still has large systematic uncertainties due to the limitations of the current sample (e.g., sample statistics and the lack of coverage on $\sim$few cMpc scales). } 

For a power-law ACF model, $\xi(r)=(r/r_0)^{-\gamma}$, and assuming the comoving correlation length $r_0$ does not evolve strongly over the surveyed redshift range \citep[e.g.,][]{Adelberger_2005}, we have $\gamma \equiv \beta + 1=1.8$ and 
\begin{equation}\label{eqn:limber1}
A_0=\frac{r_0^\gamma B[1/2,(\gamma-1)/2]\int_0^\infty N^2(z)f^{1-\gamma} g^{-1}dz}{[\int_0^\infty N(z)dz ]^2}\ ,
\end{equation}
where $B$ is the beta function, $g(z)=c/H(z)$ is the radial comoving distance, $f(z)\equiv (1+z)D_A(z)$ with $D_A(z)$ the angular diameter distance. $N(z)$ is the redshift distribution of the sample. In practice, the integral in Eqn.~(\ref{eqn:limber1}) is calculated over the sample redshift range $[z_{\rm min},z_{\rm max}]$. For cross correlations between two samples, Eqn.~(\ref{eqn:limber1}) is modified as \citep[e.g.,][]{Croom_Shanks_1999}:
\begin{equation}\label{eqn:limber}
A_0=\frac{r_0^\gamma B[1/2,(\gamma-1)/2]\int_0^\infty N_1(z)N_2(z)f^{1-\gamma} g^{-1}dz }{[\int_0^\infty N_1(z)dz ][\int_0^\infty N_2(z)dz ]}\ ,
\end{equation}
where $N_1(z)$ and $N_2(z)$ are the normalized redshift distributions of the two samples. 

We obtain the real-space correlation lengths for the ACF and CCF samples and list them in Table~\ref{tab:clustering}. To measure the linear bias of the clustering signal, we use the integrated correlation function within $r=[5,20]\,{h^{-1}{\rm cMpc}}$ \citep[e.g.,][]{Shen_etal_2007}:
\begin{equation}\label{eqn:xi20}
    \xi_{20}=\frac{3}{r_{\rm max}^3}\int_{r_{\rm min}}^{r_{\rm max}}\xi(r)r^2dr\ ,
\end{equation}
where $r_{\rm min}=5\,h^{-1}{\rm cMpc}$ and $r_{\rm max}=20\,h^{-1}{\rm cMpc}$.

For the power-law model $\xi(r)=(r/r_0)^{-\gamma}$, Eqn.~(\ref{eqn:xi20}) reduces to 
\begin{equation}
\xi_{20}=\frac{3r_0^\gamma}{(3-\gamma)r_{\rm max}^3}\left(r_{\rm max}^{3-\gamma} - r_{\rm min}^{3-\gamma} \right)\ .
\end{equation}

The linear bias $b$ of the clustering sample is measured as 
\begin{equation}
    b^2 = \frac{\xi_{20}}{\xi_{m,20}D^2(z)}\ ,
\end{equation}
where $\xi_{m,20}$ is the present-day ($z=0$) integrated correlation function of dark matter assuming $\sigma_8=0.84$, and $D(z)$ is the linear growth factor of density fluctuations in our adopted cosmology. Finally, the linear bias of the cross-correlation sample $b_{QG}^2\sim b_Qb_G$, where $b_Q$ and $b_G$ are the linear bias for the AGN (or LRD) sample and the galaxy sample, respectively. 

Our clustering results imply a linear bias of {$b_{\rm BLAGN}=3.30_{-2.04}^{+2.88}$} for BLAGNs and {$b_{\rm LRD}=11.44_{-5.04}^{+5.58}$} for LRDs at the sample median redshift of $\left<z\right>\approx 4.5$. Given the large uncertainties of our clustering measurements, we defer a more careful comparison with theoretical predictions to future work. Here, we simply use the \citet{SMT01} fitting formula for the halo linear bias to roughly infer the typical host dark matter halo masses of our sample. We find a typical halo mass of a few$\times 10^{11}\,h^{-1}M_\odot$ for BLAGNs, which is consistent with recent measurements of JWST BLAGNs at $z\sim 4-6$ \citep[e.g.,][]{Arita_etal_2025,Lin_etal_2025b}. However, the implied linear bias for the LRD sample is too large to be compatible with the high abundance of these objects. Even at the $1\sigma$ lower limit of the LRD linear bias, the inferred halo mass would be $\sim 10^{12}\,h^{-1}M_\odot$, which are too rare at $z\sim 4$ to host a large number of LRDs. We further discuss this point in Section~\ref{sec4}. 

\begin{figure}
  \includegraphics[width=0.48\textwidth]{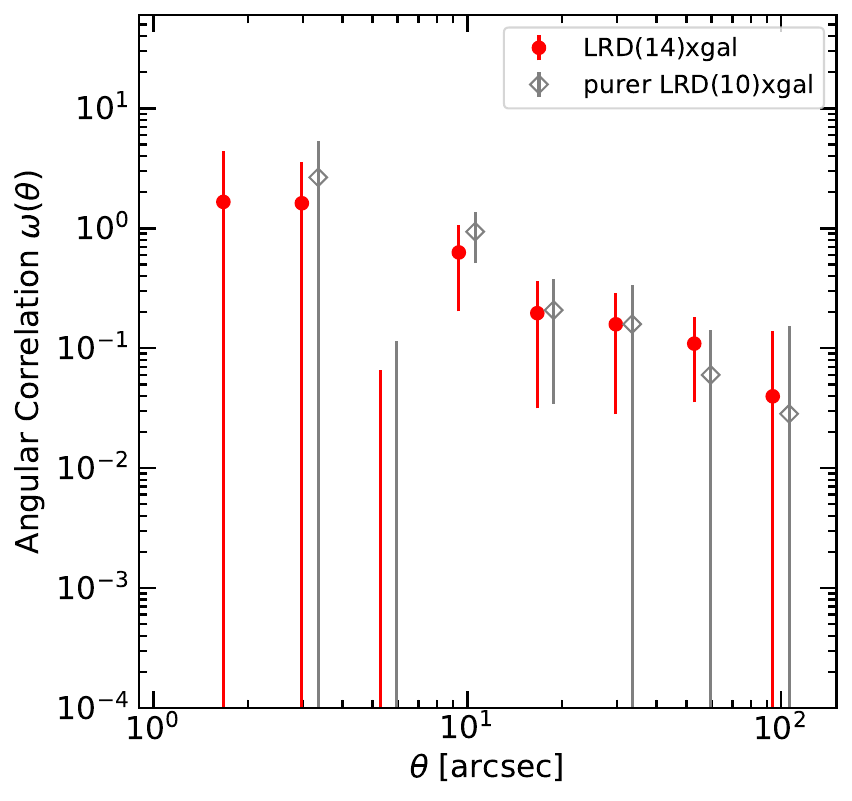}
  \caption{Test on our LRD clustering measurements. The gray points represent the cross-correlation between galaxies and the subset of 10 high-fidelity LRDs in our sample. The results are similar to our fiducial clustering measurements with the full clustering sample of 14 LRDs.}
  \label{fig:acf_ccf_pure}
\end{figure}

As a sanity check, we remove a handful of questionable LRDs from our clustering sample and repeat the cross-correlation analysis. Specifically we excluded NX29853/NX12300 (with broad FWHM$<900\,{\rm km\,s^{-1}}$), and NX8933/NX12143 (no \Ha\ coverage). However we kept NX15499 even though its broad FWHM is $<900\,{\rm km\,s^{-1}}$, because it shows blueward \Ha\ absorption rarely seen in normal BLAGNs. The results are shown in Figure~\ref{fig:acf_ccf_pure}. We find nearly identical results albeit with larger uncertainties, compared with our fiducial results. We also verify that none of our LRDs are in apparent galaxy cluster environments that would bias the clustering measurements.

\begin{table}
\caption{Clustering Measurements}\label{tab:clustering}
\centering
%\scalebox{0.8}{
%\resizebox{\columnwidth}{!}{%
\begin{tabular}{llll}
\hline\hline
Sample & $\theta_0$ ($\beta=0.8$) & $r_0$ ($\gamma=1.8$) & linear bias \\
 & & $h^{-1}{\rm cMpc}$ & \\
(1) & (4) & (5) & (6) \\
\hline
gal & $0.40_{-0.14}^{+0.15}$ & $3.04_{-0.51}^{+0.47}$ &  $2.15_{-0.39}^{+0.55}$ \\
LRDxgal  & $1.77_{-0.91}^{+1.07}$ & $6.90_{-1.92}^{+1.62}$ & $4.90_{-1.40}^{+1.49}$ \\
BLAGNxgal & $0.41_{-0.29}^{+0.49}$ & $3.53_{-1.45}^{+1.51}$ & $2.62_{-1.03}^{+1.16}$ \\
\hline
\hline\\
\end{tabular}
%}
{\raggedright Notes. We caution that the linear bias is derived by extrapolating the power-law model measured on the one-halo scales to the much larger linear scales. The extrapolation may lead to significantly overestimated linear bias factors. }
\end{table}

\begin{figure}
\centering
  \includegraphics[width=0.5\textwidth]{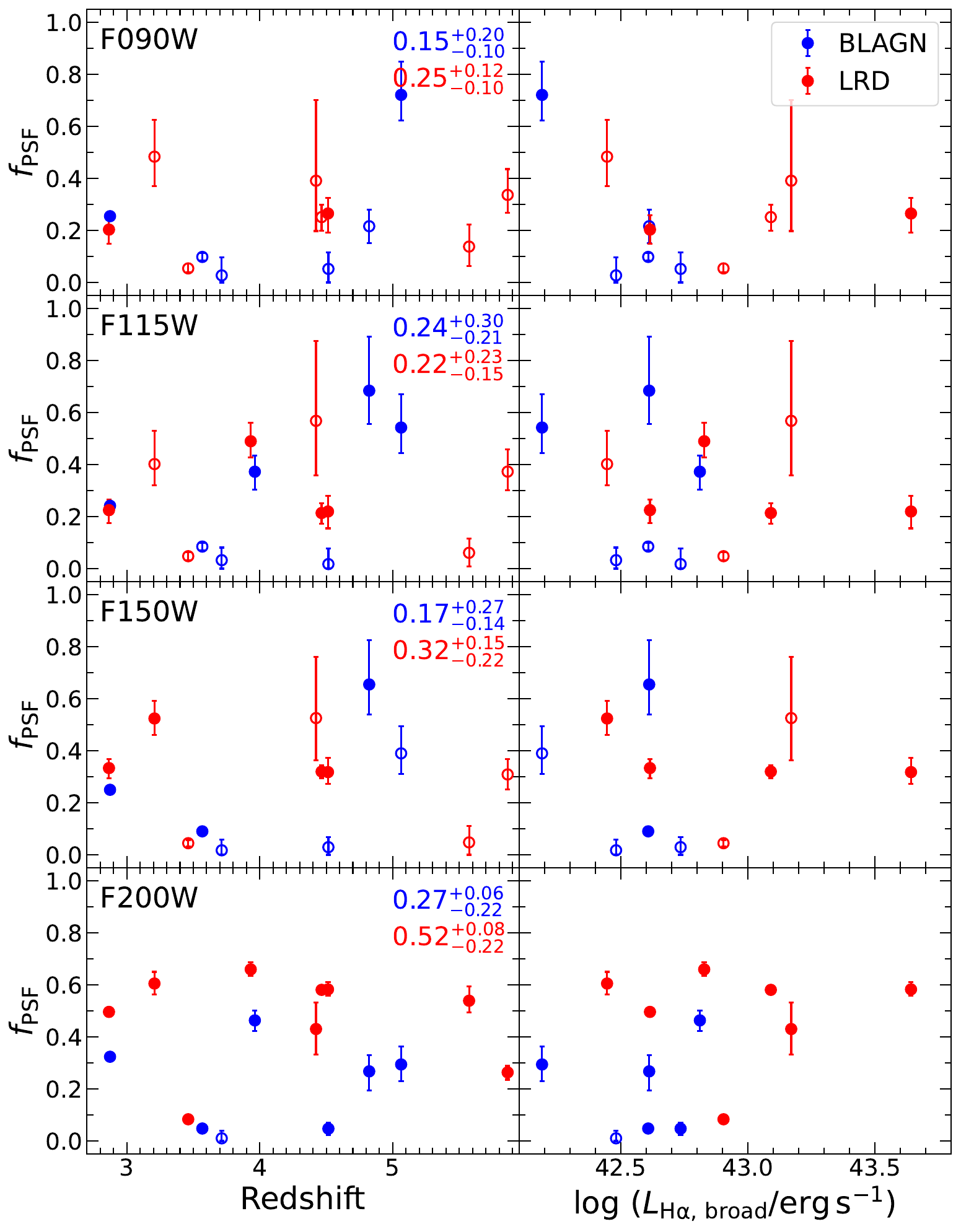}
  \caption{The fractional contribution of the PSF component to the total (PSF+\sersic) flux from the PSF+\sersic\ model set ($f_{\rm PSF}$) versus redshift (left column) and $L_{\rm H\alpha,\, broad}$ (right column). Rows from top to the bottom show the results in the F090W, F115W, F150W, and F200W filters, respectively. Blue and red dots represent BLAGNs and LRDs, respectively. Filled (open) dots indicate PSF magnitude brighter (fainter) than the $3\sigma$-depth of the image. The median value and 16th and 84th percentiles of $f_{\rm PSF}$ for BLAGNs and LRDs are shown at the upper-right corner of each panel in the left column. 
  \label{fig:fPSF}}
\end{figure}

\section{Discussion}\label{sec4}

Our findings with the NEXUS BLAGN/LRD sample reaffirm many observed properties in previous studies \citep{Greene+2024, Matthee+24, Lin_etal_2025b, JADES_BLAGN, Lin+2024, Matthee_etal_2024b, Taylor+2024, Juodzbalis2024rosetta}. In particular, we find that half (6/12) of LRDs show obvious Balmer absorption, which suggests high gas densities surrounding the emitting region. These dense gas clouds could also attenuate UV emission and produce the prominent Balmer break featured in LRDs \citep[e.g.,][]{Inayoshi&Maiolino2025, Naidu_BHstar_2025, Ji+2025}. Unfortunately, our WFSS spectral SNR is insufficient to study the Balmer line profile to investigate the origin of line broadening.

{The number density of BLAGNs remains roughly constant around a few times $10^{-5}\,{\rm cMpc}^{-1}$ over the redshift range $3\lesssim z\lesssim 6$, while the number density of LRDs tends to decline towards lower redshift as suggested in previous works \citep[e.g.,][]{Kocevski_LRD_selection, Ma+2025}. The decline of LRD density may be related to the cosmic evolution of accretion mode and ambient environment of BHs, with more frequent super-Eddington accretion triggered by surrounding dense gas at higher redshift \citep{Inayoshi2025firstbh}. Our results suggest that LRDs seem to dominate the BLAGN populations ($\sim70$\%) at $L_{\rm H\alpha}\gtrsim 10^{42}$ erg s$^{-1}$. However, towards lower luminosity ranges, \citet{Taylor+2024} find that their NIRSpec BLAGNs, which go down to $L_{\rm H\alpha}=10^{41.4}$ erg s$^{-1}$, have a diverse range of UV and optical slopes and only $\sim$20\% meet the LRD selection criteria. }

\subsection{Extended rest-UV emission}

The origin of the rest UV emission of LRDs has been debated \citep[e.g.,][]{Greene+2024, Barro+2024, Perez-Gonzalez+2024, Rinaldi+2024, Zhang2025sed, Inayoshi2025binary}. While the F444W emission is dominated by the point source, a significant fraction of LRDs are found to have complicated morphologies at rest-UV wavelengths \citep[e.g.,][]{Matthee+24, Rinaldi+2024, Chen+2025a, Hainline+2025_LRD_selection}. In Section~\ref{sec3.2}, we show that nearly 70\% of NEXUS LRDs show extended emission in the F200W filter, suggesting significant contribution from the underlying host galaxies. Figure~\ref{fig:fPSF} shows the fractional contribution of the PSF component to the total (PSF+\sersic) flux from the PSF+\sersic\ model set in four SW filters. We find that the extended emission makes a significant contribution to the total observed rest-UV emission, with the contribution increasing from $\sim 50\%$ in the F200W filter ($\sim$3600\AA\ at $z=4.5$) to $\sim 75\%$ in the F090W filter ($\sim$1600\AA\ at $z=4.5$). The disk-like structure ($n\approx1$) with sizes of $\sim$600 parsec and a high fraction (8/10) of asymmetric morphology or close companions strongly disfavor {an AGN accretion disk origin (e.g., scattered AGN continuum, binary SMBHs, two-component accretion disks; \citealt{Greene+2024, Zhang2025sed, Inayoshi2025binary})} for the UV emission in most LRDs -- though such models may still be viable for a small fraction of LRDs. Although host galaxy continuum emission from stellar populations can describe the observed SED of the extended emission, \citet{Chen+2025b} suggest that nebular emission from ionized gas in the host galaxy can also provide a viable explanation for the extended rest-UV emission in a handful of LRDs studied therein. 

\begin{figure*}
\centering
  \includegraphics[width=\textwidth]{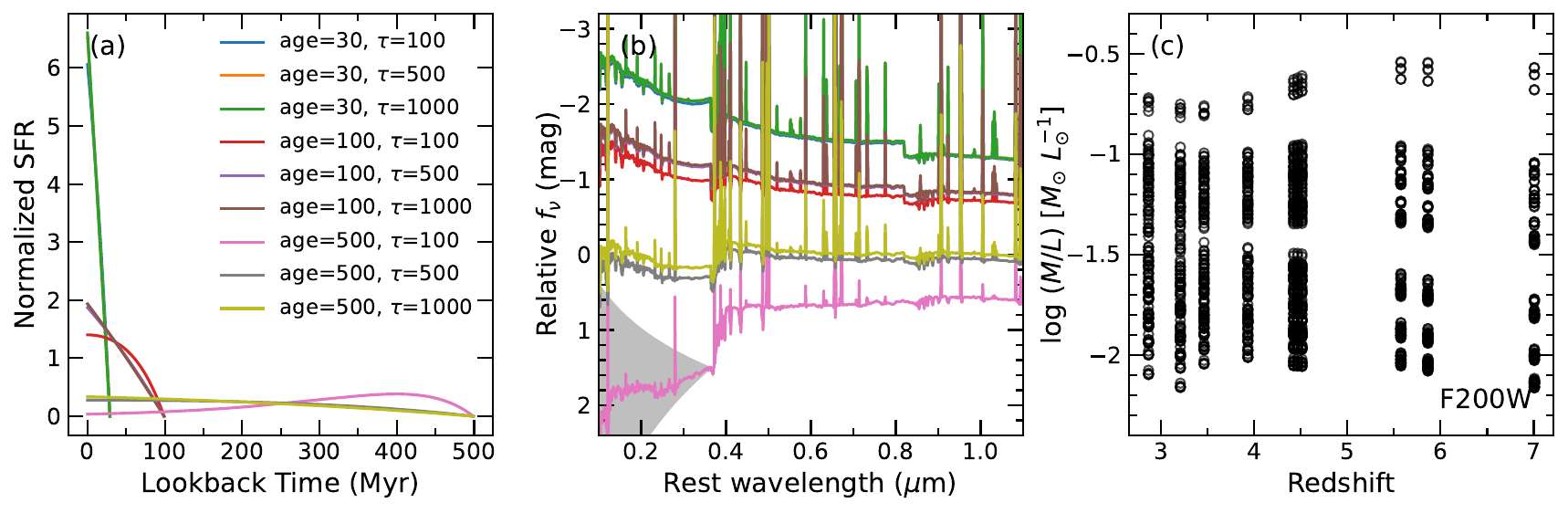}
  \caption{Star formation history (a), galaxy spectra (b), and mass-to-light ratio ($M/L$) in the observed F200W filter (c) of mock galaxies generated using \texttt{CIGALE}. Colors represent different combinations of stellar age and e-folding time $\tau$ in unit of Myr. Star formation histories are normalized by the integration of star formation rate (SFR) over time. Mock galaxy spectra are normalized by their current stellar masses. Only a subset of combinations are shown here for clarity in panels (a) and (b). Since stellar metallicity has much smaller influence on the overall shape of galaxy spectra probed here, we only show galaxy spectra with stellar metallicity of $0.4\times Z_{\odot}$ in panel (b). Gray shaded region in panel (b) shows the LRD UV spectral slope selection criterion ($-2.8<\beta_{\rm UV}<-0.37$).
  \label{fig:mock_galaxy}}
\end{figure*}

\begin{figure*}
\centering
  \includegraphics[width=0.8\textwidth]{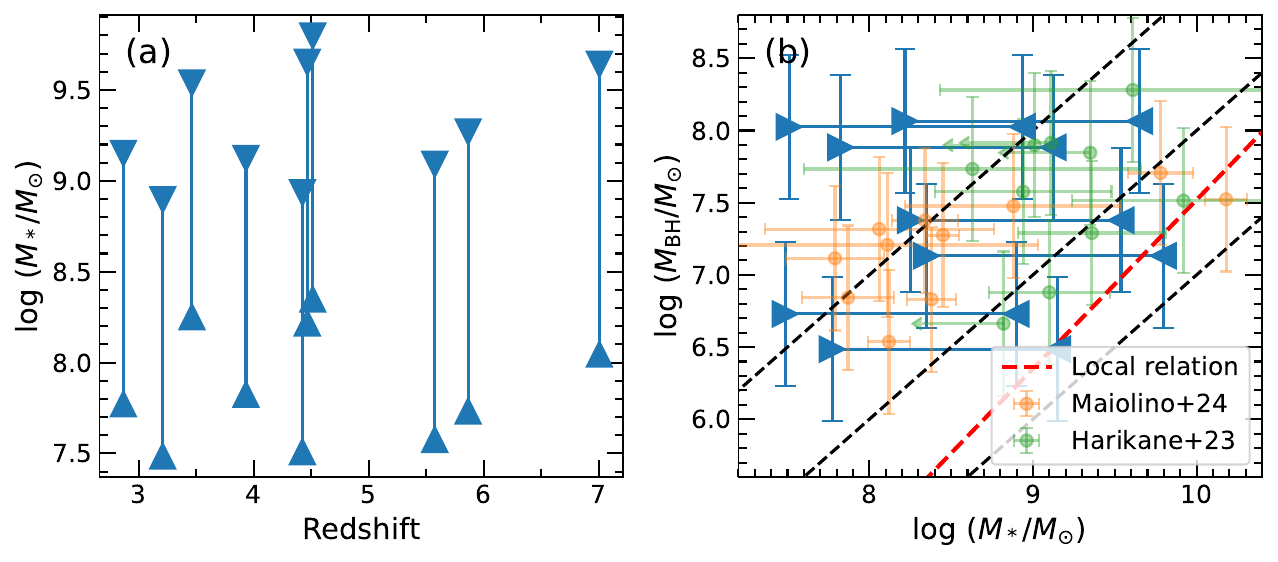}
  \caption{Stellar mass versus redshift (a) and BH mass versus stellar mass (b) for the LRDs with extended emission detection in this paper. Blue triangle pairs indicate the stellar mass range of NEXUS LRDs. Red dashed line represent the local $M_{\rm BH}-M_*$ relation for classical bulges and elliptical galaxies from \citet{Kormendy&Ho2013}, while black dashed lines from top to bottom represent $M_{\rm BH}/M_*=0.1, 0.01,$ and $0.001$, respectively. Orange and blue dots represent $4\lesssim z\lesssim 7$ BLAGNs from JADES \citep{Maiolino+2024_JADES_BLAGN} and CEERS \citep{Harikane+2023}, respectively. 
  \label{fig:LRD_Mstar}}
\end{figure*}

\subsection{Host stellar masses of LRDs}

Obtaining reliable stellar masses of the host galaxies of LRDs is extremely difficult, since the rest-optical emission is mostly unresolved, making the emission from the nucleus and the host highly degenerate. Moreover, young stars can easily outshine the overall stellar populations, leading to great uncertainties and systematics in stellar mass estimates \citep[e.g.,][]{Papovich2001, Narayanan+2024, Wang_BJ+2024}. Due to the lack of deep spectroscopy and the relatively shallow imaging data in NEXUS-EDR, we provide a crude estimate of stellar masses for the ten LRDs with robustly detected extended emission in the F200W filter. 

We explore a wide range of stellar population parameters to obtain a reasonable range of mass-to-light ratio ($M/L$) in the F200W filter to bracket the stellar masses of the LRDs. We use \texttt{CIGALE} \citep{CIGALE} to generate mock galaxy spectra at the redshift of each LRD. We adopt a delayed-$\tau$ star formation history with stellar age ranging from 30, 50, 100, 300, and 500 Myr and an e-folding time ranging from 100, 250, 500, 750, and 1000 Myr; a \citet{BC03} single stellar population model with a \citet{Chabrier2003} initial mass function and a stellar metallicity ranging from $0.02\times$, $0.2\times$, and $0.4\times$ solar metallicity ($Z_{\odot}$); a nebular emission model with gas metallicity tied to stellar metallicity and ionization parameter ranging from $10^{-4}$ and $10^{-3}$. All the other parameters are left at their default values. As we are mainly investigating the effect of different star formation histories (SFHs) on stellar mass estimates, we do not further consider the complications from extinction (such as the amount of extinction and the variation of extinction curve), which would generally result in underestimate of stellar mass. 

Figure~\ref{fig:mock_galaxy} shows the SFHs, mock galaxy spectra, and $M/L$ in the F200W filter in observed-frame of our mock galaxies. Our mock galaxies span from extremely young and bursty systems to more mildly star-forming galaxies with relatively flat or declining SFHs. At fixed stellar mass, the bursty galaxies outshine their older counterparts by up to approximately 3.5 mag at rest-UV wavelength. Regardless of their SFHs, all mock galaxies satisfy the LRD UV selection criterion ($-2.8<\beta_{\rm UV}<-0.37$), which is essential for ensuring a physically reasonable stellar mass range. The $M/L$ of the observed F200W filter has a large spread ($\sim$1.4 dex) across the entire redshift range, with small systematic variations due to the shift of the F200W filter in rest-frame wavelength. 

Applying the derived $M/L$ range to the ten LRDs with extended rest-UV emission, we find that these objects span a large range in stellar mass from $\sim10^{7.5}$ to $10^{9.8}\, M_{\odot}$. Combined with $M_{\rm BH}$ measurements, they tend to lie systematically above the local relation for classical bulges and elliptical galaxies \citep{Kormendy&Ho2013}. However, the upper boundary of the stellar mass estimates significantly relieve the tension in some of the $4\lesssim z\lesssim 7$ BLAGNs from JADES \citep{Maiolino+2024_JADES_BLAGN} and CEERS \citep{Harikane+2023}. Notably, NX7607 and NX5732 have highly overmassive BH with $M_{\rm BH}/M_*\approx0.1$ even adopting the upper boundary of stellar mass. Such objects may be real, which could be the tip of the iceberg owing to the selection bias in flux-limited AGN samples \citep{Li2025a}. On the other hand, the great uncertainty ($\sim0.5$~dex) in the single-epoch $M_{\rm BH}$ estimates may also contribute to the extreme mass ratio. 

Bearing in mind that the stellar mass ranges estimated here are only based on limited combinations of galaxy parameters and the mass-to-light ratio in a single band that probes rest-frame $<4000$\AA\ at $z>4$, the actual stellar masses of LRDs may still lie outside these boundaries. Fortunately, these objects are high priority NIRSpec spectroscopic follow-up targets in the NEXUS Deep field. The high SNR NIRSpec PRISM spectroscopy will enable more robust stellar mass estimation with modeling of the stellar and AGN continuum emission. Furthermore, additional epochs in the NEXUS Wide field \citep{nexus} will provide more constraints on the extended emission. 

{As mentioned in Section~\ref{sec3.2}, the companions of NX8933 and NX12143 (denoted as NX8933-P and NX12143-P) are at the same redshift as the LRDs. We derive the stellar masses of the companions from six band NIRCam photometry using \texttt{CIGALE}. We adopt the same set of modules as those for the mock galaxies, but use a denser grid in stellar age, a larger range of e-folding time by adding two values at 1250 and 1500 Myr, and add a dust attenuation module \texttt{dustatt\_modified\_starburst} with emission-line \texttt{E(B-V)} of 0.0, 0.01, 0.05, 0.1, 0.2, and 0.3. We find that both of the companions are relatively massive galaxies at $z\approx6$ with $\log\, M_*=9.4\pm0.2\, M_{\odot}$ for NX8933-P and $\log\, M_*=9.7\pm0.1\, M_{\odot}$ for NX12143-P, which are $\sim0.4$ dex higher than the upper boundary of the stellar masses of the LRDs ($\log\, M_*=9.1\, M_{\odot}$ for NX8933 and $\log\, M_*=9.3\, M_{\odot}$ for NX12143). Despite the large uncertainties in their stellar mass estimates, the low stellar masses of the two LRDs suggest that they are unlikely to be the central galaxy in their group environments.}

\subsection{Enhanced small-scale clustering?}

While the current NEXUS sample is still limited in statistics, we were able to measure the small-scale clustering of BLAGNs and LRDs over $3\lesssim z\lesssim 6$, albeit with substantial uncertainties. If we assume the measured small-scale power-law correlation function can be extrapolated to large (linear) scales, we infer a typical host halo mass of a few $10^{11}\,h^{-1}M_\odot$ for BLAGNs. This result is consistent with recent measurements of JWST BLAGNs at similar or higher redshifts \citep[e.g.,][]{Arita_etal_2025, Matthee_etal_2024b, Lin_etal_2025b}. 

However, the inferred linear bias and hence host halo masses ($\gtrsim 10^{12}\,h^{-1}M_\odot$ at $1\sigma$) for the spectroscopic LRD sample are too large. At face value, the measured strong LRD clustering is incompatible with the high number density of LRDs \citep[e.g.,][]{Pizzati_etal_2025}, given the rareness of these massive halos at high redshift. One solution is that LRDs are associated with more abundant satellite halos in these massive central halos. A more likely interpretation is that the power-law extrapolation from the one-halo term to larger (linear) scales is problematic, resulting in overestimated linear bias and host halo masses. It is possible that LRDs have strong excess clustering on small scales compared with the power-law extrapolation from large scales. For instance, \citet{Tanaka_etal_2024} reported cases of candidate dual LRDs on kpc-scales, which suggests strong excess over the clustering of BLAGNs that include both LRDs and non-LRDs \citep{Arita_etal_2025}. Future measurements using more JWST data covering the linear clustering scales are required to validate these initial clustering measurements for LRDs. 

\section{Conclusions}

We report a spectroscopic search for broad Balmer emitters and little red dots using data from the first set of NIRCam imaging and 2.4--5\,\micron WFSS spectroscopy of the NEXUS field. We present a spectroscopic sample of 23 BLAGNs, including 15 classified as LRDs, and one luminous quasar. These BLAGNs span a redshift range of $3\lesssim z\lesssim 6$ (with one at $z\approx 7$), and have similar properties to those of previous low-luminosity AGNs discovered by JWST. {The number density of the BLAGNs remain roughly constant around a few times $10^{-5}\,{\rm cMpc^{-1}}$ over this redshift range, while the number density of LRDs tend to decline towards the lower end of our redshift range albeit with large uncertainties.}

Our imaging analysis revealed significant host emission in rest-frame UV and optical of these LRDs. The morphology of extended host emission on a scale of hundreds of parsecs suggests they are most likely from star formation instead of scattered AGN light. The extended host emission contributes significantly or even dominantly to the total rest-UV flux of these LRDs, which can explain the UV upturn of the peculiar SED shape of LRDs. The large fraction (50\%) of LRDs displaying Balmer absorption implies large gas column densities, and is generally consistent with the gas enshrouded BH scenario to produce the red rest-optical spectral slope \citep[e.g.,][]{Naidu_BHstar_2025}. 

Finally, we perform clustering measurements of these BLAGNs and LRDs, using cross-correlation with a photometric galaxy sample in the NEXUS field (Section~\ref{sec3.4}). The clustering analyses suggest that JWST BLAGNs at $3\lesssim z\lesssim 6$ reside in dark matter halos with typical masses of several $10^{11}\,h^{-1}M_\odot$. However, the measured formal LRD linear bias is too large to be compatible with the abundance of LRDs. It is possible that LRDs have excess small-scale clustering, and therefore the extrapolation to larger scales significantly overestimates the linear bias. A larger LRD clustering sample that covers a wider survey footprint is required to fully address this issue.  

The current NEXUS WFSS spectroscopic sample represents one quarter of the final sample in terms of size and survey area \citep{nexus}. In addition, half of the current area will be revisited by the NEXUS Deep tier (NIRSpec MSA spectroscopy and NIRCam imaging) with a cadence of 2 months over 2025-2028. These future NEXUS observations will provide light curves for covered LRDs to investigate their variability properties, as well as high SNR NIRSpec 0.6--5.3\,\micron\ spectra to cover a broad range of spectral features.

\begin{acknowledgments}

We thank Jenny Greene for useful discussions that significantly improved the manuscript. Based on observations with the NASA/ESA/CSA James Webb Space Telescope obtained from the Barbara A. Mikulski Archive at the Space Telescope Science Institute, which is operated by the Association of Universities for Research in Astronomy, Incorporated, under NASA contract NAS5-03127. Support for Program numbers JWST-GO-02057, JWST-AR-03038, and JWST-GO-05105 was provided through a grant from the STScI under NASA contract NAS5-03127. 

\end{acknowledgments}


\vspace{5mm}
\facilities{JWST (NIRCam)}


\software{
\texttt{Astropy} \citep{2013A&A...558A..33A,2018AJ....156..123A}, 
\texttt{CIGALE} \citep{CIGALE}, 
\texttt{GALFITM} \citep{Haussler+2013MNRAS}, 
\texttt{Matplotlib} \citep{Hunter2007}, 
\texttt{Numpy} \citep{Harris2020}, 
\texttt{photutils} \citep{photutils}, 
\texttt{scipy} \citep{scipy}
          }



\appendix
\section{Image Decomposition Results in the F444W Filter}\label{appendix}
Figure~\ref{fig:Decomp_F444W} shows the image decomposition results in the F444W filter. We find that the \sersic\ component from the PSF+\sersic\ set is either very weak ($\gtrsim2$ mag fainter than the PSF component) or resembles a PSF model (barely resolved) in all the LRDs except for NX28630 and NX7680. As demonstrated in \citet{Zhuang&Shen2024}, the host and nucleus contributions in such cases are highly degenerated and the output magnitudes from the decomposition are not reliable. For NX28630, although its nucleus dominates the total flux, we can clearly see the underlying extended emission, which likely comes from the clumps seen in the F200W filter (Figure~\ref{fig:img_decomp1}). NX7680 also exhibits evidence of extended emission with F444W size larger than F200W size. We argue that it may be due to the extended \OIII\ emission falling in the bandpass of F444W and resulting in the more extended F444W morphology.

\begin{figure*}[t]
\centering
  \includegraphics[width=\textwidth]{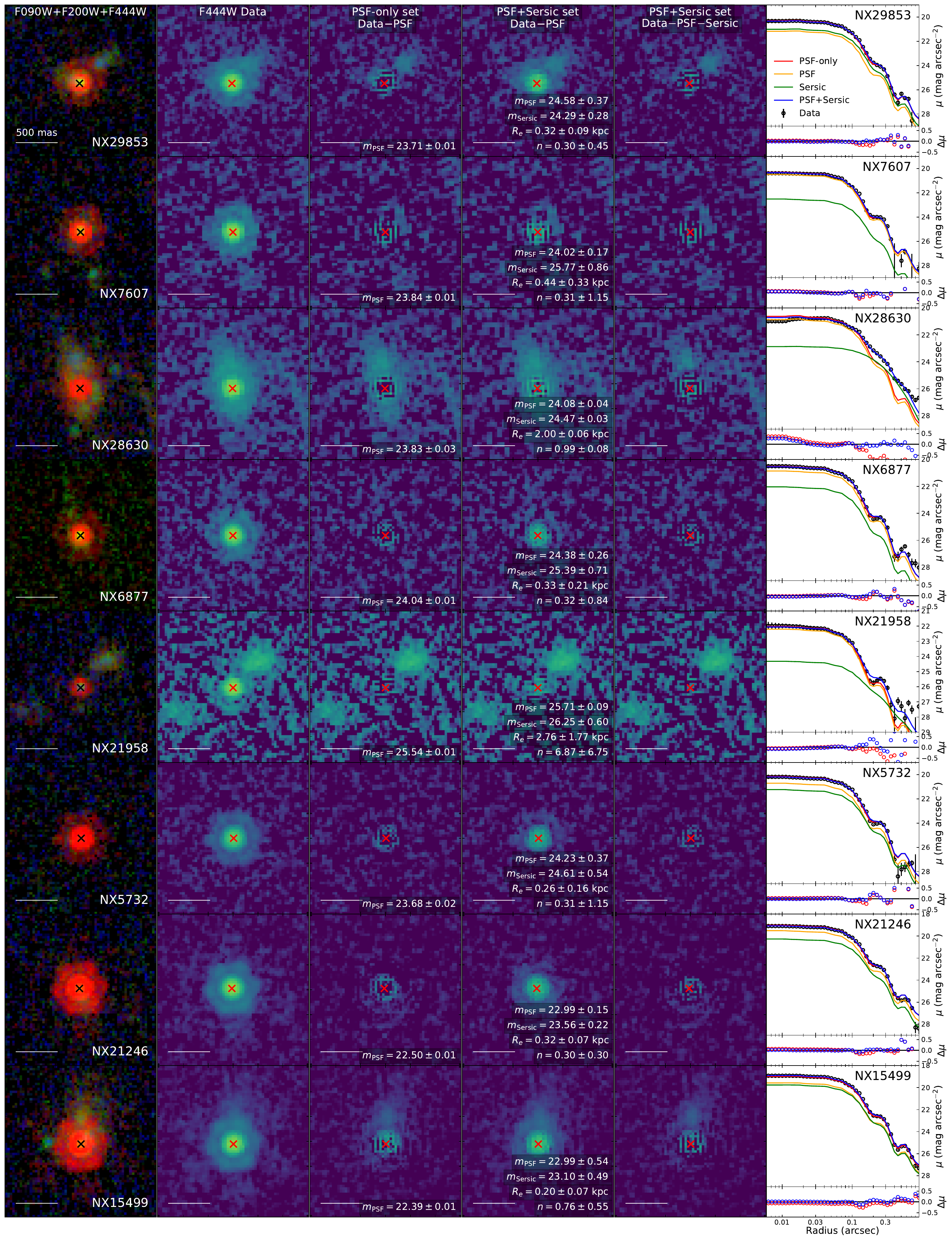}
  \caption{Same as Figure~\ref{fig:img_decomp1} but for the F444W filter.}\label{fig:Decomp_F444W}
\end{figure*}

\begin{figure*}[t]
\figurenum{17}
\centering
  \includegraphics[width=\textwidth]{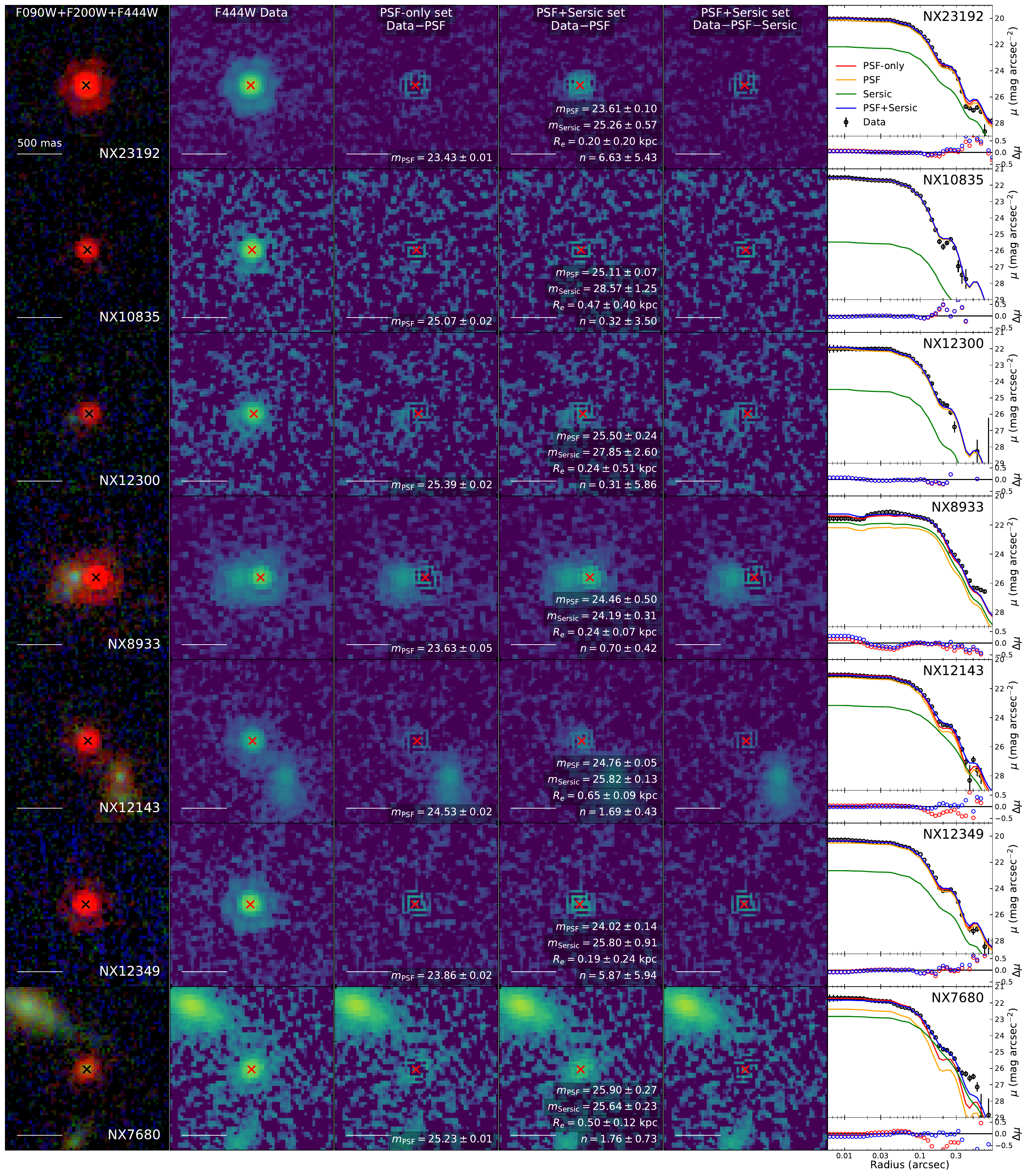}
  \caption{Continued.}
\end{figure*}


\bibliography{sample631}{}

\begin{thebibliography}{}
\expandafter\ifx\csname natexlab\endcsname\relax\def\natexlab#1{#1}\fi
\providecommand{\url}[1]{\href{#1}{#1}}
\providecommand{\dodoi}[1]{doi:~\href{http://doi.org/#1}{\nolinkurl{#1}}}
\providecommand{\doeprint}[1]{\href{http://ascl.net/#1}{\nolinkurl{http://ascl.net/#1}}}
\providecommand{\doarXiv}[1]{\href{https://arxiv.org/abs/#1}{\nolinkurl{https://arxiv.org/abs/#1}}}

\bibitem[{{Adelberger} {et~al.}(2005){Adelberger}, {Steidel}, {Pettini},
  {Shapley}, {Reddy}, \& {Erb}}]{Adelberger_2005}
{Adelberger}, K.~L., {Steidel}, C.~C., {Pettini}, M., {et~al.} 2005, \apj, 619,
  697, \dodoi{10.1086/426580}

\bibitem[{{Ananna} {et~al.}(2024){Ananna}, {Bogd{\'a}n}, {Kov{\'a}cs},
  {Natarajan}, \& {Hickox}}]{Ananna2024}
{Ananna}, T.~T., {Bogd{\'a}n}, {\'A}., {Kov{\'a}cs}, O.~E., {Natarajan}, P., \&
  {Hickox}, R.~C. 2024, \apjl, 969, L18, \dodoi{10.3847/2041-8213/ad5669}

\bibitem[{{Arita} {et~al.}(2025){Arita}, {Kashikawa}, {Onoue}, {Yoshioka},
  {Takeda}, {Hoshi}, \& {Shimizu}}]{Arita_etal_2025}
{Arita}, J., {Kashikawa}, N., {Onoue}, M., {et~al.} 2025, \mnras, 536, 3677,
  \dodoi{10.1093/mnras/stae2765}

\bibitem[{{Astropy Collaboration} {et~al.}(2013){Astropy Collaboration},
  {Robitaille}, {Tollerud}, {Greenfield}, {Droettboom}, {Bray}, {Aldcroft},
  {Davis}, {Ginsburg}, {Price-Whelan}, {Kerzendorf}, {Conley}, {Crighton},
  {Barbary}, {Muna}, {Ferguson}, {Grollier}, {Parikh}, {Nair}, {Unther},
  {Deil}, {Woillez}, {Conseil}, {Kramer}, {Turner}, {Singer}, {Fox}, {Weaver},
  {Zabalza}, {Edwards}, {Azalee Bostroem}, {Burke}, {Casey}, {Crawford},
  {Dencheva}, {Ely}, {Jenness}, {Labrie}, {Lim}, {Pierfederici}, {Pontzen},
  {Ptak}, {Refsdal}, {Servillat}, \& {Streicher}}]{2013A&A...558A..33A}
{Astropy Collaboration}, {Robitaille}, T.~P., {Tollerud}, E.~J., {et~al.} 2013,
  \aap, 558, A33, \dodoi{10.1051/0004-6361/201322068}

\bibitem[{{Astropy Collaboration} {et~al.}(2018){Astropy Collaboration},
  {Price-Whelan}, {Sip{\H{o}}cz}, {G{\"u}nther}, {Lim}, {Crawford}, {Conseil},
  {Shupe}, {Craig}, {Dencheva}, {Ginsburg}, {VanderPlas}, {Bradley},
  {P{\'e}rez-Su{\'a}rez}, {de Val-Borro}, {Aldcroft}, {Cruz}, {Robitaille},
  {Tollerud}, {Ardelean}, {Babej}, {Bach}, {Bachetti}, {Bakanov}, {Bamford},
  {Barentsen}, {Barmby}, {Baumbach}, {Berry}, {Biscani}, {Boquien}, {Bostroem},
  {Bouma}, {Brammer}, {Bray}, {Breytenbach}, {Buddelmeijer}, {Burke},
  {Calderone}, {Cano Rodr{\'\i}guez}, {Cara}, {Cardoso}, {Cheedella}, {Copin},
  {Corrales}, {Crichton}, {D'Avella}, {Deil}, {Depagne}, {Dietrich}, {Donath},
  {Droettboom}, {Earl}, {Erben}, {Fabbro}, {Ferreira}, {Finethy}, {Fox},
  {Garrison}, {Gibbons}, {Goldstein}, {Gommers}, {Greco}, {Greenfield},
  {Groener}, {Grollier}, {Hagen}, {Hirst}, {Homeier}, {Horton}, {Hosseinzadeh},
  {Hu}, {Hunkeler}, {Ivezi{\'c}}, {Jain}, {Jenness}, {Kanarek}, {Kendrew},
  {Kern}, {Kerzendorf}, {Khvalko}, {King}, {Kirkby}, {Kulkarni}, {Kumar},
  {Lee}, {Lenz}, {Littlefair}, {Ma}, {Macleod}, {Mastropietro}, {McCully},
  {Montagnac}, {Morris}, {Mueller}, {Mumford}, {Muna}, {Murphy}, {Nelson},
  {Nguyen}, {Ninan}, {N{\"o}the}, {Ogaz}, {Oh}, {Parejko}, {Parley}, {Pascual},
  {Patil}, {Patil}, {Plunkett}, {Prochaska}, {Rastogi}, {Reddy Janga},
  {Sabater}, {Sakurikar}, {Seifert}, {Sherbert}, {Sherwood-Taylor}, {Shih},
  {Sick}, {Silbiger}, {Singanamalla}, {Singer}, {Sladen}, {Sooley},
  {Sornarajah}, {Streicher}, {Teuben}, {Thomas}, {Tremblay}, {Turner},
  {Terr{\'o}n}, {van Kerkwijk}, {de la Vega}, {Watkins}, {Weaver}, {Whitmore},
  {Woillez}, {Zabalza}, \& {Astropy Contributors}}]{2018AJ....156..123A}
{Astropy Collaboration}, {Price-Whelan}, A.~M., {Sip{\H{o}}cz}, B.~M., {et~al.}
  2018, \aj, 156, 123, \dodoi{10.3847/1538-3881/aabc4f}

\bibitem[{{Baggen} {et~al.}(2024){Baggen}, {van Dokkum}, {Brammer}, {de
  Graaff}, {Franx}, {Greene}, {Labb{\'e}}, {Leja}, {Maseda}, {Nelson}, {Rix},
  {Wang}, \& {Weibel}}]{Baggen2024}
{Baggen}, J. F.~W., {van Dokkum}, P., {Brammer}, G., {et~al.} 2024, \apjl, 977,
  L13, \dodoi{10.3847/2041-8213/ad90b8}

\bibitem[{{Barro} {et~al.}(2024){Barro}, {P{\'e}rez-Gonz{\'a}lez}, {Kocevski},
  {McGrath}, {Trump}, {Simons}, {Somerville}, {Yung}, {Arrabal Haro}, {Akins},
  {Bagley}, {Cleri}, {Costantin}, {Davis}, {Dickinson}, {Finkelstein},
  {Giavalisco}, {G{\'o}mez-Guijarro}, {Hathi}, {Hirschmann}, {Holwerda},
  {Huertas-Company}, {Kartaltepe}, {Koekemoer}, {Lucas}, {Papovich}, {Pirzkal},
  {Seill{\'e}}, {Tacchella}, {Wuyts}, {Wilkins}, {de la Vega}, {Yang}, \&
  {Zavala}}]{Barro+2024}
{Barro}, G., {P{\'e}rez-Gonz{\'a}lez}, P.~G., {Kocevski}, D.~D., {et~al.} 2024,
  \apj, 963, 128, \dodoi{10.3847/1538-4357/ad167e}

\bibitem[{{Bellovary}(2025)}]{Bellovary2025}
{Bellovary}, J. 2025, \apjl, 984, L55, \dodoi{10.3847/2041-8213/adce6c}

\bibitem[{{Bertemes} {et~al.}(2025){Bertemes}, {Wylezalek}, {Rupke},
  {Zakamska}, {Veilleux}, {Beckmann}, {Vayner}, {Sankar}, {Ishikawa},
  {Diachenko}, {Liu}, {Chen}, {Seebeck}, {Lutz}, \& {Liu}}]{Bertemes2025}
{Bertemes}, C., {Wylezalek}, D., {Rupke}, D.~S.~N., {et~al.} 2025, \aap, 693,
  A176, \dodoi{10.1051/0004-6361/202450451}

\bibitem[{{Bertin} \& {Arnouts}(1996)}]{1996A&AS..117..393B}
{Bertin}, E., \& {Arnouts}, S. 1996, \aaps, 117, 393,
  \dodoi{10.1051/aas:1996164}

\bibitem[{{Bhowmick} {et~al.}(2024){Bhowmick}, {Blecha}, {Torrey},
  {Somerville}, {Kelley}, {Vogelsberger}, {Weinberger}, {Hernquist}, \&
  {Sivasankaran}}]{Bhowmick2024}
{Bhowmick}, A.~K., {Blecha}, L., {Torrey}, P., {et~al.} 2024, \mnras, 533,
  1907, \dodoi{10.1093/mnras/stae1819}

\bibitem[{{Bogd{\'a}n} {et~al.}(2024){Bogd{\'a}n}, {Goulding}, {Natarajan},
  {Kov{\'a}cs}, {Tremblay}, {Chadayammuri}, {Volonteri}, {Kraft}, {Forman},
  {Jones}, {Churazov}, \& {Zhuravleva}}]{Bogdan2024}
{Bogd{\'a}n}, {\'A}., {Goulding}, A.~D., {Natarajan}, P., {et~al.} 2024, Nature
  Astronomy, 8, 126, \dodoi{10.1038/s41550-023-02111-9}

\bibitem[{{Boquien} {et~al.}(2019){Boquien}, {Burgarella}, {Roehlly}, {Buat},
  {Ciesla}, {Corre}, {Inoue}, \& {Salas}}]{CIGALE}
{Boquien}, M., {Burgarella}, D., {Roehlly}, Y., {et~al.} 2019, \aap, 622, A103,
  \dodoi{10.1051/0004-6361/201834156}

\bibitem[{Bradley {et~al.}(2022)Bradley, Sip{\H o}cz, Robitaille, Tollerud,
  Vin{\'{\i}}cius, Deil, Barbary, Wilson, Busko, G{\"u}nther, Cara, Conseil,
  Bostroem, Droettboom, Bray, Bratholm, Lim, Barentsen, Craig, Pascual, Perren,
  Greco, Donath, de~Val-Borro, Kerzendorf, Bach, Weaver, D'Eugenio, Souchereau,
  \& Ferreira}]{photutils}
Bradley, L., Sip{\H o}cz, B., Robitaille, T., {et~al.} 2022, astropy/photutils:
  1.5.0, 1.5.0,  Zenodo, \dodoi{10.5281/zenodo.6825092}

\bibitem[{{Brammer} {et~al.}(2008){Brammer}, {van Dokkum}, \& {Coppi}}]{EAZY}
{Brammer}, G.~B., {van Dokkum}, P.~G., \& {Coppi}, P. 2008, \apj, 686, 1503,
  \dodoi{10.1086/591786}

\bibitem[{{Bruzual} \& {Charlot}(2003)}]{BC03}
{Bruzual}, G., \& {Charlot}, S. 2003, \mnras, 344, 1000,
  \dodoi{10.1046/j.1365-8711.2003.06897.x}

\bibitem[{{Casey} {et~al.}(2024){Casey}, {Akins}, {Kokorev}, {McKinney},
  {Cooper}, {Long}, {Franco}, \& {Manning}}]{Casey2024}
{Casey}, C.~M., {Akins}, H.~B., {Kokorev}, V., {et~al.} 2024, \apjl, 975, L4,
  \dodoi{10.3847/2041-8213/ad7ba7}

\bibitem[{{Chabrier}(2003)}]{Chabrier2003}
{Chabrier}, G. 2003, \pasp, 115, 763, \dodoi{10.1086/376392}

\bibitem[{{Chen} {et~al.}(2025{\natexlab{a}}){Chen}, {Ho}, {Li}, \&
  {Inayoshi}}]{Chen+2025b}
{Chen}, C.-H., {Ho}, L.~C., {Li}, R., \& {Inayoshi}, K. 2025{\natexlab{a}},
  arXiv e-prints, arXiv:2505.03183, \dodoi{10.48550/arXiv.2505.03183}

\bibitem[{{Chen} {et~al.}(2025{\natexlab{b}}){Chen}, {Ho}, {Li}, \&
  {Zhuang}}]{Chen+2025a}
{Chen}, C.-H., {Ho}, L.~C., {Li}, R., \& {Zhuang}, M.-Y. 2025{\natexlab{b}},
  \apj, 983, 60, \dodoi{10.3847/1538-4357/ada93a}

\bibitem[{{Croom} \& {Shanks}(1999)}]{Croom_Shanks_1999}
{Croom}, S.~M., \& {Shanks}, T. 1999, \mnras, 303, 411,
  \dodoi{10.1046/j.1365-8711.1999.02232.x}

\bibitem[{{de Graaff} {et~al.}(2025){de Graaff}, {Rix}, {Naidu}, {Labbe},
  {Wang}, {Leja}, {Matthee}, {Katz}, {Greene}, {Hviding}, {Baggen}, {Bezanson},
  {Boogaard}, {Brammer}, {Dayal}, {van Dokkum}, {Goulding}, {Hirschmann},
  {Maseda}, {McConachie}, {Miller}, {Nelson}, {Oesch}, {Setton}, {Shivaei},
  {Weibel}, {Whitaker}, \& {Williams}}]{deGraaff2025}
{de Graaff}, A., {Rix}, H.-W., {Naidu}, R.~P., {et~al.} 2025, arXiv e-prints,
  arXiv:2503.16600, \dodoi{10.48550/arXiv.2503.16600}

\bibitem[{{D'Eugenio} {et~al.}(2025){D'Eugenio}, {Maiolino}, {Perna}, {Uebler},
  {Ji}, {McClymont}, {Koudmani}, {Sijacki}, {Juod{\v{z}}balis}, {Scholtz},
  {Bennett}, {Bunker}, {Carniani}, {Charlot}, {Cresci}, {Curtis-Lake}, {Dalla
  Bont{\`a}}, {Jones}, {Lyu}, {Marconi}, {Mazzolari}, {Nelson}, {Parlanti},
  {Robertson}, {Schneider}, {Simmonds}, {Tacchella}, {Venturi}, {Willott},
  {Witstok}, \& {Witten}}]{DEugenio2025}
{D'Eugenio}, F., {Maiolino}, R., {Perna}, M., {et~al.} 2025, arXiv e-prints,
  arXiv:2503.11752, \dodoi{10.48550/arXiv.2503.11752}

\bibitem[{{Furtak} {et~al.}(2025){Furtak}, {Secunda}, {Greene}, {Zitrin},
  {Labb{\'e}}, {Golubchik}, {Bezanson}, {Kokorev}, {Atek}, {Brammer},
  {Chemerynska}, {Cutler}, {Dayal}, {Feldmann}, {Fujimoto}, {Glazebrook},
  {Leja}, {Ma}, {Matthee}, {Naidu}, {Nelson}, {Oesch}, {Pan}, {Price}, {Suess},
  {Wang}, {Weaver}, \& {Whitaker}}]{Furtak2025}
{Furtak}, L.~J., {Secunda}, A.~R., {Greene}, J.~E., {et~al.} 2025, arXiv
  e-prints, arXiv:2502.07875, \dodoi{10.48550/arXiv.2502.07875}

\bibitem[{{Greene} \& {Ho}(2005)}]{Greene_Ho_2005}
{Greene}, J.~E., \& {Ho}, L.~C. 2005, \apj, 630, 122, \dodoi{10.1086/431897}

\bibitem[{{Greene} {et~al.}(2024){Greene}, {Labbe}, {Goulding}, {Furtak},
  {Chemerynska}, {Kokorev}, {Dayal}, {Volonteri}, {Williams}, {Wang}, {Setton},
  {Burgasser}, {Bezanson}, {Atek}, {Brammer}, {Cutler}, {Feldmann}, {Fujimoto},
  {Glazebrook}, {de Graaff}, {Khullar}, {Leja}, {Marchesini}, {Maseda},
  {Matthee}, {Miller}, {Naidu}, {Nanayakkara}, {Oesch}, {Pan}, {Papovich},
  {Price}, {van Dokkum}, {Weaver}, {Whitaker}, \& {Zitrin}}]{Greene+2024}
{Greene}, J.~E., {Labbe}, I., {Goulding}, A.~D., {et~al.} 2024, \apj, 964, 39,
  \dodoi{10.3847/1538-4357/ad1e5f}

\bibitem[{{Greene} {et~al.}(2017){Greene}, {Kelly}, {Stansberry}, {Leisenring},
  {Egami}, {Schlawin}, {Chu}, {Hodapp}, \& {Rieke}}]{2017JATIS...3c5001G}
{Greene}, T.~P., {Kelly}, D.~M., {Stansberry}, J., {et~al.} 2017, Journal of
  Astronomical Telescopes, Instruments, and Systems, 3, 035001,
  \dodoi{10.1117/1.JATIS.3.3.035001}

\bibitem[{{Hainline} {et~al.}(2025){Hainline}, {Maiolino}, {Juod{\v{z}}balis},
  {Scholtz}, {{\"U}bler}, {D'Eugenio}, {Helton}, {Sun}, {Sun}, {Robertson},
  {Tacchella}, {Bunker}, {Carniani}, {Charlot}, {Curtis-Lake}, {Egami},
  {Johnson}, {Lin}, {Lyu}, {P{\'e}rez-Gonz{\'a}lez}, {Rinaldi}, {Silcock},
  {Venturi}, {Williams}, {Willmer}, {Willott}, {Zhang}, \&
  {Zhu}}]{Hainline+2025_LRD_selection}
{Hainline}, K.~N., {Maiolino}, R., {Juod{\v{z}}balis}, I., {et~al.} 2025, \apj,
  979, 138, \dodoi{10.3847/1538-4357/ad9920}

\bibitem[{{Hall}(2007)}]{Hall_2007}
{Hall}, P.~B. 2007, \aj, 133, 1271, \dodoi{10.1086/511272}

\bibitem[{{Harikane} {et~al.}(2023){Harikane}, {Zhang}, {Nakajima}, {Ouchi},
  {Isobe}, {Ono}, {Hatano}, {Xu}, \& {Umeda}}]{Harikane+2023}
{Harikane}, Y., {Zhang}, Y., {Nakajima}, K., {et~al.} 2023, \apj, 959, 39,
  \dodoi{10.3847/1538-4357/ad029e}

\bibitem[{Harris {et~al.}(2020)Harris, Millman, van~der Walt, Gommers,
  Virtanen, Cournapeau, Wieser, Taylor, Berg, Smith, Kern, Picus, Hoyer, van
  Kerkwijk, Brett, Haldane, del R{\'{i}}o, Wiebe, Peterson,
  G{\'{e}}rard-Marchant, Sheppard, Reddy, Weckesser, Abbasi, Gohlke, \&
  Oliphant}]{Harris2020}
Harris, C.~R., Millman, K.~J., van~der Walt, S.~J., {et~al.} 2020, Nature, 585,
  357, \dodoi{10.1038/s41586-020-2649-2}

\bibitem[{{H{\"a}u{\ss}ler} {et~al.}(2013){H{\"a}u{\ss}ler}, {Bamford}, {Vika},
  {Rojas}, {Barden}, {Kelvin}, {Alpaslan}, {Robotham}, {Driver}, {Baldry},
  {Brough}, {Hopkins}, {Liske}, {Nichol}, {Popescu}, \&
  {Tuffs}}]{Haussler+2013MNRAS}
{H{\"a}u{\ss}ler}, B., {Bamford}, S.~P., {Vika}, M., {et~al.} 2013, \mnras,
  430, 330, \dodoi{10.1093/mnras/sts633}

\bibitem[{{He} {et~al.}(2018){He}, {Akiyama}, {Bosch}, {Enoki}, {Harikane},
  {Ikeda}, {Kashikawa}, {Kawaguchi}, {Komiyama}, {Lee}, {Matsuoka}, {Miyazaki},
  {Nagao}, {Nagashima}, {Niida}, {Nishizawa}, {Oguri}, {Onoue}, {Oogi},
  {Ouchi}, {Schulze}, {Shirasaki}, {Silverman}, {Tanaka}, {Tanaka}, {Toba},
  {Uchiyama}, \& {Yamashita}}]{He_etal_2018}
{He}, W., {Akiyama}, M., {Bosch}, J., {et~al.} 2018, \pasj, 70, S33,
  \dodoi{10.1093/pasj/psx129}

\bibitem[{Hunter(2007)}]{Hunter2007}
Hunter, J.~D. 2007, Computing in Science \& Engineering, 9, 90,
  \dodoi{10.1109/MCSE.2007.55}

\bibitem[{{Inayoshi}(2025)}]{Inayoshi2025firstbh}
{Inayoshi}, K. 2025, arXiv e-prints, arXiv:2503.05537,
  \dodoi{10.48550/arXiv.2503.05537}

\bibitem[{{Inayoshi} {et~al.}(2024){Inayoshi}, {Kimura}, \&
  {Noda}}]{Inayoshi2024xray}
{Inayoshi}, K., {Kimura}, S., \& {Noda}, H. 2024, arXiv e-prints,
  arXiv:2412.03653, \dodoi{10.48550/arXiv.2412.03653}

\bibitem[{{Inayoshi} \& {Maiolino}(2025)}]{Inayoshi&Maiolino2025}
{Inayoshi}, K., \& {Maiolino}, R. 2025, \apjl, 980, L27,
  \dodoi{10.3847/2041-8213/adaebd}

\bibitem[{{Inayoshi} {et~al.}(2025){Inayoshi}, {Shangguan}, {Chen}, {Ho}, \&
  {Haiman}}]{Inayoshi2025binary}
{Inayoshi}, K., {Shangguan}, J., {Chen}, X., {Ho}, L.~C., \& {Haiman}, Z. 2025,
  arXiv e-prints, arXiv:2505.05322, \dodoi{10.48550/arXiv.2505.05322}

\bibitem[{{Jeon} {et~al.}(2025){Jeon}, {Bromm}, {Liu}, \&
  {Finkelstein}}]{Jeon2025}
{Jeon}, J., {Bromm}, V., {Liu}, B., \& {Finkelstein}, S.~L. 2025, \apj, 979,
  127, \dodoi{10.3847/1538-4357/ad9f3a}

\bibitem[{{Ji} {et~al.}(2025){Ji}, {Maiolino}, {{\"U}bler}, {Scholtz},
  {D'Eugenio}, {Sun}, {Perna}, {Turner}, {Arribas}, {Bennett}, {Bunker},
  {Carniani}, {Charlot}, {Cresci}, {Curti}, {Egami}, {Fabian}, {Inayoshi},
  {Isobe}, {Jones}, {Juod{\v{z}}balis}, {Kumari}, {Lyu}, {Mazzolari},
  {Parlanti}, {Robertson}, {Rodr{\'\i}guez Del Pino}, {Schneider}, {Sijacki},
  {Tacchella}, {Trinca}, {Valiante}, {Venturi}, {Volonteri}, {Willott},
  {Witten}, \& {Witstok}}]{Ji+2025}
{Ji}, X., {Maiolino}, R., {{\"U}bler}, H., {et~al.} 2025, arXiv e-prints,
  arXiv:2501.13082, \dodoi{10.48550/arXiv.2501.13082}

\bibitem[{{Jiang} {et~al.}(2019){Jiang}, {Stone}, \&
  {Davis}}]{2019ApJ...880...67J}
{Jiang}, Y.-F., {Stone}, J.~M., \& {Davis}, S.~W. 2019, \apj, 880, 67,
  \dodoi{10.3847/1538-4357/ab29ff}

\bibitem[{{Juod{\v{z}}balis} {et~al.}(2024){Juod{\v{z}}balis}, {Ji},
  {Maiolino}, {D'Eugenio}, {Scholtz}, {Risaliti}, {Fabian}, {Mazzolari},
  {Gilli}, {Prandoni}, {Arribas}, {Bunker}, {Carniani}, {Charlot},
  {Curtis-Lake}, {de Graaff}, {Hainline}, {Parlanti}, {Perna},
  {P{\'e}rez-Gonz{\'a}lez}, {Robertson}, {Tacchella}, {{\"U}bler}, {Williams},
  {Willott}, \& {Witstok}}]{Juodzbalis2024rosetta}
{Juod{\v{z}}balis}, I., {Ji}, X., {Maiolino}, R., {et~al.} 2024, \mnras, 535,
  853, \dodoi{10.1093/mnras/stae2367}

\bibitem[{{Juod{\v{z}}balis} {et~al.}(2025){Juod{\v{z}}balis}, {Maiolino},
  {Baker}, {Lake}, {Scholtz}, {D'Eugenio}, {Trefoloni}, {Isobe}, {Tacchella},
  {Bunker}, {Carniani}, {Charlot}, {Jones}, {Parlanti}, {Perna}, {Rinaldi},
  {Robertson}, {{\"U}bler}, {Venturi}, \& {Willott}}]{JADES_BLAGN}
{Juod{\v{z}}balis}, I., {Maiolino}, R., {Baker}, W.~M., {et~al.} 2025, arXiv
  e-prints, arXiv:2504.03551, \dodoi{10.48550/arXiv.2504.03551}

\bibitem[{{Killi} {et~al.}(2024){Killi}, {Watson}, {Brammer}, {McPartland},
  {Antwi-Danso}, {Newshore}, {Coe}, {Allen}, {Fynbo}, {Gould}, {Heintz},
  {Rusakov}, \& {Vejlgaard}}]{Killi2024}
{Killi}, M., {Watson}, D., {Brammer}, G., {et~al.} 2024, \aap, 691, A52,
  \dodoi{10.1051/0004-6361/202348857}

\bibitem[{{Kocevski} {et~al.}(2024){Kocevski}, {Finkelstein}, {Barro},
  {Taylor}, {Calabr{\`o}}, {Laloux}, {Buchner}, {Trump}, {Leung}, {Yang},
  {Dickinson}, {P{\'e}rez-Gonz{\'a}lez}, {Pacucci}, {Inayoshi}, {Somerville},
  {McGrath}, {Akins}, {Bagley}, {Bisigello}, {Bowler}, {Carnall}, {Casey},
  {Cheng}, {Cleri}, {Costantin}, {Cullen}, {Davis}, {Donnan}, {Dunlop},
  {Ellis}, {Ferguson}, {Fujimoto}, {Fontana}, {Giavalisco}, {Grazian},
  {Grogin}, {Hathi}, {Hirschmann}, {Huertas-Company}, {Holwerda},
  {Illingworth}, {Juneau}, {Kartaltepe}, {Koekemoer}, {Li}, {Lucas}, {Magee},
  {Mason}, {McLeod}, {McLure}, {Napolitano}, {Papovich}, {Pirzkal},
  {Rodighiero}, {Santini}, {Wilkins}, \& {Yung}}]{Kocevski_LRD_selection}
{Kocevski}, D.~D., {Finkelstein}, S.~L., {Barro}, G., {et~al.} 2024, arXiv
  e-prints, arXiv:2404.03576, \dodoi{10.48550/arXiv.2404.03576}

\bibitem[{{Kokorev} {et~al.}(2024){Kokorev}, {Chisholm}, {Endsley},
  {Finkelstein}, {Greene}, {Akins}, {Bromm}, {Casey}, {Fujimoto}, {Labb{\'e}},
  \& {Larson}}]{Kokorev2024z4LRD}
{Kokorev}, V., {Chisholm}, J., {Endsley}, R., {et~al.} 2024, \apj, 975, 178,
  \dodoi{10.3847/1538-4357/ad7d03}

\bibitem[{{Kokubo} \& {Harikane}(2024)}]{Kokubo2024}
{Kokubo}, M., \& {Harikane}, Y. 2024, arXiv e-prints, arXiv:2407.04777,
  \dodoi{10.48550/arXiv.2407.04777}

\bibitem[{{Kormendy} \& {Ho}(2013)}]{Kormendy&Ho2013}
{Kormendy}, J., \& {Ho}, L.~C. 2013, \araa, 51, 511,
  \dodoi{10.1146/annurev-astro-082708-101811}

\bibitem[{{Labbe} {et~al.}(2024){Labbe}, {Greene}, {Matthee}, {Treiber},
  {Kokorev}, {Miller}, {Kramarenko}, {Setton}, {Ma}, {Goulding}, {Bezanson},
  {Naidu}, {Williams}, {Atek}, {Brammer}, {Cutler}, {Chemerynska}, {Cloonan},
  {Dayal}, {de Graaff}, {Fudamoto}, {Fujimoto}, {Furtak}, {Glazebrook},
  {Heintz}, {Leja}, {Marchesini}, {Nanayakkara}, {Nelson}, {Oesch}, {Pan},
  {Price}, {Shivaei}, {Sobral}, {Suess}, {van Dokkum}, {Wang}, {Weaver},
  {Whitaker}, \& {Zitrin}}]{Labbe2024}
{Labbe}, I., {Greene}, J.~E., {Matthee}, J., {et~al.} 2024, arXiv e-prints,
  arXiv:2412.04557, \dodoi{10.48550/arXiv.2412.04557}

\bibitem[{{Labbe} {et~al.}(2025){Labbe}, {Greene}, {Bezanson}, {Fujimoto},
  {Furtak}, {Goulding}, {Matthee}, {Naidu}, {Oesch}, {Atek}, {Brammer},
  {Chemerynska}, {Coe}, {Cutler}, {Dayal}, {Feldmann}, {Franx}, {Glazebrook},
  {Leja}, {Maseda}, {Marchesini}, {Nanayakkara}, {Nelson}, {Pan}, {Papovich},
  {Price}, {Suess}, {Wang}, {Weaver}, {Whitaker}, {Williams}, \&
  {Zitrin}}]{Labbe+2025ApJ}
{Labbe}, I., {Greene}, J.~E., {Bezanson}, R., {et~al.} 2025, \apj, 978, 92,
  \dodoi{10.3847/1538-4357/ad3551}

\bibitem[{{Lambrides} {et~al.}(2024){Lambrides}, {Garofali}, {Larson}, {Ptak},
  {Chiaberge}, {Long}, {Hutchison}, {Norman}, {McKinney}, {Akins}, {Berg},
  {Chisholm}, {Civano}, {Cloonan}, {Endsley}, {Faisst}, {Gilli}, {Gillman},
  {Hirschmann}, {Kartaltepe}, {Kocevski}, {Kokorev}, {Pacucci}, {Richardson},
  {Stiavelli}, \& {Whalen}}]{Lambrides2024}
{Lambrides}, E., {Garofali}, K., {Larson}, R., {et~al.} 2024, arXiv e-prints,
  arXiv:2409.13047, \dodoi{10.48550/arXiv.2409.13047}

\bibitem[{{Landy} \& {Szalay}(1993)}]{Landy_Szalay}
{Landy}, S.~D., \& {Szalay}, A.~S. 1993, \apj, 412, 64, \dodoi{10.1086/172900}

\bibitem[{{Laor}(2006)}]{Laor_2006}
{Laor}, A. 2006, \apj, 643, 112, \dodoi{10.1086/502798}

\bibitem[{{Lauer} {et~al.}(2007){Lauer}, {Tremaine}, {Richstone}, \&
  {Faber}}]{Lauer_2007}
{Lauer}, T.~R., {Tremaine}, S., {Richstone}, D., \& {Faber}, S.~M. 2007, \apj,
  670, 249, \dodoi{10.1086/522083}

\bibitem[{{Leung} {et~al.}(2024){Leung}, {Finkelstein},
  {P{\'e}rez-Gonz{\'a}lez}, {Morales}, {Taylor}, {Barro}, {Kocevski}, {Akins},
  {Carnall}, {Ch{\'a}vez Ortiz}, {Cleri}, {Cullen}, {Donnan}, {Dunlop},
  {Ellis}, {Grogin}, {Hirschmann}, {Koekemoer}, {Kokorev}, {Lucas}, {McLeod},
  {Papovich}, \& {Yung}}]{Leung2024}
{Leung}, G. C.~K., {Finkelstein}, S.~L., {P{\'e}rez-Gonz{\'a}lez}, P.~G.,
  {et~al.} 2024, arXiv e-prints, arXiv:2411.12005,
  \dodoi{10.48550/arXiv.2411.12005}

\bibitem[{{Li} {et~al.}(2025{\natexlab{a}}){Li}, {Shen}, \& {Zhuang}}]{Li2025b}
{Li}, J., {Shen}, Y., \& {Zhuang}, M.-Y. 2025{\natexlab{a}}, arXiv e-prints,
  arXiv:2502.05048, \dodoi{10.48550/arXiv.2502.05048}

\bibitem[{{Li} {et~al.}(2025{\natexlab{b}}){Li}, {Silverman}, {Shen},
  {Volonteri}, {Jahnke}, {Zhuang}, {Scoggins}, {Ding}, {Harikane}, {Onoue}, \&
  {Tanaka}}]{Li2025a}
{Li}, J., {Silverman}, J.~D., {Shen}, Y., {et~al.} 2025{\natexlab{b}}, \apj,
  981, 19, \dodoi{10.3847/1538-4357/ada603}

\bibitem[{{Li} {et~al.}(2025{\natexlab{c}}){Li}, {Ho}, \&
  {Chen}}]{Li_Ruancun+2025}
{Li}, R., {Ho}, L.~C., \& {Chen}, C.-H. 2025{\natexlab{c}}, arXiv e-prints,
  arXiv:2505.12867, \dodoi{10.48550/arXiv.2505.12867}

\bibitem[{{Li} {et~al.}(2025{\natexlab{d}}){Li}, {Inayoshi}, {Chen},
  {Ichikawa}, \& {Ho}}]{LiZR2025}
{Li}, Z., {Inayoshi}, K., {Chen}, K., {Ichikawa}, K., \& {Ho}, L.~C.
  2025{\natexlab{d}}, \apj, 980, 36, \dodoi{10.3847/1538-4357/ada5fb}

\bibitem[{{Liddle}(2007)}]{BIC}
{Liddle}, A.~R. 2007, \mnras, 377, L74,
  \dodoi{10.1111/j.1745-3933.2007.00306.x}

\bibitem[{{Limber}(1953)}]{Limber}
{Limber}, D.~N. 1953, \apj, 117, 134, \dodoi{10.1086/145672}

\bibitem[{{Lin} {et~al.}(2024){Lin}, {Wang}, {Fan}, {Cai}, {Champagne}, {Sun},
  {Volonteri}, {Yang}, {Hennawi}, {Ba{\~n}ados}, {Barth}, {Eilers}, {Farina},
  {Liu}, {Jin}, {Jun}, {Lupi}, {Kakiichi}, {Mazzucchelli}, {Onoue}, {Pan},
  {Pizzati}, {Rojas-Ruiz}, {Schindler}, {Trakhtenbrot}, {Shen}, {Trebitsch},
  {Zhuang}, {Endsley}, {Meyer}, {Li}, {Li}, {Pudoka}, {Tee}, {Wu}, \&
  {Zhang}}]{Lin+2024}
{Lin}, X., {Wang}, F., {Fan}, X., {et~al.} 2024, \apj, 974, 147,
  \dodoi{10.3847/1538-4357/ad6565}

\bibitem[{{Lin} {et~al.}(2025{\natexlab{a}}){Lin}, {Fan}, {Wang}, {Sun},
  {Champagne}, {Egami}, {Kakiichi}, {Lyu}, {Tee}, {Yang}, {Bian}, {Bosman},
  {Cai}, {Casey}, {Decarli}, {Faisst}, {Fujimoto}, {Harish}, {Ilbert}, {Inoue},
  {Jin}, {Kartaltepe}, {Kocevski}, {Li}, {Liu}, {Liu}, {Schindler}, {Shuntov},
  {Tanaka}, {Vestergaard}, {Wu}, {Zhang}, \& {Zhang}}]{Lin+2025_C3D_BLAGN}
{Lin}, X., {Fan}, X., {Wang}, F., {et~al.} 2025{\natexlab{a}}, arXiv e-prints,
  arXiv:2504.08039, \dodoi{10.48550/arXiv.2504.08039}

\bibitem[{{Lin} {et~al.}(2025{\natexlab{b}}){Lin}, {Fan}, {Sun}, {Zhang},
  {Egami}, {Helton}, {Wang}, {Zhang}, {Bunker}, {Cai}, {Ji}, {Jin}, {Maiolino},
  {Pudoka}, {Rinaldi}, {Robertson}, {Tacchella}, {Tee}, {Sun}, {Willmer},
  {Willott}, \& {Zhu}}]{Lin_etal_2025b}
{Lin}, X., {Fan}, X., {Sun}, F., {et~al.} 2025{\natexlab{b}}, arXiv e-prints,
  arXiv:2505.02896, \dodoi{10.48550/arXiv.2505.02896}

\bibitem[{{Ma} {et~al.}(2025){Ma}, {Greene}, {Setton}, {Goulding},
  {Annunziatella}, {Fan}, {Kokorev}, {Labbe}, {Li}, {Lin}, {Marchesini},
  {Matthee}, {Robbins}, {Sajina}, {Sawicki}, \& {Telford}}]{Ma+2025}
{Ma}, Y., {Greene}, J.~E., {Setton}, D.~J., {et~al.} 2025, arXiv e-prints,
  arXiv:2504.08032, \dodoi{10.48550/arXiv.2504.08032}

\bibitem[{{Madau} \& {Haardt}(2024)}]{Madau2024}
{Madau}, P., \& {Haardt}, F. 2024, \apjl, 976, L24,
  \dodoi{10.3847/2041-8213/ad90e1}

\bibitem[{{Maiolino} {et~al.}(2024){Maiolino}, {Scholtz}, {Curtis-Lake},
  {Carniani}, {Baker}, {de Graaff}, {Tacchella}, {{\"U}bler}, {D'Eugenio},
  {Witstok}, {Curti}, {Arribas}, {Bunker}, {Charlot}, {Chevallard},
  {Eisenstein}, {Egami}, {Ji}, {Jones}, {Lyu}, {Rawle}, {Robertson},
  {Rujopakarn}, {Perna}, {Sun}, {Venturi}, {Williams}, \&
  {Willott}}]{Maiolino+2024_JADES_BLAGN}
{Maiolino}, R., {Scholtz}, J., {Curtis-Lake}, E., {et~al.} 2024, \aap, 691,
  A145, \dodoi{10.1051/0004-6361/202347640}

\bibitem[{{Maiolino} {et~al.}(2025){Maiolino}, {Risaliti}, {Signorini},
  {Trefoloni}, {Juod{\v{z}}balis}, {Scholtz}, {{\"U}bler}, {D'Eugenio},
  {Carniani}, {Fabian}, {Ji}, {Mazzolari}, {Bertola}, {Brusa}, {Bunker},
  {Charlot}, {Comastri}, {Cresci}, {DeCoursey}, {Egami}, {Fiore}, {Gilli},
  {Perna}, {Tacchella}, \& {Venturi}}]{Maiolino2025}
{Maiolino}, R., {Risaliti}, G., {Signorini}, M., {et~al.} 2025, \mnras, 538,
  1921, \dodoi{10.1093/mnras/staf359}

\bibitem[{{Matthee} {et~al.}(2024{\natexlab{a}}){Matthee}, {Naidu}, {Brammer},
  {Chisholm}, {Eilers}, {Goulding}, {Greene}, {Kashino}, {Labbe}, {Lilly},
  {Mackenzie}, {Oesch}, {Weibel}, {Wuyts}, {Xiao}, {Bordoloi}, {Bouwens}, {van
  Dokkum}, {Illingworth}, {Kramarenko}, {Maseda}, {Mason}, {Meyer}, {Nelson},
  {Reddy}, {Shivaei}, {Simcoe}, \& {Yue}}]{Matthee+24}
{Matthee}, J., {Naidu}, R.~P., {Brammer}, G., {et~al.} 2024{\natexlab{a}},
  \apj, 963, 129, \dodoi{10.3847/1538-4357/ad2345}

\bibitem[{{Matthee} {et~al.}(2024{\natexlab{b}}){Matthee}, {Naidu}, {Kotiwale},
  {Furtak}, {Kramarenko}, {Mackenzie}, {Greene}, {Adamo}, {Bouwens}, {Di
  Cesare}, {Eilers}, {de Graaff}, {Heintz}, {Kashino}, {Maseda}, {Tacchella},
  \& {Torralba}}]{Matthee_etal_2024b}
{Matthee}, J., {Naidu}, R.~P., {Kotiwale}, G., {et~al.} 2024{\natexlab{b}},
  arXiv e-prints, arXiv:2412.02846, \dodoi{10.48550/arXiv.2412.02846}

\bibitem[{{Naidu} {et~al.}(2025){Naidu}, {Matthee}, {Katz}, {de Graaff},
  {Oesch}, {Smith}, {Greene}, {Brammer}, {Weibel}, {Hviding}, {Chisholm},
  {Labb\textbackslash'e}, {Simcoe}, {Witten}, {Atek}, {Baggen}, {Belli},
  {Bezanson}, {Boogaard}, {Bose}, {Covelo-Paz}, {Dayal}, {Fudamoto}, {Furtak},
  {Giovinazzo}, {Goulding}, {Gronke}, {Heintz}, {Hirschmann}, {Illingworth},
  {Inoue}, {Johnson}, {Leja}, {Leonova}, {McConachie}, {Maseda}, {Natarajan},
  {Nelson}, {Setton}, {Shivaei}, {Sobral}, {Stefanon}, {Tacchella}, {Toft},
  {Torralba}, {van Dokkum}, {van der Wel}, {Volonteri}, {Walter}, {Wang}, \&
  {Watson}}]{Naidu_BHstar_2025}
{Naidu}, R.~P., {Matthee}, J., {Katz}, H., {et~al.} 2025, arXiv e-prints,
  arXiv:2503.16596, \dodoi{10.48550/arXiv.2503.16596}

\bibitem[{{Narayanan} {et~al.}(2024){Narayanan}, {Lower}, {Torrey}, {Brammer},
  {Cui}, {Dav{\'e}}, {Iyer}, {Li}, {Lovell}, {Sales}, {Stark}, {Marinacci}, \&
  {Vogelsberger}}]{Narayanan+2024}
{Narayanan}, D., {Lower}, S., {Torrey}, P., {et~al.} 2024, \apj, 961, 73,
  \dodoi{10.3847/1538-4357/ad0966}

\bibitem[{{Natarajan} {et~al.}(2024){Natarajan}, {Pacucci}, {Ricarte},
  {Bogd{\'a}n}, {Goulding}, \& {Cappelluti}}]{Natarajan2024}
{Natarajan}, P., {Pacucci}, F., {Ricarte}, A., {et~al.} 2024, \apjl, 960, L1,
  \dodoi{10.3847/2041-8213/ad0e76}

\bibitem[{{Pacucci} {et~al.}(2023){Pacucci}, {Nguyen}, {Carniani}, {Maiolino},
  \& {Fan}}]{Pacucci2023}
{Pacucci}, F., {Nguyen}, B., {Carniani}, S., {Maiolino}, R., \& {Fan}, X. 2023,
  \apjl, 957, L3, \dodoi{10.3847/2041-8213/ad0158}

\bibitem[{{Papovich} {et~al.}(2001){Papovich}, {Dickinson}, \&
  {Ferguson}}]{Papovich2001}
{Papovich}, C., {Dickinson}, M., \& {Ferguson}, H.~C. 2001, \apj, 559, 620,
  \dodoi{10.1086/322412}

\bibitem[{{Peng} {et~al.}(2002){Peng}, {Ho}, {Impey}, \& {Rix}}]{Peng+2002AJ}
{Peng}, C.~Y., {Ho}, L.~C., {Impey}, C.~D., \& {Rix}, H.-W. 2002, \aj, 124,
  266, \dodoi{10.1086/340952}

\bibitem[{{Peng} {et~al.}(2010){Peng}, {Ho}, {Impey}, \& {Rix}}]{Peng+2010AJ}
---. 2010, \aj, 139, 2097, \dodoi{10.1088/0004-6256/139/6/2097}

\bibitem[{{P{\'e}rez-Gonz{\'a}lez} {et~al.}(2024){P{\'e}rez-Gonz{\'a}lez},
  {Barro}, {Rieke}, {Lyu}, {Rieke}, {Alberts}, {Williams}, {Hainline}, {Sun},
  {Pusk{\'a}s}, {Annunziatella}, {Baker}, {Bunker}, {Egami}, {Ji}, {Johnson},
  {Robertson}, {Rodr{\'\i}guez Del Pino}, {Rujopakarn}, {Shivaei}, {Tacchella},
  {Willmer}, \& {Willott}}]{Perez-Gonzalez+2024}
{P{\'e}rez-Gonz{\'a}lez}, P.~G., {Barro}, G., {Rieke}, G.~H., {et~al.} 2024,
  \apj, 968, 4, \dodoi{10.3847/1538-4357/ad38bb}

\bibitem[{{Perger} {et~al.}(2025){Perger}, {Fogasy}, {Frey}, \&
  {Gab{\'a}nyi}}]{Perger2025}
{Perger}, K., {Fogasy}, J., {Frey}, S., \& {Gab{\'a}nyi}, K.~{\'E}. 2025, \aap,
  693, L2, \dodoi{10.1051/0004-6361/202452422}

\bibitem[{{Pizzati} {et~al.}(2025){Pizzati}, {Hennawi}, {Schaye}, {Eilers},
  {Huang}, {Schindler}, \& {Wang}}]{Pizzati_etal_2025}
{Pizzati}, E., {Hennawi}, J.~F., {Schaye}, J., {et~al.} 2025, \mnras,
  \dodoi{10.1093/mnras/staf660}

\bibitem[{{Reines} \& {Volonteri}(2015)}]{Reines2015}
{Reines}, A.~E., \& {Volonteri}, M. 2015, \apj, 813, 82,
  \dodoi{10.1088/0004-637X/813/2/82}

\bibitem[{{Rinaldi} {et~al.}(2024){Rinaldi}, {Bonaventura}, {Rieke}, {Alberts},
  {Caputi}, {Baker}, {Baum}, {Bhatawdekar}, {Bunker}, {Carniani},
  {Curtis-Lake}, {D'Eugenio}, {Egami}, {Ji}, {Hainline}, {Helton}, {Lin},
  {Lyu}, {Johnson}, {Ma}, {Maiolino}, {P{\'e}rez-Gonz{\'a}lez}, {Rieke},
  {Robertson}, {Shivaei}, {Stone}, {Sun}, {Tacchella}, {{\"U}bler}, {Williams},
  {Willmer}, {Willott}, {Zhang}, \& {Zhu}}]{Rinaldi+2024}
{Rinaldi}, P., {Bonaventura}, N., {Rieke}, G.~H., {et~al.} 2024, arXiv
  e-prints, arXiv:2411.14383, \dodoi{10.48550/arXiv.2411.14383}

\bibitem[{{Rusakov} {et~al.}(2025){Rusakov}, {Watson}, {Nikopoulos}, {Brammer},
  {Gottumukkala}, {Harvey}, {Heintz}, {Nielsen}, {Sim}, {Sneppen}, {Vijayan},
  {Adams}, {Austin}, {Conselice}, {Goolsby}, {Toft}, \&
  {Witstok}}]{Rusakov2025}
{Rusakov}, V., {Watson}, D., {Nikopoulos}, G.~P., {et~al.} 2025, arXiv
  e-prints, arXiv:2503.16595, \dodoi{10.48550/arXiv.2503.16595}

\bibitem[{{Setton} {et~al.}(2024){Setton}, {Greene}, {de Graaff}, {Ma}, {Leja},
  {Matthee}, {Bezanson}, {Boogaard}, {Cleri}, {Katz}, {Labbe}, {Maseda},
  {McConachie}, {Miller}, {Price}, {Suess}, {van Dokkum}, {Wang}, {Weibel},
  {Whitaker}, \& {Williams}}]{Setton2024}
{Setton}, D.~J., {Greene}, J.~E., {de Graaff}, A., {et~al.} 2024, arXiv
  e-prints, arXiv:2411.03424, \dodoi{10.48550/arXiv.2411.03424}

\bibitem[{{Shen}(2013)}]{Shen_2013}
{Shen}, Y. 2013, Bulletin of the Astronomical Society of India, 41, 61,
  \dodoi{10.48550/arXiv.1302.2643}

\bibitem[{{Shen} {et~al.}(2007){Shen}, {Strauss}, {Oguri}, {Hennawi}, {Fan},
  {Richards}, {Hall}, {Gunn}, {Schneider}, {Szalay}, {Thakar}, {Vanden Berk},
  {Anderson}, {Bahcall}, {Connolly}, \& {Knapp}}]{Shen_etal_2007}
{Shen}, Y., {Strauss}, M.~A., {Oguri}, M., {et~al.} 2007, \aj, 133, 2222,
  \dodoi{10.1086/513517}

\bibitem[{{Shen} {et~al.}(2024){Shen}, {Zhuang}, {Li}, {Burgasser}, {Fan},
  {Greene}, {Narayan}, {Shapley}, {Sun}, {Wang}, \& {Yang}}]{nexus}
{Shen}, Y., {Zhuang}, M.-Y., {Li}, J., {et~al.} 2024, arXiv e-prints,
  arXiv:2408.12713, \dodoi{10.48550/arXiv.2408.12713}

\bibitem[{{Sheth} {et~al.}(2001){Sheth}, {Mo}, \& {Tormen}}]{SMT01}
{Sheth}, R.~K., {Mo}, H.~J., \& {Tormen}, G. 2001, \mnras, 323, 1,
  \dodoi{10.1046/j.1365-8711.2001.04006.x}

\bibitem[{{Stern} \& {Laor}(2012)}]{Stern&Laor}
{Stern}, J., \& {Laor}, A. 2012, \mnras, 423, 600,
  \dodoi{10.1111/j.1365-2966.2012.20901.x}

\bibitem[{{Sun} {et~al.}(2025){Sun}, {Rieke}, {Lyu}, {Stone}, {Ji}, {Rinaldi},
  {Willmer}, \& {Zhu}}]{SunY2025}
{Sun}, Y., {Rieke}, G.~H., {Lyu}, J., {et~al.} 2025, \apj, 983, 165,
  \dodoi{10.3847/1538-4357/adc250}

\bibitem[{{Tanaka} {et~al.}(2024){Tanaka}, {Silverman}, {Shimasaku}, {Arita},
  {Akins}, {Inayoshi}, {Ding}, {Onoue}, {Liu}, {Casey}, {Lambrides}, {Kokorev},
  {Jin}, {Faisst}, {Drakos}, {Shen}, {Li}, {Zhuang}, {Fei}, {Ito}, {Ren},
  {Matsui}, {Ando}, {Hatano}, {Fujii}, {Kartaltepe}, {Koekemoer}, {Liu},
  {McCracken}, {Rhodes}, {Robertson}, {Franco}, {Andika}, {Cloonan}, {Fan},
  {Gozaliasl}, {Harish}, {Hayward}, {Huertas-Company}, {Kakkad}, {Kinugawa},
  {Roy}, {Shuntov}, {Talia}, {Toft}, {Vijayan}, \& {Zhang}}]{Tanaka_etal_2024}
{Tanaka}, T.~S., {Silverman}, J.~D., {Shimasaku}, K., {et~al.} 2024, arXiv
  e-prints, arXiv:2412.14246, \dodoi{10.48550/arXiv.2412.14246}

\bibitem[{{Taylor} {et~al.}(2023){Taylor}, {Barger}, {Cowie}, {Hasinger}, {Hu},
  \& {Songaila}}]{HEROES}
{Taylor}, A.~J., {Barger}, A.~J., {Cowie}, L.~L., {et~al.} 2023, \apjs, 266,
  24, \dodoi{10.3847/1538-4365/accd70}

\bibitem[{{Taylor} {et~al.}(2024){Taylor}, {Finkelstein}, {Kocevski}, {Jeon},
  {Bromm}, {Amorin}, {Arrabal Haro}, {Backhaus}, {Bagley}, {Ba{\~n}ados},
  {Bhatawdekar}, {Brooks}, {Calabro}, {Chavez Ortiz}, {Cheng}, {Cleri}, {Cole},
  {Davis}, {Dickinson}, {Donnan}, {Dunlop}, {Ellis}, {Fernandez}, {Fontana},
  {Fujimoto}, {Giavalisco}, {Grazian}, {Guo}, {Hathi}, {Holwerda},
  {Hirschmann}, {Inayoshi}, {Kartaltepe}, {Khusanova}, {Koekemoer}, {Kokorev},
  {Larson}, {Leung}, {Lucas}, {McLeod}, {Napolitano}, {Onoue}, {Pacucci},
  {Papovich}, {P{\'e}rez-Gonz{\'a}lez}, {Pirzkal}, {Somerville}, {Trump},
  {Wilkins}, {Yung}, \& {Zhang}}]{Taylor+2024}
{Taylor}, A.~J., {Finkelstein}, S.~L., {Kocevski}, D.~D., {et~al.} 2024, arXiv
  e-prints, arXiv:2409.06772, \dodoi{10.48550/arXiv.2409.06772}

\bibitem[{{Taylor} {et~al.}(2025){Taylor}, {Kokorev}, {Kocevski}, {Akins},
  {Cullen}, {Dickinson}, {Finkelstein}, {Arrabal Haro}, {Bromm}, {Giavalisco},
  {Inayoshi}, {Juneau}, {Leung}, {Perez-Gonzalez}, {Somerville}, {Trump},
  {Amorin}, {Barro}, {Burgarella}, {Brooks}, {Carnall}, {Casey}, {Cheng},
  {Chisholm}, {Chworowsky}, {Davis}, {Donnan}, {Dunlop}, {Ellis}, {Fernandez},
  {Fujimoto}, {Grogin}, {Gupta}, {Hathi}, {Jung}, {Hirschmann}, {Kartaltepe},
  {Koekemoer}, {Larson}, {Leung}, {Llerena}, {Lucas}, {McLeod}, {McLure},
  {Napolitano}, {Papovich}, {Stanton}, {Tripodi}, {Wang}, {Wilkins}, {Yung}, \&
  {Zavala}}]{Taylor2025}
{Taylor}, A.~J., {Kokorev}, V., {Kocevski}, D.~D., {et~al.} 2025, arXiv
  e-prints, arXiv:2505.04609, \dodoi{10.48550/arXiv.2505.04609}

\bibitem[{{Tee} {et~al.}(2025){Tee}, {Fan}, {Wang}, \& {Yang}}]{Tee2025}
{Tee}, W.~L., {Fan}, X., {Wang}, F., \& {Yang}, J. 2025, \apjl, 983, L26,
  \dodoi{10.3847/2041-8213/adc5e3}

\bibitem[{{Virtanen} {et~al.}(2020){Virtanen}, {Gommers}, {Oliphant},
  {Haberland}, {Reddy}, {Cournapeau}, {Burovski}, {Peterson}, {Weckesser},
  {Bright}, {van der Walt}, {Brett}, {Wilson}, {Millman}, {Mayorov}, {Nelson},
  {Jones}, {Kern}, {Larson}, {Carey}, {Polat}, {Feng}, {Moore}, {VanderPlas},
  {Laxalde}, {Perktold}, {Cimrman}, {Henriksen}, {Quintero}, {Harris},
  {Archibald}, {Ribeiro}, {Pedregosa}, {van Mulbregt}, \& {SciPy 1. 0
  Contributors}}]{scipy}
{Virtanen}, P., {Gommers}, R., {Oliphant}, T.~E., {et~al.} 2020, Nature
  Methods, 17, 261, \dodoi{10.1038/s41592-019-0686-2}

\bibitem[{{Volonteri} {et~al.}(2025){Volonteri}, {Trebitsch}, {Greene},
  {Dubois}, {Dong-Paez}, {Habouzit}, {Lupi}, {Ma}, {Beckmann}, {Dayal}, \&
  {Schneider}}]{Volonteri2025}
{Volonteri}, M., {Trebitsch}, M., {Greene}, J.~E., {et~al.} 2025, \aap, 695,
  A33, \dodoi{10.1051/0004-6361/202451963}

\bibitem[{{Wang} {et~al.}(2024){Wang}, {Leja}, {Atek}, {Labb{\'e}}, {Li},
  {Bezanson}, {Brammer}, {Cutler}, {Dayal}, {Furtak}, {Greene}, {Kokorev},
  {Pan}, {Price}, {Suess}, {Weaver}, {Whitaker}, \& {Williams}}]{Wang_BJ+2024}
{Wang}, B., {Leja}, J., {Atek}, H., {et~al.} 2024, \apj, 963, 74,
  \dodoi{10.3847/1538-4357/ad187c}

\bibitem[{{Wang} {et~al.}(2025){Wang}, {de Graaff}, {Davies}, {Greene}, {Leja},
  {Brammer}, {Goulding}, {Miller}, {Suess}, {Weibel}, {Williams}, {Bezanson},
  {Boogaard}, {Cleri}, {Hirschmann}, {Katz}, {Labb{\'e}}, {Maseda}, {Matthee},
  {McConachie}, {Naidu}, {Oesch}, {Rix}, {Setton}, \&
  {Whitaker}}]{WangBJ2025lrd}
{Wang}, B., {de Graaff}, A., {Davies}, R.~L., {et~al.} 2025, \apj, 984, 121,
  \dodoi{10.3847/1538-4357/adc1ca}

\bibitem[{{Wang} {et~al.}(2023){Wang}, {Yang}, {Hennawi}, {Fan}, {Sun},
  {Champagne}, {Costa}, {Habouzit}, {Endsley}, {Li}, {Lin}, {Meyer},
  {Schindler}, {Wu}, {Ba{\~n}ados}, {Barth}, {Bhowmick}, {Bieri}, {Blecha},
  {Bosman}, {Cai}, {Colina}, {Connor}, {Davies}, {Decarli}, {De Rosa}, {Drake},
  {Egami}, {Eilers}, {Evans}, {Farina}, {Haiman}, {Jiang}, {Jin}, {Jun},
  {Kakiichi}, {Khusanova}, {Kulkarni}, {Li}, {Liu}, {Loiacono}, {Lupi},
  {Mazzucchelli}, {Onoue}, {Pudoka}, {Rojas-Ruiz}, {Shen}, {Strauss}, {Tee},
  {Trakhtenbrot}, {Trebitsch}, {Venemans}, {Volonteri}, {Walter}, {Xie}, {Yue},
  {Zhang}, {Zhang}, \& {Zou}}]{ASPIRE}
{Wang}, F., {Yang}, J., {Hennawi}, J.~F., {et~al.} 2023, \apjl, 951, L4,
  \dodoi{10.3847/2041-8213/accd6f}

\bibitem[{{Wu} \& {Shen}(2022)}]{Wu_DR16Q}
{Wu}, Q., \& {Shen}, Y. 2022, \apjs, 263, 42, \dodoi{10.3847/1538-4365/ac9ead}

\bibitem[{{Yue} {et~al.}(2024){Yue}, {Eilers}, {Ananna}, {Panagiotou}, {Kara},
  \& {Miyaji}}]{Yue2024xray}
{Yue}, M., {Eilers}, A.-C., {Ananna}, T.~T., {et~al.} 2024, \apjl, 974, L26,
  \dodoi{10.3847/2041-8213/ad7eba}

\bibitem[{{Zhang} {et~al.}(2025{\natexlab{a}}){Zhang}, {Wu}, {Fan}, {Ho}, {Wu},
  {Zhang}, {Lyu}, {Cao}, \& {Wang}}]{Zhang2025sed}
{Zhang}, C., {Wu}, Q., {Fan}, X., {et~al.} 2025{\natexlab{a}}, arXiv e-prints,
  arXiv:2505.12719.
\newblock \doarXiv{2505.12719}

\bibitem[{{Zhang} {et~al.}(2025{\natexlab{b}}){Zhang}, {Egami}, {Sun}, {Lin},
  {Lyu}, {Zhu}, {Rinaldi}, {Sun}, {Bunker}, {Bhatawdekar}, {Helton},
  {Maiolino}, {Ma}, {Robertson}, {Tacchella}, {Venturi}, {Williams}, \&
  {Willott}}]{ZhangJY2025}
{Zhang}, J., {Egami}, E., {Sun}, F., {et~al.} 2025{\natexlab{b}}, arXiv
  e-prints, arXiv:2505.02895, \dodoi{10.48550/arXiv.2505.02895}

\bibitem[{{Zhang} {et~al.}(2024){Zhang}, {Jiang}, {Liu}, \& {Ho}}]{ZhangZJ2024}
{Zhang}, Z., {Jiang}, L., {Liu}, W., \& {Ho}, L.~C. 2024, arXiv e-prints,
  arXiv:2411.02729, \dodoi{10.48550/arXiv.2411.02729}

\bibitem[{{Zhuang} \& {Ho}(2022)}]{Zhuang&Ho2022}
{Zhuang}, M.-Y., \& {Ho}, L.~C. 2022, \apj, 934, 130,
  \dodoi{10.3847/1538-4357/ac7aaf}

\bibitem[{{Zhuang} \& {Ho}(2023)}]{Zhuang&Ho2023}
---. 2023, NatAs, \dodoi{10.1038/s41550-023-02051-4}

\bibitem[{{Zhuang} {et~al.}(2024{\natexlab{a}}){Zhuang}, {Li}, \&
  {Shen}}]{Zhuang+2024}
{Zhuang}, M.-Y., {Li}, J., \& {Shen}, Y. 2024{\natexlab{a}}, \apj, 962, 93,
  \dodoi{10.3847/1538-4357/ad1517}

\bibitem[{{Zhuang} \& {Shen}(2024)}]{Zhuang&Shen2024}
{Zhuang}, M.-Y., \& {Shen}, Y. 2024, \apj, 962, 139,
  \dodoi{10.3847/1538-4357/ad1183}

\bibitem[{{Zhuang} {et~al.}(2024{\natexlab{b}}){Zhuang}, {Wang}, {Sun}, {Shen},
  {Li}, {Burgasser}, {Fan}, {Greene}, {Narayan}, {Shapley}, \&
  {Yang}}]{nexus-edr}
{Zhuang}, M.-Y., {Wang}, F., {Sun}, F., {et~al.} 2024{\natexlab{b}}, arXiv
  e-prints, arXiv:2411.06372, \dodoi{10.48550/arXiv.2411.06372}

\end{thebibliography}
\bibliographystyle{aasjournal}



\end{document}